\documentclass[acmsmall]{acmart} % CSCW camera ready
\AtBeginDocument{%
  \providecommand\BibTeX{{%
    \normalfont B\kern-0.5em{\scshape i\kern-0.25em b}\kern-0.8em\TeX}}}

\setcopyright{acmcopyright}
\copyrightyear{2024}
\acmYear{2024}
\acmDOI{XXXXXXX.XXXXXXX}

\acmPrice{15.00}
\acmISBN{978-1-4503-XXXX-X/18/06}

\usepackage{csquotes}
\usepackage{tabularx}
\usepackage{multirow}
\usepackage{multicol}
\usepackage{colortbl}
\usepackage{arydshln}
\usepackage{dblfloatfix}
\usepackage{float}

\begin{document}

\title[Thoughtful Adoption of NLP for Civic Participation]{Thoughtful Adoption of NLP for Civic Participation:\\Understanding Differences Among Policymakers}


\author{Jose A. Guridi}
\email{jg2222@cornell.edu}
\orcid{0000-0003-0543-699X}
\affiliation{%
  \institution{Cornell University}
  \city{Ithaca}
  \country{United States}}

\author{Cristobal Cheyre}
\email{cac555@cornell.edu}
\orcid{0000-0002-1221-1978}
\affiliation{%
  \institution{Cornell University}
  \city{Ithaca}
  \country{United States}
}

\author{Qian Yang}
\email{qy242@cornell.edu}
\orcid{0000-0002-3548-2535}
\affiliation{%
 \institution{Cornell University}
 \city{Ithaca}
 \country{United States}}

%%
%% By default, the full list of authors will be used on the page
%\renewcommand{\shortauthors}{Guridi et al.}

\begin{abstract}
Natural language processing (NLP) tools have the potential to boost civic participation and enhance democratic processes because they can significantly increase governments’ capacity to gather and analyze citizen opinions. 
However, their adoption in government remains limited, and harnessing their benefits while preventing unintended consequences remains a challenge.
While prior work has focused on improving NLP performance, this work examines how different internal government stakeholders influence NLP tools' thoughtful adoption.
We interviewed seven politicians (politically appointed officials as heads of government institutions) and thirteen public servants (career government employees who design and administrate policy interventions), inquiring how they choose whether and how to use NLP tools to support civic participation processes.
The interviews suggest that policymakers across both groups focused on their needs for career advancement and the need to showcase the legitimacy and fairness of their work when considering NLP tool adoption and use.
Because these needs vary between politicians and public servants, their preferred NLP features and tool designs also differ.
Interestingly, despite their differing needs and opinions, neither group clearly identifies who should advocate for NLP adoption to enhance civic participation or address the unintended consequences of a poorly considered adoption. This lack of clarity in responsibility might have caused the governments' low adoption of NLP tools.
We discuss how these findings reveal new insights for future HCI research. They inform the design of NLP tools for increasing civic participation efficiency and capacity, the design of other tools and methods that ensure thoughtful adoption of AI tools in government, and the design of NLP tools for collaborative use among users with different incentives and needs.
\end{abstract}





\received{15 January 2024}
\received[revised]{16 July 2024}
 \received[accepted]{}

\maketitle

\section{Introduction}
Natural language processing (NLP) systems can increase policymakers' ability to analyze citizens' comments on policy issues or drafts, a process known as civic participation in policymaking. 
In prior research, NLP systems have shown promising results in helping policymakers to organize and make sense of large volumes of citizen comments and opinions and even in helping policymakers to respond to constituents promptly~\cite{arana-catania_citizen_2021, chen_barriers_2019, romberg_automated_2022, hagen_content_2018, romberg_making_2023,guridi_supporting_2024}.
However, NLP tool adoption in government remains limited, perhaps because of the concerns that careless deployment of NLP tools can harm the democratic processes and erode public trust in government~\cite{chen_barriers_2019, birhane_algorithmic_2021, sloane_participation_2022,delgado_participatory_2023,jasim_communitypulse_2021}. 
How can the design of NLP tools and their use protocols effectively enhance civic participation capacity and efficiency while mitigating potential risks and unintended consequences?

CSCW research has long focused on how technology influences collaborative work~\cite{kozlowski_taxonomy_2012} and how the design of new socio-technical systems can address conflicting stakeholder needs and enhance teamwork~\cite[e.g., ][]{wong_tactics_2021, yildirim_how_2022, madaio_co-designing_2020, passi_trust_2018, quinones_cultivating_2014, rose_arguing_2016}.
These complexities also apply to adopting and using NLP tools in government.
Prior research has highlighted the sometimes conflicting interest among government's internal and external stakeholders~\cite{scholl_involving_2004, flak_stakeholders_2006, bretschneider_management_1990, boyne_public_2002, persson_government_2010, de_rituals_2010}. 
These organizational dynamics differ from those in the private sector because value within government is primarily associated with how citizens perceive and evaluate policy outcomes rather than corporate interest~\cite{boyne_public_2002, bretschneider_management_1990, rose_managing_2015, bonina_public_2009, persson_government_2010}.
% Moreover, the existence of multiple political visions makes legitimacy a particularly contested arena since what the government should do and how it is done depends on different stakeholders' normative perspectives~\cite{rose_stakeholder_2018, rowley_e-government_2011, rose_managing_2015, dean_deliberating_2023}. % Qian: Removing this sentence since it seems to give away our findings.
Interestingly, despite CSCW's increasing interest in civic tech and governments~\cite{aragon_civic_2020, stapleton_who_2022}, little research has studied the internal dynamics of governments around NLP tools' adoption and use.

This paper asks: \textit{How do politicians and public servants---the two primary groups involved in policymaking within government---consider whether and how to use NLP tools to enhance civic participation? How do they consider possible ways to address potential risks and challenges?}
We focused on two Latin American countries--Chile and Uruguay--as our study sites.
Latin America has a long-standing civic participatory tradition. As most countries transitioned to democracy in the last 30 years, it is a fertile ground for studying the use and innovation of technology for democratic processes~\cite{pogrebinschi_thirty_2021}. 
Yet, the canon research literature---in HCI and beyond---has rarely studied these countries~\cite{Reynolds-Cuellar2022, wong-villacres_lessons_2021,pogrebinschi_thirty_2021}.
With HCI and CSCW communities increasingly advocating for the inclusion of the Global South in their research agendas~\cite{Reynolds-Cuellar2022, wong-villacres_lessons_2021, talhouk_re-articulating_2023, alvarado_garcia_fostering_2020, reynolds-cuellar_para_2023}, now is an opportune time to address this critical gap.

We conducted an interview study with seven politicians and thirteen public servants from five ministries in Chile and a public agency in Uruguay. 
Our findings suggest that policymakers across both groups focused on their needs for career advancement and the need to strengthen the legitimacy of their work when considering NLP tools' adoption and use.
Because these needs vary between politicians and public servants, their preferred NLP features and tool designs also differ.
Politicians' legitimacy was rooted in their constituents' trust and approval; therefore, they wanted NLP features that could showcase the fairness and capacity of their public participation processes and the use of cutting-edge technologies.
Public servants, in contrast, were internally oriented. Their jobs' legitimacy came from the approval of their superiors; consequently, they looked for NLP features that could help them become both more efficient and more empathetic in analyzing citizen comments at the current volume. 
Interestingly, despite their differing needs and opinions, neither group clearly identified who should advocate for NLP adoption to enhance civic participation or address the unintended consequences of a poorly considered adoption. Instead, they blamed each other for NLP tools' low adoption in government.

Our study makes two key contributions. 
First, it provides a rich description of how different policymakers in Latin America considered the adoption of NLP tools and their use in civic participation processes. It offers a valuable reference point for future research on similar topics in other government contexts and for innovating NLP tools for the public sector.
Second, it challenges the common assumptions in HCI and AI literature that the adoption of AI tools in government is solely about algorithmic performance or fairness and that "policymakers" are a homogeneous group. By illuminating the complexities and differences among various types of policymakers, this work opens up new research and design opportunities for AI tools in the public sector.

\section{Related work}

\subsection{Civic Participation in Policymaking}
In this paper, we use the term ``\textit{civic participation}'' to broadly refer to the processes by which individuals and organizations can influence the design and implementation of governmental policies~\cite{irvin_citizen_2004, baum_citizen_2015, arnstein_ladder_1969, delgado_participatory_2023}. The level of empowerment and mode of engagement may vary based on the government and policy contexts~\cite{Yang_HCIPolicy_CHI24}.

Civic participation in policy rule-making and implementation offers many important benefits. For citizens, it serves as a crucial component of civic education and engagement and increases trust in governmental processes~\cite{aitamurto_value_2017, koc-michalska_digital_2017}.
For policymakers, civic participation processes provide access to a broad range of experience-based knowledge at a low cost, enabling them to create better policy~\cite{aitamurto_value_2017, liu_crowdsourcing_2017, pecaric_can_2017, aitamurto_unmasking_2017, arana-catania_citizen_2021}.
At a higher level, civic engagement enhances the inclusiveness, transparency, and accountability of democratic processes; it is a pillar of democracy itself~\cite{arana-catania_citizen_2021,aitamurto_value_2017,arnstein_ladder_1969}.

However, poorly implemented civic participation processes can cause more harm than good. 
For instance, when the participation process---or the use of AI tools within it---is unfair or opaque, it could inadvertently amplify the voices of powerful organizations or individuals, harming the less powerful citizens~\cite{arnstein_ladder_1969,birhane_power_2022,delgado_participatory_2023}.
Over time, if citizens perceive the government as a passive listener that neither properly analyzes information nor responds and acts promptly, it could erode the public trust in both the participatory process and the government agency itself~\cite{aitamurto_crowdsourced_2016, chen_barriers_2019, esaiasson_will_2010, strebel_importance_2019, schmidt_democracy_2013, mendelson_should_2012,birhane_power_2022,sloane_participation_2020}. 
In summary, the benefits of incorporating rich, diverse citizen input in policymaking must be balanced with the need to maintain transparency and fairness.

This dual requirement poses significant challenges to policymakers and public institutions, who often struggle to read or synthesize the large volume and wide variety of citizen inputs promptly~\cite{aitamurto_crowdsourced_2016, arana-catania_citizen_2021, jasim_communitypulse_2021, livermore_computationally_2017, mahyar_civic_2019, reynante_framework_2021, simonofski_supporting_2021, chen_barriers_2019}.
This issue is especially pronounced when dealing with unstructured citizen input, such as free texts~\cite{chun_government_2010, janssen_benefits_2012, font_tracing_2016}.
To address these challenges, policymakers use heuristics to filter citizen inputs and prioritize those deemed more substantive. However, this approach has been criticized for potentially dismissing too many contributions, thereby biasing policy outcomes~\cite{perez_complexity_2008,chen_barriers_2019, roetzel_information_2019}.

\subsection{NLP Tools that Facilitate Civic Participation Processes}

% This paragraph: The exciting promises of NLP tools
With the rapid advances in AI in recent years, NLP tools have become capable of assisting policymakers in processing and synthesizing citizen inputs in many valuable ways~\cite[e.g., ][]{romberg_making_2023, romberg_automated_2022,weng_ai_2021,simonofski_supporting_2021,guridi_supporting_2024,kim_improving_2021}.
Prior NLP research has demonstrated these tools' usefulness in pre-processing data (e.g., voice-to-text transcription, detecting duplicates, identifying spams), summarizing data (e.g., identifying topics and themes in citizen comments), and clustering data (e.g., grouping citizen comments with similar sentiments) within civic participation contexts~\cite{chen_barriers_2019,romberg_automated_2022,romberg_making_2023,arana-catania_citizen_2021}.
In recent years, Large Language Models (LLMs) have greatly improved NLP models' performances across various tasks, further increasing the potential of NLP tools to assist policymakers~\cite{lam_concept_2024}.
However, adopting NLP tools in civic participation involves complexities beyond model performance or tool usefulness~\cite{tangi_challenges_2023, wirtz_artificial_2019}.
To ensure public trust in the civic participation process, NLP tools need to ensure that they offer sufficient transparency to the public~\cite{edwards_slave_2017, liao_designing_2022, wang_designing_2019, rechkemmer_when_2022, tangi_challenges_2023}, treat diverse citizen comments fairly~\cite{tangi_challenges_2023, levy_algorithms_2021, jacobs_meaning_2020}, and protect citizen privacy~\cite{levy_algorithms_2021, wirtz_artificial_2019, tangi_challenges_2023}.
% Adopting technology in government is different from the private sector. ensure sufficient human policymaker involvement~\cite{chalmers_seamful_2004, susser_invisible_2019, tolmie_unremarkable_2002}, 
Even after these basic requirements are met, the decisions on whether governments should adopt NLP tools and how to use those tools in practice are contentious ones, highly dependent on different stakeholders' normative perspectives~\cite{rose_stakeholder_2018, rowley_e-government_2011, rose_managing_2015}.
After all, the legitimacy of democratic governments fundamentally rests on the intrinsic right of citizens to be served by their representatives~\cite{rose_managing_2015,saebo_understanding_2011}. 
Therefore, how citizens perceive NLP tool use and the outcomes of policymaking processes is paramount~\cite{boyne_public_2002, bretschneider_management_1990, rose_managing_2015, bonina_public_2009, persson_government_2010}.

% This paragraph: The risks of using NLP tools in the public sector (2) organizational complexities within gov (i.e., internal stakeholders)
Using NLP tools in civic participation also involves significant complexities within government institutions.
Prior research suggests that governments often lack the expertise, infrastructure, and adoption mechanisms for thoughtfully adopting AI systems themselves~\cite{wirtz_artificial_2019, tangi_challenges_2023, giest_big_2017, poel_big_2018}. Consequently, they often rely on third-party vendors to provide NLP systems through procurement processes. Even so, policymakers within government institutions frequently lack the expertise to effectively evaluate these systems and do not have access to proper guidelines for doing so~\cite{nagitta_human-centered_2022, dor_procurement_2021, sloane_ai_2021, madan_ai_2023}.

Moreover, internal tensions within government institutions can further complicate the adoption and use of NLP tools~\cite{scholl_involving_2004, flak_stakeholders_2006, bretschneider_management_1990, boyne_public_2002, persson_government_2010, de_rituals_2010}. 
Policymakers may have differing opinions based on their roles and affiliations~\cite{rowley_e-government_2011, rose_stakeholder_2018}. Prior research on policymaking processes suggests at least four types of internal stakeholders in government~\cite{kawakami_studying_2024, orange_care_2007,rowley_e-government_2011,saebo_understanding_2011,dean_deliberating_2023}:

\begin{itemize}
    \item Politicians~are authorities within government institutions elected or designated by a central elected authority such as the president, governor, or mayor. They are usually the heads of government institutions;
    \item Public servants~are administrators and advisors working at public institutions whose primary function is to design and implement policies following their institution's mandate and the government's political agenda. They may be hired by the politician and/or share their political agenda, but this is only sometimes the case, depending on the position, role, and type of contract.
    \item Support staff~are public employees performing day-to-day administrative tasks such as running administrative procedures, maintaining the IT infrastructure, running legal and financial controls, etc.
    \item Front-line workers~are government employees that usually have occupations that bring them in direct contact with the communities their institutions serve. Their roles are not designing policies but interacting with the public to provide their institutions' services.
\end{itemize}

Interestingly, despite CSCW's increasing interest in civic tech and governments~\cite{aragon_civic_2020, stapleton_who_2022}, little research has studied these internal dynamics of governments around NLP tools' adoption and use~\cite{kawakami_situate_2024, kawakami_studying_2024, kawakami_improving_2022, saxena_framework_2021, levy_algorithms_2021}.
One exception is Kawakami et al.~\cite{kawakami_studying_2024}, who analyzed how the interactions between front-line workers and agency leaders influence AI design and adoption decisions in public agencies.
This paper adds to this emergent line of research.

% Moreover, our work examines central government institutions, which have different political dynamics than prior works' settings, such as public agencies~\cite{kawakami_studying_2024,dean_deliberating_2023}.
% The current lack of knowledge can lead to ill-implemented participatory processes that harm people~\cite{delgado_participatory_2023,jasim_communitypulse_2021,sloane_participation_2022} and can prevent the implementation of HCI approaches to responsible AI into real-world practice in the public sector~\cite{kawakami_studying_2024, kawakami_situate_2024}.

\section{Method}

\subsection{Participants}
We wanted to understand how different stakeholders within government institutions consider whether and how to use NLP tools to enhance civic participation, as well as how they consider possible ways to address potential risks and challenges. We hope these insights will inform better NLP tool design and illuminate new social practices to ensure thoughtful adoption and use of these tools.
To extend prior research that focused on public agencies, we chose to focus on the internal stakeholder dynamics within central government institutions.
Instead of studying front-line workers, we focus on politicians and public servants---the two primary groups involved in policymaking within government.
Additionally, we focused on two Latin American countries--Chile and Uruguay--as our study sites.

\begin{figure*}
    \centering
    \includegraphics[width = 0.95\textwidth]{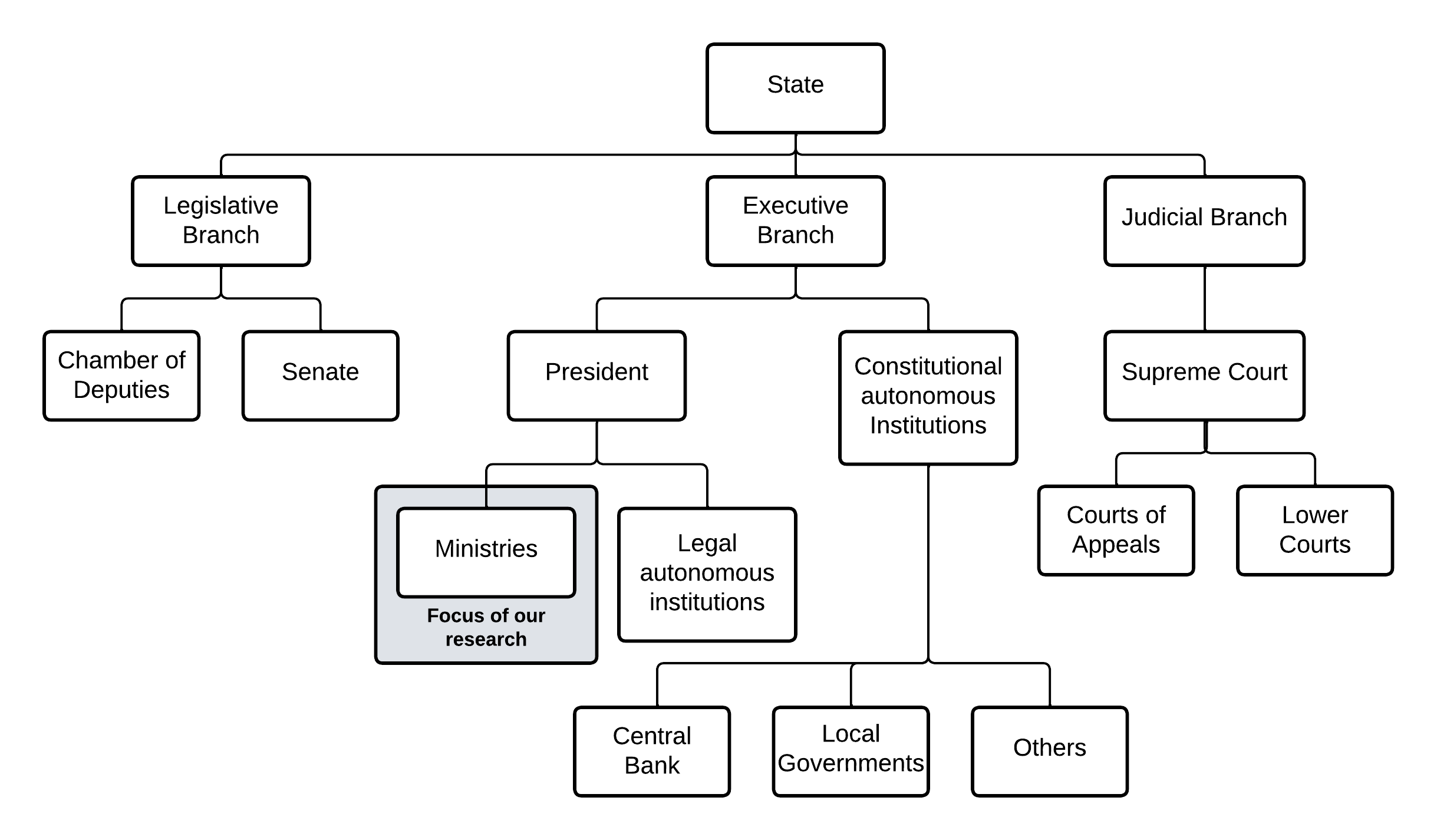}
    \caption{An illustration of the role of interviewees' institutions within the overall structure of Chile's and Uruguay's government frameworks.}
    \label{fig:state}
    \Description{An illustration of the role of interviewees' institutions within the overall structure of Chile's and Uruguay's government frameworks.} 
    %At the top is the State, which is divided into three branches: legislative, executive, and judicial. The legislative branch is divided into a Chamber of Deputies and a Senate. The judicial branch has two levels, on the top one, it has the Supreme Court. On the bottom one, there are Courts of Appeal and Lower Courts. The executive branch is divided into the president, who oversees ministries, which are the focus of the study, and legal autonomous institutions linked to the central government. Moreover, the executive branch has constitutional autonomous institutions such as the Central Bank and Local Governments.
\end{figure*}

\begin{figure*}%[h]
    \centering
    \includegraphics[width = 0.9\textwidth]{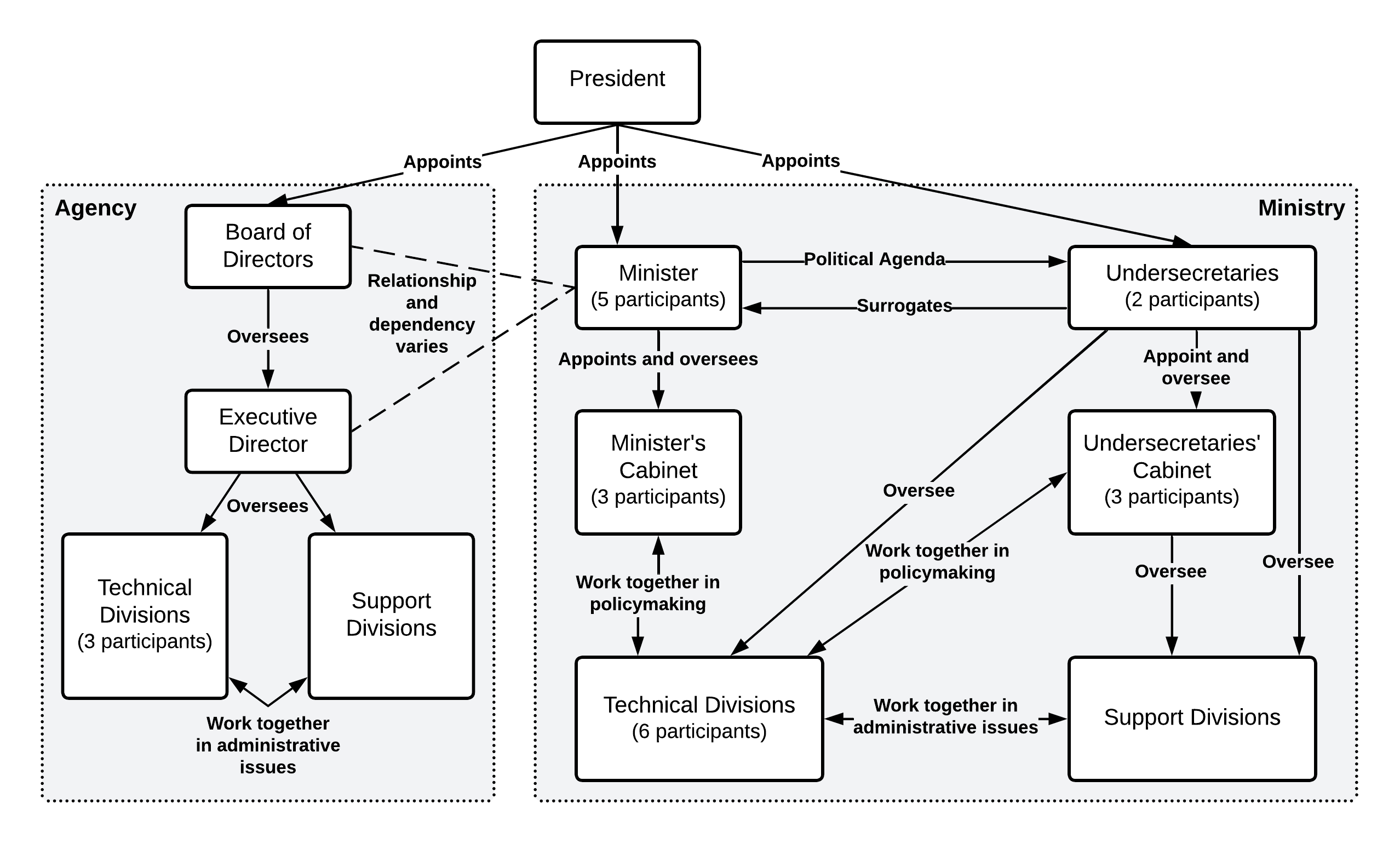}
    \caption{A simplified and generalized illustration of the Ministries and Agencies structure in Chile and Uruguay. Details can vary depending on the specific Ministry and Agency.}
    \Description{This figure zooms into the generic structure of agencies and ministries. At the top is the president, who appoints boards of directors in agencies, ministers in ministries, and undersecretaries within ministries. In agencies, the board oversees an executive director who, in turn, oversees technical divisions and support divisions that work together on administrative issues. The technical division works on policymaking, and three participants were in an Agency's technical division. In ministries, both ministries and undersecretaries have their cabinets. Ministers define the political agenda for undersecretaries and are surrogated by them. We include five ministers and two undersecretaries. Moreover, we included three participants from Ministers' cabinets and three from undersecretary cabinets. Policymakers in cabinets work with the technical divisions in policymaking. We included six participants from technical divisions. Finally, the support division is oversaw by the undersecretary and its cabinet and work together with the technical division in administrative issues.}
    \vspace{-0.2cm}
    % \caption{Institutional structure}
    \label{fig:institutions}
\end{figure*}

% This paragraph: Describe where we recruited participants from.
We conducted IRB-approved semi-structured interviews with 20 politicians and policymakers. 
We recruited participants from five ministries in Chile and one agency in Uruguay. 
These institutions are all under the executive branch of government, which more often manages civic participation processes than other branches.
The institutions vary in terms of legal mandate, the type of policies they develop, and how they conduct participation processes. 
Interviewees were initially recruited from contacts of one of the authors who served in a fraction of both government periods in Chile and increased through snowball sampling. Table~\ref{tab:institutions} details the institutions from which we recruited the participants. Figure~\ref{fig:state} illustrates the role of these institutions within the overall structure of Chile's and Uruguay's government frameworks. Appendix~\ref{appendix_chile_uruguay} offers additional context about the countries, their institutions, and their civic participation processes. Within the appendix, figure~\ref{fig:institutions} provides an overview of how institutions are organized and where our participants worked within them.
% Within ministries, we have a varied pool of participants from institutions working on different topics (see table~\ref{tab:institutions}) and working both at cabinets and technical divisions.
% We decided not to include people from support divisions since they are not directly involved in policymaking or conducting the participation process.
% Figure~\ref{fig:institutions} provides an overview of how institutions work and where our participants were when we interviewed them.  
\begin{table*}[h!]
\centering
    \begin{tabular}{p{0.6\textwidth}p{0.15\textwidth}<{
\centering} p{0.18\textwidth}<{
\centering} }
    \toprule
     \textbf{Policymaking Institution} & \textbf{Politician \newline Interviewees} & \textbf{Public Servant \newline Interviewees}\\
     \midrule
     Uruguay - Agency for Electronic Government and Information and Knowledge Society & 0 & 3 \\ % 3 in total
     Chile - Ministry General Secretary of the Presidency & 1 & 3 \\  % 4 in total
     Chile - Ministry of Economy, Development, and Tourism & 2 & 1\\  % 3 in total
     Chile - Ministry of Education & 1 & 1\\ % 2 in total
     Chile - Ministry General Secretary of Government & 1 & 1 \\ % 2 in total
     Chile - Ministry of Science, Technology, Knowledge and Innovation & 2 & 4 \\  % 6 in total
     \bottomrule
    \end{tabular}
\vspace{0.2cm}
\caption{The institutions our interviewees come from. We interviewed 20 politicians and public servants from six different government agencies across two Latin American countries.}~\label{tab:institutions}
%\description{The institutions our interviewees come from. We interviewed 20 politicians and public servants from six different government agencies across two Latin America countries. One participant served as a public servant and politician at different times.}
\end{table*}
  %     \begin{tabular}{c c c}
  %  \toprule
  %    & \textbf{Experience in Government} & \\
  %   \midrule
  %   0-5 years &  & 11\\
  %   5-10 years &  & 5\\
  %   10+ years &  & 4\\
  %   \midrule
  %    & \textbf{Education Level} & \\
  %    \midrule
  %   Bachelor &  & 5\\
  %   Master &  & 11\\
  %   Ph.D. &  & 4\\
  %  \midrule
  %    & \textbf{Previous Experience Using NLP (yes/no)} & \\
  %    \midrule
  %   Politicians  & & 4/7\\
  %   Policymakers &  & 4/13\\
  %   \bottomrule
  %  \end{tabular}

% This paragraph: Describe how we selected the participants.
We selected participants who served during the last two government periods in Chile and Uruguay from these institutions.
To collect diverse opinions, we selected participants from opposing political coalitions\footnote{The last two governments in both countries have been from opposing political coalitions. Some participants were part of only one of the two, and others served in both. We cannot provide specific details about who is who to protect their anonymity, but there were no differences in results when comparing both groups.} and across cabinets and technical divisions.
Additionally, we selected politicians and policymakers who worked concurrently for at least part of their appointments, which allowed us to contrast their visions and experiences. 
Table~\ref{tab:participants} provides details on the participants. 

\begin{table*}[ht]
    \centering
% Politicians
\resizebox{0.85\textwidth}{!}{
    \begin{tabular}{p{0.15\linewidth} p{0.15\linewidth} p{0.1\linewidth} p{0.20\linewidth}<{
\centering} p{0.25\linewidth}<{
\centering}}
    \toprule
       \multicolumn{5}{c}{\cellcolor{lightgray!20!}\textbf{Politicians}}\\
       \hline
       \textbf{Participant} & \textbf{Education} & \textbf{Gender} & \textbf{YRs in Gov} & \textbf{Experience Using NLP} \\
       \hline
       PT1 & Master & Male & 5-10 & No\\
       PT2 & Master & Male & 10+ & No \\
       PT3 & Ph.D. & Male & 0-5 & Yes \\
       PT4 \& PT5 & Master & Male & 5-10 & Yes \\
       PT6 & Ph.D. & Male & 0-5 & No \\
       PT7 & Ph.D. & Female & 0-5 & Yes \\
    \bottomrule
    \end{tabular}
}
    \vspace{0.1cm}
%Policymakers

\resizebox{0.85\textwidth}{!}{
    \begin{tabular}{p{0.15\linewidth} p{0.15\linewidth} p{0.1\linewidth} p{0.2\linewidth}<{
\centering} p{0.25\linewidth}<{
\centering}}
    \toprule
       \multicolumn{5}{c}{\cellcolor{lightgray!20!}\textbf{Public Servants}}\\
       \hline
       \textbf{Participant} & \textbf{Education} & \textbf{Gender} & \textbf{YRs in Gov} & \textbf{Experience Using NLP} \\
       \hline
       PS1 & Master & Female & 5-10 & No\\
       PS2 \& PS9 & Master & Female & 10+ & No \\
       PS3 & Master & Male & 0-5 & Yes \\
       PS4 & Bachelor & Female & 0-5 & Yes \\
       PS5 \& PS10 & Master & Female & 0-5 & Yes \\
       PS6 & Bachelor & Female & 10+ & No \\
       PS7 \& PS13 & Bachelor & Female & 0-5 & No \\
       PS8 & Ph.D. & Female & 0-5 & No \\
       PS11 & Master & Female & 0-5 & No \\
       PS12 & Bachelor & Male & 5-10 & No \\
    \bottomrule
    \end{tabular}
}
\vspace{0.2cm}
    \caption{Interviewees description. Our interviewees come from varied professional backgrounds, have diverse experience in the public sector, possess different levels of familiarity with NLP tools, and hold different political affiliations.}
    \label{tab:participants}
\end{table*}

% We focus on the executive branch of government since public participation processes are more common there than in the legislative or judicial branches. 
% Despite having three participants from an agency, we mainly focus on ministries at the central government, which are different from local governments or other branches of government.
% Figure~\ref{fig:state} provides a generalized and simplified overview of the institutional structure of Chile and Uruguay.
% Within ministries, we have a varied pool of participants from institutions working on different topics (see table~\ref{tab:institutions}) and working both at cabinets and technical divisions.
% We decided not to include people from support divisions since they are not directly involved in policymaking or conducting the participation process.
% Figure~\ref{fig:institutions} provides an overview of how institutions work and where our participants were when we interviewed them.  

\subsection{Interview Data Collection}

We conducted the interviews remotely in Spanish, the participants' native language. Each interview lasted about about 60 minutes.
In each interview, we started by inviting the interviewees to describe their experience in government briefly, choosing at least one participation process and describing it from conception to implementation in detail.
We asked follow-up questions to better understand how they engaged with citizens and how they recorded, systematized, and analyzed the information to produce policy outputs.
For participants who had used NLP tools in civic participatory processes, we proceeded to ask how they had used NLP tools in their participatory processes, how they perceived the advantages and challenges of using NLP tools, and changes compared to how they work without NLP.
For participants without related NLP use experience,  we invited them to describe their rationale and how they considered whether or not to use NLP tools in their future work. Based on the tools and processes they described, we asked follow-up questions to distill perceived barriers and risks and how they would approach them when using NLP in participation processes. 
We focused on NLP tools performing the functions related to the Related Work section.
We recorded and transcribed all interviews. 

% We focused on three types of tools involving NLP.
% First, we considered support} tools that involve pre-processing data, such as speech-to-text, detection of machine-produced text, detection of comments and duplicates, etc.
% Second, we considered summarization} tools, which can organize data and provide general insights, such as topic modeling and entities extraction.
% Finally, we considered analysis} tools, which could provide additional insights from the data and/or complementary sources, such as sentiment analysis, sentiment classification, relation extraction, etc. % Qian: Strike this part because this is arealdy covered in RW.

% During the first set of questions in the interviews, we provided little information about NLP's specific algorithms or tools nor what they could do.
% We expected participants to vocalize their wishes and requirements based on their experiences and backgrounds.
% However, after their first set of responses, we provided specific examples of NLP applications to probe about their (potential) use in the processes our interviewees were describing.
% When needed, we provided generic explanations of what different types of NLP applications could do (e.g., sentiment analysis, topic modeling, relation extraction).
% When participants mentioned applications or systems beyond NLP, we only explained that after they finished explaining their ideas and probed about the potential implications of their thoughts.
\subsection{Interview Data Analysis}
We analyzed the interview recordings using a two-phase coding technique and the qualitative data analysis software MaxQDA. 
The first phase of open coding contrasts the data and builds a broad set of codes, which are then connected in the second phase of axial coding~\cite{corbin_basics_2014}.
The first author conducted the open coding stage, producing 73 codes with 735 coded sentences.
All the authors discussed the findings and related work, and the first author refined the coding, classifying them into 24 second-level codes, which were grouped into four categories that are presented in the Findings section~\ref{sec:findings} in Tables~\ref{tab:motivations},~\ref{tab:risks},~\ref{tab:considerations}, and~\ref{tab:barriers}.

\section{Findings}
\label{sec:findings}
Our interviews revealed three major findings. The first two findings are interrelated.
First, both groups of policymakers---politicians and public servants--focused on their needs for career advancement and the need to strengthen the legitimacy of their work when considering NLP tools' adoption and use.
Second, because these needs varied between politicians and public servants, their preferred NLP features and tool designs also differed.
Politicians' legitimacy was rooted in their constituents' trust and approval; therefore, they wanted NLP features that could showcase the fairness and capacity of their public participation process and the use of cutting-edge technologies.
Public servants, in contrast, were internally oriented. Their jobs' legitimacy came from the approval of their superiors; consequently, they looked for NLP features that could help them become both more efficient and more empathetic in analyzing citizen comments at the current volume.
Finally, despite their differing needs and opinions, neither group clearly identified who should advocate for NLP adoption to enhance civic participation or address the unintended consequences of a poorly considered adoption. Instead, they blamed each other for NLP tools' low adoption in government.

We compared politicians and public servants in different categories (e.g., general and institutional level, experience using NLP) but only found relevant differences at a general level (politicians-public servants) and some particular cases for experience using NLP, which we highlight. 
Below, we report our findings using textual quotes from the interviews translated from Spanish by the authors, paraphrasing multiple quotes, and summarizing exemplary cases narrated by the interviewees. 
When using quotes within paragraphs, fragments are selected without changing the original quote's intention, context, and meaning. 

\subsection{Differing Motivations of Using NLP Tools}
Politicians and public servants differed in their incentives when deciding whether to use NLP to support participatory processes. 
Politicians focused on increasing legitimacy towards their constituents and managing power, while public servants were primarily interested in reducing their workload, increasing efficiency, and avoiding being replaced. 
Table~\ref{tab:motivations} summarizes the percentage of interviewees in each group that discussed each motivation.

\begin{table}[h!]
\centering

    \begin{tabular}{p{0.3\columnwidth} p{0.3\columnwidth}<{
\centering} p{0.3\columnwidth}<{
\centering}}
    \toprule
     \textbf{Motivation} & \textbf{Public Servants (13)} & \textbf{Politicians (7)}\\
     \midrule
     Increasing legitimacy & 23\% & 100\%\\
     Managing power & 0\% & 86\%\\
     Increasing efficiency & 85\% & 29\%\\
     Reducing workload & 100\% & 14\%\\
     Human core role in the process & 77\% & 29\%\\
     \bottomrule
    \end{tabular}
 \caption{Motivations to Use NLP tools}~\label{tab:motivations} 
\end{table}

%Objectivity
\subsubsection{Politicians prioritized building legitimacy towards external stakeholders}
Politicians highlighted that using NLP could increase objectivity and rigor, increasing their legitimacy by promoting constituents' trust. 
There was a shared belief among politicians that \blockquote[PT1]{\textit{when there is human intervention concerning the comments and responses of those who participated, there is always manipulation stemming from who takes notes and then informs the authority}}.
Using NLP tools, politicians could \blockquote[PT4]{\textit{identify the true conversations and the words people used to refer to the different topics without the filters and biases of who listened and systematized}}. 
Politicians tended to believe that NLP tools were \blockquote[PT1]{\textit{not interpreting, only processing data}}, which was essential to achieve their goal of \blockquote[PT3]{\textit{estimating frequencies of converging responses to identify topics you can incorporate with more assurance of people adhering to them [...] to give the process a scientific approach through technology}}.
PT5 summarized this idea well:

\begin{displayquote}[PT5]
\textit{I think we can establish mechanisms [when using NLP tools] that [...] make the process more objective, leaving aside note-takers and moderators' manipulations in real-time and eliminating bias when systematizing. Maybe we could even do something like the academia, have more representative samples, or fix them ex-post.}
\end{displayquote}

Although some politicians recognized that \blockquote[PT4]{\textit{NLP tools might have some bias}}, they believed AI's biases could be identified more precisely and efficiently than biases from people taking notes~(PT4, PT5). 
Politicians considered machines superior to humans, so using them would increase their constituents' trust in the process. 
For example, PT4 explained that \blockquote[PT4]{\textit{there is no mechanism without bias [...], for example, how a topic modeling algorithm groups topics can be directed. However, eliminating the first intermediary, the policymaker, allows the raw data to be considered better when first processed by a machine}}. 

%Press
Some politicians argued that using NLP tools could increase their public approval through positive press: \blockquote[PT2]{\textit{Technology has good selling, and politicians like that. Authorities would implement technology to support participation because it will sell; it has a certain glamour and attractiveness}}. 
Politicians argued that although some citizens were afraid of AI, most people believed that using it would improve the quality of policymaking, especially since trust in governments in Latin America was shallow. 
One public servant mentioned that emphasizing the public image benefits of AI adoption might help convince authorities to implement technology: 
\blockquote[PS2]{\textit{Politicians would use AI because they think it is interesting. They would be able to say they are at the frontier, and that is something you can communicate. Authorities will always consider how they benefit from the method, so using technology can be an element of marketing}}.

%Manage power
\subsubsection{Politicians perceived NLP could help them manage power dynamics.}
Politicians thought NLP tools could help them manage power among stakeholders and align the processes to their political agendas by including new groups that could change power dynamics. 
The role of NLP tools here was dual. 
On the one hand, politicians saw NLP as a tool to identify trends in data, increase the salience of marginalized stakeholders, and build a case against stakeholders that are traditionally more vocal and visible to the public in traditional participatory processes. 
PT1 explained their needs: \blockquote[PT1]{\textit{We could identify trends from majorities that usually don't voice their demands or are not able to articulate as a group, that would give us more support for certain policies, distributing power}}. 
On the other hand, politicians saw NLP tools as capable of decreasing biases in participation processes, which suffer from self-selection. 
Politicians believed working with NLP could decrease barriers for underrepresented groups by enabling them to work with raw data and a larger pool of participants. 
The latter would dilute specific interest groups' power, making processes more objective and legitimate. 
PT4 described an example during the COVID-19 pandemic:

\begin{displayquote}[PT4]
\textit{Our problem was that, although we had scientific evidence for our policies, the most recognized voices were attacking our ideas ad hominem: anything proposed by the government was wrong. We had to find and give voice to people suffering from COVID-19 daily, so we gathered information from a broader set of stakeholders. Every opinion was considered, and we analyzed the data using NLP techniques such as topic modeling. We built the Opening Plan based on that input. Without our methodology, including NLP, we would not have had the same legitimacy, and the Opening Plan could have failed. By gathering more people, we had more legitimate voices supporting our policy.}
\end{displayquote}

Politicians would only implement NLP tools if they considered it would not harm their legitimacy with relevant external stakeholders.
For example, PS5 explained how NLP was discarded from a massive consultation in education because politicians considered that using AI would not be accepted by core stakeholders (i.e., teachers' unions) and thus reduce the process' legitimacy.
PS5 argued that NLP tools could enable them to consult a broader set of stakeholders, but politicians preferred to reduce the number of participants and ask less specific questions while keeping the analysis manual.

\subsubsection{Public servants prioritized efficiency to build their legitimacy towards their superiors. }
Public servants wanted NLP to alleviate their workload and make the processes more efficient by \blockquote[PS3]{\textit{broadening the reach while cutting down time demands}}. 
Public servants argued that analyzing participatory data manually was hard or impossible since it took too much time and they had to fulfill multiple roles in government~(PS3, PS5, PS10, PS12). 
Public servants were usually asked to deliver quickly, which they argued to be incompatible with an in-depth analysis of qualitative information~(PS3, PS4, PS5, PS8, PS9). 
Many times, public servants had to narrow down questions to collect less information to be able to analyze and respond to citizens promptly~(PS5), which they believed could change using NLP tools~(PS3, PS5).

Public servants believed NLP tools could support organizing, filtering, navigating, and summarizing information. 
Interviewees wanted NLP tools to efficiently identify the most frequent topics and core ideas~(PS7, PS9, PS10, PS12), propose relationships between topics, build clusters, and extract the most relevant quotes~(PS1, PS5, PS8) that they could connect to political priorities~(PS8). 

Public servants approached the representation issue from an efficiency perspective: they needed more efficient strategies to engage with the right stakeholders and find missing minority groups in less time~(PS7).
Public servants needed support from NLP to \blockquote[PS5]{\textit{identify actors}} and \blockquote[PS10]{\textit{be able to map stakeholders because the public is so broad that it is hard to identify which groups are relevant}} or \blockquote[PS12]{\textit{which institutions need to be involved}}. 
Additionally, public servants discussed the possibility of using AI tools to \blockquote[PS4]{\textit{have more representative samples of participants or warn potential missing groups from alternative sources such as social media data}}. 
The latter was not necessarily related to NLP tools, but they did mention analyzing textual data to identify similar stakeholders within the data and compare it to textual data in other sources, such as social media.

\subsubsection{Public servants' use of NLP was mediated by their perception of their role in the process.}
Public servants did not see NLP tools replacing their role in analyzing participatory data or policy design. 
Public servants wanted to have a tool that \blockquote[PS4]{\textit{could automate systematizing and summarizing the information, not to a final document, to something where we can explore the main ideas efficiently}}. 
Public servants envisioned NLP as \blockquote[PS1]{\textit{a team member with better capacities to systematize and summarize the information and that then I can interrogate}}.

Public servants who did not have previous experience with NLP tools insisted that there were elements of the participatory process that machines should never perform. 
This group argued they had to use the data to \blockquote[PS2]{\textit{build the public policy document that can be communicated, and that they would not trust an AI system to build the political narrative}}.
They argued that NLP tools could never replace human analysis since humans were the ones who could provide theoretical explanations for correlations~(PS7) and understand human sensitivities~(PS11).

\subsection{Differing Considerations on How to Use NLP Tools}
Politicians and public servants differed in how they perceived and approached risks from NLP, depending on their legitimacy-building orientation (i.e., from whom they draw legitimacy) and relationship with the NLP tools.
Politicians' interaction with external stakeholders was more direct and core to their legitimacy, so their orientation was often external.
However, politicians' interaction with the NLP tools was usually indirect, so they were less conscious of the risks. 
In contrast, public servants' legitimacy depended less on external stakeholders than on their superiors' opinions about their role, so their orientation was often internal.
Moreover, public servants usually interacted more intensively with the NLP system, which put them at greater risk of being replaced and often made them more conscious of the risks.
Table~\ref{tab:risks} summarizes the percentage of interviewees that discussed each risk, and table~\ref{tab:considerations} summarizes the percentage of interviewees that discussed different mechanisms for a thoughtful adoption of NLP.

\begin{table}[h!]
\centering
    \begin{tabular}{p{0.3\columnwidth} p{0.3\columnwidth}<{
\centering} p{0.3\columnwidth}<{
\centering}}
    \toprule
     \textbf{Risk} & \textbf{Public Servants (13)} & \textbf{Politicians (7)}\\
     \midrule
     Manipulation & 15\% & 43\%\\
     Black box & 23\% & 14\%\\
     Data protection & 46\% & 14\%\\
     Errors in analysis & 46\% & 0\%\\
     Illusion of objectivity & 23\% & 57\%\\
     Surveillance & 23\% & 0\%\\
     Biases & 46\% & 57\%\\
     \bottomrule
    \end{tabular}
     \caption{Perceived risks in using NLP tools. %*Politicians do not identify this risk explicitly by this term; we distill they have the illusion of objectivity from what they say.
     }~\label{tab:risks} 
\end{table}

\begin{table}[h!]
\centering
    \begin{tabular}{p{0.25\columnwidth} p{0.25\columnwidth} p{0.2\columnwidth}<{
\centering} p{0.2\columnwidth}<{
\centering}}
    \toprule
     \textbf{Orientation} & \textbf{Mechanism} & \textbf{Public Servants (13)} & \textbf{Politicians (7)}\\
     \midrule
     & Provider reputation & 38\% & 71\%\\
     External & Replicability & 8\% & 71\%\\
     &Audit mechanisms & 31\% & 43\%\\
     \hline
     Internal & Human validation & 69\% & 14\%\\
     & Traceability & 77\% & 43\%\\
     \hline
     Both&Tailored explanations & 62\% & 71\%\\
     &Transparency & 100\% & 71\%\\
     \bottomrule
    \end{tabular}
 \caption{Considerations for thoughtful adoption}~\label{tab:considerations} 
\end{table}

\subsubsection{Politicians were less aware of risks than public servants.}
Relying on external validation and seeing themselves far from the tool, politicians were less aware of the risks of using NLP in participatory processes. 
In contrast, public servants saw themselves as key actors in implementing NLP tools, so they were more conscious and knowledgeable. 
Public servants identified risks and challenges such as errors in analysis (e.g., hallucinations, misclassification), lack of explainability, surveillance, biases, and privacy breaches. 
The only risk both groups were equally aware of was potential biases. 
However, politicians tended to think that although NLP tools could be biased, they would probably be less biased than humans (PT1 \& PT4) and that reducing bias was just a technical problem (PT3). 

Politicians tended to believe that \blockquote[PT1]{\textit{AI could help since it provides an objective perspective}}, revealing an illusion of objectivity. 
Public servants recognized the risk that politicians could argue that the machine was objective because it is not human and use NLP systems to claim neutrality and manipulate citizens (intentionally or unintentionally). 
PS7 voiced this fear: 
\begin{displayquote}[PS7]
    \textit{Some politicians believe the machine is objective because it is not human. Thus, it is trustworthy. I fear that if we cannot explain why algorithms are not necessarily better than humans, they will replace people with no considerations}.
\end{displayquote}

Politicians discussed the risks of political manipulation from a different perspective. 
For politicians, the risk was present if someone intentionally altered the code or tampered with the results, which could put democracy at risk and harm citizens~(PT2, PT4, PT5). 
Politicians analyzed the risk, considering that it did not stem from the system's characteristics or use but from the users' direct intervention.
The latter was consistent with their external orientation, which usually expected attacks from opposing political forces.

\subsubsection{Politicians favored external-facing mechanisms towards responsible AI}
Politicians built legitimacy towards the public (external orientation), favoring external-facing mechanisms. 
Politicians considered providers' reputation and replicability as the most critical elements in thoughtfully adopting NLP.
Politicians explained they needed information about providers' previous experience and team composition to assess their reputation and fit to safely develop and implement an NLP~(PT1, PT3, PT4).
PT1 explained that providers had to \blockquote[PT1]{\textit{prove neutrality and capacities before the project implementation}}.

Politicians highlighted the role of civil society and academia in replicating and challenging results, so data needed to be made available along with well-crafted documentation so that external reviewers could audit the processes and results~(PT2, PT3, PT5, PT6). 
PT3 argued that involvement from academia could increase trust in the system: \blockquote[PT3]{\textit{Despite academia not being a default honest broker, under certain conditions, it can be trustworthy and increase trust in the process and systems when involved}}.

Regarding the system, politicians sought signals and information about the systems' trustworthiness in standards, certifications, and compliance with existing regulations~(PT1, PT3, PT4, PT5). 
PT1, PT3, and PT4 mentioned how providing certifications such as ISO would help increase citizens' trust and reliance on AI. 
Moreover, certifications and standards could be linked to the legal process, such as procurements: \blockquote[PT6]{\textit{We have to incorporate technical requirements and certifications in the procurement process to support institutions with no technical capacities}}.

\subsubsection{Public servants favored human-in-the-loop internal-facing approaches towards responsible AI}
Public servants built legitimacy within the organization (internal orientation) by validating their roles and capacities towards their superiors. 
Thus, they tended to favor internal-facing mechanisms where they played a relevant role: human validation, piloting periods, improving their knowledge to serve as internal counterparts, and using guidelines/standards.

Public servants favored internal controls through human involvement (humans-in-the-loop) to ensure thoughtful adoption processes by validating the algorithms throughout their life cycle.
Public servants argued the need to \blockquote[PS7]{\textit{access the raw data and check if the analysis made sense and correct when it is wrong}}. 
Even in non-analytical tasks like transcriptions, they believed they should review the results: \blockquote[PS9]{\textit{When there is too much jargon, we need to check the automated transcription and correct it}}.

Public servants discussed that NLP systems would not replace them but change their role into acting more as reviewers. 
Interviewees argued that the processes should have \blockquote[PS6]{\textit{stages where the human has the role of revising, supervising, and ensuring the data quality and analysis is right}}.
The associated challenge was that public servants had to be competent counterparts for third parties providing the NLP systems, which was not always the case.
To this point, public servants believed they \blockquote[PS3]{\textit{did not need to understand the algorithm in-depth but to have some basic understanding and a critical approach}} to work \blockquote[PS8]{\textit{hand in hand with the provider to understand all the decisions and results}}.

Public servants recognized the lack of knowledge and considered either bringing technical people to the team~(PS7), using existing guidelines and standards~(PS7 \& PS12), or connecting to third parties~(PS9 \& PS12).
Thus, external validation was an option when the expertise was unavailable, but public servants still required strong connections and involvement with the review board. 
Some mechanisms were discussed, such as having third parties \blockquote[PS12]{\textit{randomly revising a sample of raw quotes}}, adapting \blockquote[PS9]{\textit{universities ethics revision boards and revision processes [...] or creating ad honorem advisory boards}}.

Public servants emphasized that adoption processes should include piloting periods and strong collaborations with the providers~(PS1 \& PS12). To do so, public servants agreed they needed \blockquote[PS13]{\textit{guidelines to know what to do when evaluating [NLP] projects}} and \blockquote[PS7]{\textit{sectorial standards which lower the knowledge asymmetries [with providers] when implementing NLP systems}}.

\subsubsection{Politicians and public servants manifested different transparency and explainability needs}
Transparency and explainability were operationalized differently depending on which stakeholder was relevant to the politician or public servant.
Both groups agreed that data should be open when it was allowed by regulation and privacy was adequately protected. 
Moreover, most participants emphasized that traceability from results to the raw data should be easy and interactive.
However, they disagreed on how, to whom, and how much data/code to open.

Politicians were prone to argue about opening data/code and making it explainable to the public or external reviewers. 
Politicians argued that tailored explanations were needed for citizens to understand, replicate, and audit the analysis~(PT2, PT5): 
\blockquote[PT6]{\textit{authorities need to explain to citizens how they did the analysis, so they need a semi-technical explanation of the system they can understand to have a legitimate process}}.
Thus, politicians' core goal was to increase trust in the process among external stakeholders.

Public servants usually discussed transparency and explainability when working with the system and explaining results to their superiors. 
Public servants prioritized explanations from providers to understand the process themselves in elements such as \blockquote[PS1]{\textit{how the insights were extracted}} and \blockquote[PS4]{\textit{how the system filtered and classified information}}.
Public servants argued that despite explanations to external stakeholders being needed, they would not completely open the code since it could increase costs and decrease providers, making the process less efficient (PS1, PS3, PS12).
For example, PS12 argued that \blockquote[PS12]{\textit{providers will not open the code of their systems, and probably it is not viable nor desirable because that would increase costs and providers' availability}}.

\subsection{Who is Responsible for NLP Tool Adoption and Proper Use?}
Politicians and public servants blamed each other on who was responsible for the lack of widespread thoughtful use of NLP for participation processes. 
Both groups highlighted bureaucracy and lack of knowledge about NLP as the main barriers to its use.
However, there were subtleties related to how they interpreted the implications of these barriers.
Politicians believed that lack of knowledge made public servants resist change, blaming them for not exploring and not being open to using new technology. 
Public servants countered, blaming politicians’ lack of leadership for their lack of knowledge, infrastructure, and support.
Table~\ref{tab:barriers} summarizes the percentage of interviewees who discussed each barrier.

\begin{table}[h!]
\centering

    \begin{tabular}{p{0.5\columnwidth} p{0.2\columnwidth}<{
\centering} p{0.2\columnwidth}<{
\centering}}
    \toprule
     \textbf{Barrier} & \textbf{Public Servants (13)} & \textbf{Politicians (7)}\\
     \midrule
     Bureaucracy & 46\% & 71\%\\
     Lack of knowledge & 69\% & 57\%\\
     Resistance to change in policymakers & 8\% & 57\%\\
     Lack of infrastructure & 46\% & 0\%\\
     Lack of leadership in politicians & 31\% & 0\%\\         
     \bottomrule
    \end{tabular}
     \caption{Barriers for adoption of NLP tools}~\label{tab:barriers} 
\end{table}

\subsubsection{Politicians and public servants agreed bureaucracy was a relevant burden to adopt NLP tools}
Public servants and politicians agreed that excessive or unfit bureaucracy and inadequate procurement processes were relevant barriers to adoption. 
Interviewees believed \blockquote[PT2]{\textit{the problem is institutional; it is bureaucracy}} because in the public sector \blockquote[PS12]{\textit{you will always have a limitation which is that you need an enabling regulation}}. 
Procurement processes are not suited for technology, as one participant explained: \blockquote[PS1]{\textit{we were trying to hire a massive mailing service for over a year because paying a monthly license did not fit the procurement processes}}.
Moreover, interviewees complained that many \blockquote[PS5]{\textit{regulation frameworks are outdated; when they were created, technology had other business models. We have issues even for maintaining web pages}}.

\subsubsection{There was a lack of knowledge about NLP, but no clarity on who was responsible for it}
The most relevant consensual barrier was the lack of knowledge within the government, but the two groups perceived it differently. 
Politicians argued that public servants resisted change because they lacked the technical knowledge about what NLP could do and how to implement it. 
Politicians highlighted the importance of fostering \blockquote[PT3]{\textit{an educational effort on understanding AI systems for public servants}}. 
PT2 explained:

\begin{displayquote}[PT2]
\textit{NLP tools are complex systems that people in, for example, the acquisitions department will not understand. They will not risk authorizing anything they do not understand that could lead to potential sanctions. Public servants resist change, especially when it will affect their comfort and way of doing things.}
\end{displayquote}

Public servants agreed that the government lacked knowledge about NLP, but this was not necessarily a reason for resisting change. 
Public servants acknowledged that background heterogeneity could create barriers (e.g., PS8 highlighted legal and procurement teams) or particular sectors that did not trust technology (e.g., interviewees exemplified education and culture public servants). 
Still, it was not usually the reason for stopping the adoption of NLP. 
Lack of knowledge was often a barrier because public servants did not know what NLP tools could do nor how to design and implement them~(PS4, PS10, PS12).
Moreover, governments usually did not have the required technological infrastructure regarding computation capacities or budget~(PS7, PS8, PS10). 
Public servants said they did not necessarily resist; they did not know the available tools and had no time to explore.

Public servants argued that the lack of knowledge was partly due to politicians' lack of leadership, which made adoption harder. 
They believed that \blockquote[PS2]{\textit{educating people who will implement NLP tools depends on the head of the organization, but bosses often do not see the relevance of using them}}. 
Public servants contended that having the time and flexibility to explore methodologies and tools was critical to acquiring knowledge and successfully implementing new tools, but politicians often did not allow that.
For example, PS1 explained that \blockquote[PS1]{\textit{exploration is incredibly valuable to learn what new technologies could do, now I can tell because I could do it with my former supervisor, but now I have to fight too hard to be able to learn and try new tools}}. 
Public servants argued that resistance to change has to be managed by politicians, which does not always happens: 
\blockquote[PS7]{\textit{Change and human management are critical, but they need the authority to be convinced. If the person in charge sees the value, they will bring the necessary people and resources to the team; if not, there is nothing to do, and change will be impossible}}.

\section{Discussion}
Our findings examined why NLP tools have failed to be widely adopted thoughtfully for civic participation processes by governments despite recent technical advancements~\cite[e.g., ][]{romberg_making_2023, romberg_is_2022, arana-catania_citizen_2021, hagen_content_2018}.
Previous research in NLP for participation has focused on efficiency and technical limitations of offering consistent and reliable support to public servants~\cite[e.g., ][]{romberg_making_2023, romberg_automated_2022, arana-catania_citizen_2021, hagen_content_2018}. 
Still, that body of work has little connection to research about the complexities of internal stakeholders when implementing technological tools in the public sector~\cite{rose_stakeholder_2018, rowley_e-government_2011, rose_managing_2015,kawakami_studying_2024} in combination with the complexities of designing human-AI interaction~\cite{yang_re-examining_2020, yang_sketching_2019}. 
While multiple value positions have been addressed in prior CSCW work~\cite[e.g., ][]{madaio_co-designing_2020, wong_tactics_2021, gray_ethical_2019}, it is not necessarily tailored and adequate to government contexts~\cite{rose_managing_2015, rose_stakeholder_2018, kawakami_studying_2024,kawakami_situate_2024} and might fail to inform thoughtful adoption of NLP tools in civic participation processes.
We contribute to a nascent line of work within CSCW that analyzes the complexities of AI adoption in government, considering different internal stakeholder groups~\cite{kawakami_studying_2024}.

Our findings inform how to approach the design and adoption of NLP tools for participation processes, considering the nuances of politicians and public servants within the public sector.
We discuss research and design implications and opportunities around three areas: 
(1) Designing NLP tools for more than efficiency in participation processes, 
(2) Designing tools and methods for a thoughtful use of NLP in participation processes and
(3) Designing tools and methods not for single-user but for collaborative use in government.

\subsection{Designing NLP Tools for More Than Efficiency in Participation Processes}
Our findings suggest that NLP tools for supporting participation processes should not focus only on efficiency, which has been the primary goal of most prior work~\cite{romberg_making_2023, arana-catania_citizen_2021, chen_barriers_2019, livermore_computationally_2017}. 
Participation processes are moments when the government faces the public, so they are at the core of politicians' legitimacy.
Politicians will accommodate the use of NLP tools to get good press and engage with more participants who share their political agendas.
Our findings suggest that prior works' single-objective approach might fail to promote adoption because the efficiency objective~\cite{romberg_making_2023, chen_barriers_2019, arana-catania_citizen_2021, romberg_automated_2022} might not be compelling to politicians.

We do not argue that increasing efficiency does not matter.
Rather, our findings surface a potential tension between politicians and public servants regarding their motivations because they construct their legitimacy differently.
Both motivations are not necessarily incompatible, but they can be misaligned.
If politicians need to efficiently process the information and showcase the results, they will probably be aligned with public servants' needs as users.
However, if politicians are not aware of public servants' needs, are only concerned with getting good press, and/or have relevant stakeholders resistant to AI, public servants probably will not be able to convince them to use NLP tools. 
As prior work in electronic participation initiatives has shown, failing to address the motivations of the most salient stakeholder (i.e., the one with the higher power, urgency, and legitimacy) can prevent adoption~\cite{saebo_understanding_2011}.

In contrast, focusing solely on how NLP contributes to politicians' legitimacy and not considering efficiency could also lead to failed adoption processes or the design of spurious tools. 
Politicians could mandate the implementation of NLP tools without acknowledging technical limitations~\cite{ma_semantic_2016, hagen_content_2018, ash_deep_2020}, infrastructural requirements~\cite{romberg_automated_2022, aitamurto_crowdsourced_2016, kim_value_2021, weng_ai_2021}, and ethical considerations~\cite{levy_algorithms_2021,kawakami_situate_2024}. 
Furthermore, politicians' drive to replace humans to increase objectivity is directly in tension with public servants' belief in humans' intrinsic value. 
Politicians could force the adoption of NLP systems, but this could worsen the participation process by making it less efficient and harming public servants and the public~\cite{delgado_participatory_2023, jasim_communitypulse_2021}.

When public servants strongly believe that humans should not be replaced, our findings suggest that efficiency gains and reduced workloads might be insufficient.
Public servants might prefer manual analysis even if the time and effort are high~\cite{panopoulou_eparticipation_2010, rose_designing_2010} since they see an intrinsic value in having humans supervising and translating data to policy. 
Our findings suggest convincing public servants to adopt NLP tools is more complex than just achieving better results and making algorithms more readily available through software~\cite{romberg_making_2023, wirtz_artificial_2019, tangi_challenges_2023, madan_ai_2023}. 
Instead, a nuanced understanding of the political and organizational context of the public institution where they will be deployed is required.
Acquiring this understanding can help design strategies that address the right mix of requirements for the AI tool.
For instance, strategically leveraging the power of politicians who wish to implement AI tools to increase their own public acceptance could help open spaces for experimentation, letting public servants explore, learn, and reduce resistance.
Incorporating early-stage deliberation about whether to adopt an NLP tool~\cite{kawakami_situate_2024} and what the right mix of functions incorporated in the system is should be at the center of future research. 

Our findings show how politicians and public servants have heterogeneous needs from the AI system that are related to their career needs.
The HCI and NLP communities could explore how NLP tools can be designed to address heterogeneous career-related needs, such as making processes more transparent, enabling a broader base of participants, increasing citizens’ engagement and ownership of the policies they participate in, improving communication with constituents, etc.
Moreover, deliberation methods and participatory tools could be explored to collaboratively design the right mix of functions to address internal stakeholders' needs.
However, building NLP tools for specific requirements is not straightforward, given all the human data required to train the models, the computing requirements, and the technical challenges of working with language. 
Furthermore, a core challenge in future research would be identifying which requirements from different internal stakeholders are to be addressed through technical solutions (e.g., developing/tailoring the algorithm) and which require designing the right human-AI interaction, organizational change, and policy intervention.

NLP tools offer limited options for politicians and public servants to engage the public. 
Most of our findings were focused on NLP tools without considering the latest advancements in large language models (LLMs).
Future research should explore how LLMs can both support participation processes considering areas such as improving existing NLP tools~\cite[e.g., ][]{lam_concept_2024,behrendt_aqua_2024}, supporting the public in deliberative processes~\cite{small_opportunities_2023}, and changing government interaction with citizens~\cite{fares_role_2023}.
Similarly, to promote tools beyond efficiency, future research should explore other AI applications such as image-generative AI~\cite{guridi_image_2024,von_brackel-schmidt_equipping_2024} and AI + Virtual reality~\cite{porwol_transforming_2022}) that could enhance the participatory design of public policies.
However, all potential new applications bring along new risks~\cite{weidinger_taxonomy_2022} that need to be explored considering the nuances of the public sector.

\subsection{Designing Tools and Methods for a Thoughtful Use of NLP in Participation Processes}
\subsubsection{Failing to understand the nuances of legitimacy can lead to a lack of use of NLP tools in the right context or its use in the wrong context.}
Lack of use of NLP in the right context can hurt citizens by preventing the government from properly implementing participation processes due to challenges such as cognitive overload that could have been addressed~\cite{chen_barriers_2019}. 
Using NLP for participation in the wrong context or for the wrong reasons can harm citizens differently, from using constituents’ money with no positive (or even a negative) impact to discrimination and political manipulation~\cite{delgado_participatory_2023, birhane_power_2022,tangi_challenges_2023,wirtz_artificial_2019,levy_algorithms_2021}.

To move towards a thoughtful adoption of NLP tools, designers must address the gap in risk perception between public servants and politicians. 
Public servants are more aware than politicians of the risks of NLP tools and their complexities but hold less power in the hierarchical government structure~\cite{kawakami_studying_2024}, and oftentimes worry more about their acceptance by their bosses, not the public~\cite{dean_deliberating_2023}. 
Our findings show the importance of considering how politicians' involvement in the design process can shape how the risks and challenges of using NLP are effectively addressed. 
However, there is a lack of guidelines to implement AI in government~\cite{nagitta_human-centered_2022, dor_procurement_2021, sloane_ai_2021} considering politicians when navigating the complexities of public institutions~\cite{kawakami_situate_2024, kawakami_studying_2024}. 
Moreover, tools to inform politicians about the risks of AI and to asses the 'right level' of unremarkableness of the tools to prevent ill-use are scarce~\cite{susser_invisible_2019, tolmie_unremarkable_2002, yang_unremarkable_2019}.

Explanations are crucial to mitigate asymmetries between stakeholders and improve understanding of NLP systems and their risks. 
Both groups asked for multiple levels of explanations and traceability from insights to the raw data. 
More research is needed following a user-centered, explainable AI approach~\cite{wang_designing_2019} to operationalize what politicians and public servants require for specific sectors and policies, acknowledging their knowledge and power asymmetries.
Moreover, future work should tailor existing research on how to open the 'black box' of algorithms~\cite[e.g., ][]{edwards_slave_2017, liao_designing_2022, ehsan_expanding_2021, wang_designing_2019, rechkemmer_when_2022, kaur_sensible_2022} to the needs of specific users within the public sector~\cite{arya_one_2019, ehsan_who_2021,kawakami_studying_2024}.

Visualization can play a relevant role in promoting the use of NLP in the right context.
Our findings highlight the importance of promoting research on how to represent the data and how to build interactive platforms that can achieve tailored explanations to different stakeholders and provide easy access to raw data to extract individual cases for political narratives or confirm insights~\cite[e.g., ][]{yovanovic_remote_2021, cai_interactive_2018, mendez_exploring_2022, mendez_legislatio_2022}.
Since legitimacy is at the center of politicians' and public servants' decisions, visualization will also be at the core of their interest since it is the vehicle through which they showcase results to their relevant stakeholders.
Future research should continue working on representing the data and building interactive platforms that incorporate politicians' and public servants' different requirements to tailor visualizations to different stakeholders. 

\subsubsection{Future research should explore how the misalignment of legitimacy could be addressed when using Responsible AI toolkits}
Our findings regarding the external/internal legitimacy-building orientation when assessing AI risks call for the CSCW community to rethink responsible AI tools and methods for government context. 
Prior work has developed audit mechanisms~\cite[e.g., ][]{vecchione_algorithmic_2021, lam_sociotechnical_2023} and toolkits~\cite[e.g., ][]{arya_one_2019, balayn_fairness_2023, madaio_co-designing_2020, shen_model_2022,kawakami_situate_2024}, but most of them have been designed for industry contexts and/or have not been implemented in practice~\cite{kawakami_situate_2024}.
As a result, most toolkits and audit mechanisms fail to provide guidance on how to navigate the organizational and institutional power dynamics~\cite{wong_seeing_2023, kawakami_situate_2024, kawakami_studying_2024}, which are at the core of adopting technologies in government~\cite{rose_managing_2015, rose_stakeholder_2018}.
Responsible AI methods should consider how internal stakeholders perceive the role of AI in their legitimacy-building efforts and how they distribute responsibility when failing to deploy AI systems. 

Responsible AI tools and methods should be adapted to suit internal processes required by public servants and external validation mechanisms sought by politicians.
Here, our findings are aligned with the emerging call for complementary tools to address the differences between internal stakeholders in the public sector~\cite{kawakami_situate_2024}.
However, future research should carefully examine if responsible AI methods through participatory design are enough when navigating the public sector~\cite{delgado_participatory_2023,birhane_power_2022}. 
Prior work has shown that participatory design is not necessarily a universal and effective fix~\cite{delgado_participatory_2023,sloane_participation_2022}, especially when elements such as power and legitimacy dynamics in government are not understood~\cite{kawakami_studying_2024,saebo_understanding_2011}.
Complementary tools and methods should be explored when participatory design is not feasible or can cause more harm than good due to asymmetries in knowledge or power between internal stakeholders. 

Future research should also consider when it is impossible to holistically address the differences between politicians and public servants.
Particular caution should be taken in exploring how power and knowledge asymmetries could lead to exploitation dynamics.
For example, politicians could force public servants into becoming mere algorithmic tools to check standards and provide information to external reviewers~\cite{meisner_labor_2022}.
Similarly, public servants could exploit politicians' distance from the tools or lack of knowledge to avoid responsibility and blame the system for mistakes to preserve their legitimacy~\cite{levy_algorithms_2021}.
Moreover, politicians were often overconfident about NLP and technical solutions to risks, which is not consistent with the state-of-the-art research in NLP and human-AI interaction. 
Even when confronted with risks, politicians still believed the systems could be better than humans (public servants) or that technical experts could easily solve risks.
Designers could explore embedded technical fixes when alignment is impossible (or too hard/costly) to prevent ill-use from one or many groups within the government. 

This research should complement analyses of how to reduce excessive or unfit bureaucratic processes, adapt AI governance, and improve procurement processes for technological tools.
Procurement processes can be a hurdle for acquiring technological tools, legal requirements, and technical guides are crucial in a thoughtful adoption of AI~\cite{dor_procurement_2021}. 
However, there is still a lack of guidelines to address challenges unique to NLP in procurement processes~\cite{nagitta_human-centered_2022, dor_procurement_2021, sloane_ai_2021}. 
Future research should work on policies to reduce bureaucratic burden and/or use procurement as tools to improve AI governance~\cite{dor_procurement_2021} and to address politicians' and public servants' heterogeneous demands and understanding of AI tools.
This is aligned with recent calls in the HCI community for researching how to prototype technology and policy simultaneously~\cite{yang_designing_2023,jackson_policy_2014,Yang_HCIPolicy_CHI24}.

Finally, there are increasing risks considering advancements in LLMs if internal legitimacy dynamics in government are poorly understood.
Our research surfaces the complexities of thoughtfully adopting NLP in a participatory process in a context where knowledge and infrastructure constraints still play a significant role.
Tools based on LLMs~\cite{lam_concept_2024, small_opportunities_2023,behrendt_aqua_2024}) could lower the technical/knowledge barriers mentioned in prior work~\cite{romberg_making_2023}.
Potential risks associated with AI, such as discrimination, privacy, and misinformation, evolve with LLMs~\cite{weidinger_taxonomy_2022} and impact both internal and external stakeholders in public institutions.

\subsection{Designing Tools and Methods Not for a Single User, but for Collaborative Use in Government}
Designers should take a strategic design approach, shifting from a solo user-centered design paradigm to an organizational perspective.
To do so, our findings suggest that a strategic approach should develop tools through processes that incorporate mechanisms to deal with shifting power and legitimacy-building dynamics.
Drawing from eGovernment literature, we suggest exploring incorporating stakeholders' salience into CSCW  methods.
Salience considers the power, legitimacy, and urgency each stakeholder group holds, which can change over time~\cite{saebo_understanding_2011,rose_stakeholder_2018}.
Future research should analyze how to plan the design, adoption, and diffusion of AI systems within government and shape them to the dynamic evolution of internal stakeholders' salience.
For example, when politicians' salience is high, designers should collaborate with public servants to develop persuasion strategies (e.g., good press, influencing power distribution between stakeholders) and narratives aligning with politicians' focus on external acceptance.

When designing NLP tools for multiple stakeholders, our findings highlight the importance of evaluating the assets, capabilities, organizational culture, and leadership~\cite{madan_ai_2023, giest_big_2017, poel_big_2018}.
Both groups perceived similar barriers to adoption; however, they approached them differently.
Designers need to assess knowledge and infrastructure gaps in both groups and design mechanisms to either reduce them or reconcile the tensions between the groups (e.g., resistance to change vs. lack of leadership). 
Future research should assess how to design AI tools and implementation guidelines that incorporate mechanisms of dealing with knowledge and infrastructure gaps in both groups alongside how they distribute responsibilities. 
For example, if resistance is mainly from politicians, using NLP as a communication asset could unlock the chicken-and-egg problem between the lack of leadership and knowledge, prompting politicians to give public servants space to learn and explore.

\section{Limitations And Conclusion}
Our study aimed to understand the nuances between politicians and public servants when deciding to use and implement NLP tools for civic participation processes. 
To do so, we conducted twenty interviews with politicians and public servants in Latin America.
Our findings extend prior work on NLP tools for participation, which focused on a singular type of user within governments and its technical challenges~\cite{romberg_making_2023}. 
Moreover, it contributes to an emerging trend in HCI analyzing the nuances of different internal stakeholders adopting AI in government~\cite{kawakami_studying_2024}.
Our work reveals new research opportunities around (1) designing NLP tools for more than efficiency in public participation processes, (2) developing tools and methods to thoughtfully use NLP in participation process considering differences between internal stakeholders, and (3) designing AI tools for organizational contexts in governments and not just a single-homogeneous user.

We acknowledge that the country-specific institutional arrangement and regulatory environment are limitations.
However, given the nature of our findings and the generality of our interviewees across institutions, experience, and professional backgrounds, our findings are likely to hold in other democratic regimes with nuances depending on how politicians and public servants develop their careers. 
Moreover, as we discussed, research on language techniques is moving quickly, so some of the details of our findings might be constrained by the current state-of-the-art NLP and LLM systems.
Moreover, our specific context provides value by bringing insights from the Global South to the Global North, an ongoing need in HCI~\cite{alvarado_garcia_fostering_2020, reynolds-cuellar_para_2023}. 
Future research could delve deeper into comparing North/South contexts and different public agencies to identify further nuances that inform how to navigate the tensions within the public sector. 
Additionally, user studies and real-world experiments could advance more specific design and process recommendations to adapt existing toolkits and guidelines for these contexts~\cite{kawakami_situate_2024}.

Finally, given the advances in AI tools of the last few years, we encourage researchers to consider the unique complexities of governments' multiple internal stakeholders. 
Developers and designers should work closely with public policy researchers and practitioners to look for pragmatic solutions that strategically address the differences between public servants and politicians. 
Current interest in the CSCW community on civic tech~\cite{aragon_civic_2020, stapleton_who_2022} requires investigating how these tools are starting to be deployed in real-world settings to prevent citizens from being intentionally or unintentionally harmed by politicians and public servants when it is done carelessly.
To do so, pragmatism and strategic approaches to AI governance and using AI tools in government call for new ways of developing tools and policies simultaneously~\cite{yang_designing_2023,Yang_HCIPolicy_CHI24}.
We expect our work to spark further interest in this area and promote future research that enables the widespread adoption of AI tools in the public sector that are thoughtfully deployed to improve democracy.

\begin{acks}
We thank the public servants and authorities of Chile and Uruguay for their dedication, time, and valuable insights. 
We are also very grateful to our colleagues who read previous versions of the paper and provided us with valuable insights to improve it.
The last author, Qian Yang’s effort in this project, was supported by the Schmidt Futures’ AI2050 Early Career Fellowship.
Include in non anonymized version.
\end{acks}

\appendix

\section*{Appendix}
\section{Study Context: Civic Participation Processes in Chile and Uruguay}
\label{appendix_chile_uruguay}

%Latin America
We conducted our study in two Latin American countries: Chile and Uruguay.
Latin America has unique characteristics such as a tense socio-political context, colonial relationships, infrastructure and institutional gaps, a more collectivist culture, a deep appreciation of participatory methods, and a deeply social orientation in HCI research~\cite{alvarado_garcia_fostering_2020,reynolds-cuellar_para_2023, ricaurte_algorithmic_2024, wong-villacres_lessons_2021}.
Despite Latin America's longstanding participatory tradition, it is often overlooked by the canon literature~\cite{Reynolds-Cuellar2022, Rappaport2020, wong-villacres_lessons_2021}.

%Chile
Chile is one of Latin America's most economically and socially stable and advanced countries.
It has a GDP per capita of US~\$17,093.2~\cite{world_bank_world_2023}, it is ranked 21st in the world in economic freedom~\cite{kim_2024_2024}, and it is ranked 25th in the Democracy Index~\cite{economist_intelligence_unit_democracy_2024}.
Moreover, Chile is one of the leading countries regarding AI in Latin America, being always in the top 3 positions across different AI indexes~\cite{centro_nacional_de_inteligencia_artificial_indice_2023, maslej_ai_2023, rogerson_government_2022}. 
In particular, Chile has a strong ICT infrastructure. 
It was the first country to finish Unesco's AI readiness assessment and it published an AI strategy in 2021, which is being implemented, and that was updated in 2024~\cite{unesco_chile_2023,centro_nacional_de_inteligencia_artificial_indice_2023,minctci_politica_2021,minctci_politica_2024}.
Chile is a democratic republic with a unitary state divided into sixteen regions.
The president leads the executive branch and is elected by direct vote every four years.
The legislative branch is constituted by the Congress, which is composed of two chambers: the Chamber of Deputies, which has 155 members elected every four years, and the Senate, which has 50 members elected every eight years.
The Chilean government (2022-2026) is led by President Gabriel Boric, who governs with his left-wing and center-left coalition.
The previous government (2018-2022) was led by Sebastian Piñera, who governed with his center-right coalition. 

%Uruguay
Uruguay is also one of the most stable and developed democracies in the region.
It has a high GDP per capita of US~\$22,564.5~\cite{world_bank_world_2023}, it is ranked 27th in economic freedom~\cite{kim_2024_2024}, and it is considered a full democracy ranked 15th in the Democracy Index~\cite{economist_intelligence_unit_democracy_2024}. 
Moreover, Uruguay has a strong performance in AI~\cite{centro_nacional_de_inteligencia_artificial_indice_2023} and developed a strategy to incorporate AI in government in 2019, which is currently being updated with the new version expected in August 2024~\cite{agesic_estrategia_2021,agesic_proceso_2023}. 
Uruguay is a democratic republic with a unitary state divided into nineteen departments.
The president leads the executive branch and is elected by direct vote every five years.
The legislative branch is constituted by the General Assembly, which consists of two chambers: the Chamber of Representatives, which has 99 members, and the Senate, which has 30 members, both elected every five years.
Uruguay's government (2020-2025) is led by President Luis Lacalle Pou, who governs with his center-right coalition.
The previous government (2015-2020) was led by Tabaré Vásquez, who governed with his center-left coalition. 

%examples of tech in gov in Chile / UY
\textit{\textbf{Technology use in civic participation processes.~}}
Chile and Uruguay have several experiences using technology (and sometimes AI) in public participation. 
For example, Chile has experimented with initiatives such as using NLP algorithms to analyze citizen dialogues to update its constitution~\cite{yovanovic_remote_2021, raveau_citizens_2022, fierro_200k_2017, cruz_measuring_2023}, to deliberate about future scenarios~\cite{goni_experiential_2023,goni_analytical_2024}, and to analyze its AI policy~\cite{guridi_supporting_2024}.
Similarly, Uruguay is well-known for its digital government practices~\cite{mendaro_uruguayan_2020} and has experimented with using different ICTs to promote civic engagement areas such as open government~\cite{rivoir_ict-mediated_2017, aguerre_open_2024}.

%\section{Interview Structure}

%\input{Tables/interview}

\bibliographystyle{ACM-Reference-Format}
\bibliography{references,additionalRef}


\begin{thebibliography}{141}

%%% the Web address.

\ifx \showCODEN    \undefined \def \showCODEN     #1{\unskip}     \fi
\ifx \showDOI      \undefined \def \showDOI       #1{#1}\fi
\ifx \showISBNx    \undefined \def \showISBNx     #1{\unskip}     \fi
\ifx \showISBNxiii \undefined \def \showISBNxiii  #1{\unskip}     \fi
\ifx \showISSN     \undefined \def \showISSN      #1{\unskip}     \fi
\ifx \showLCCN     \undefined \def \showLCCN      #1{\unskip}     \fi
\ifx \shownote     \undefined \def \shownote      #1{#1}          \fi
\ifx \showarticletitle \undefined \def \showarticletitle #1{#1}   \fi
\ifx \showURL      \undefined \def \showURL       {\relax}        \fi
% invisible to TeX
\providecommand\bibfield[2]{#2}
\providecommand\bibinfo[2]{#2}
\providecommand\natexlab[1]{#1}
\providecommand\showeprint[2][]{arXiv:#2}

\bibitem[AGESIC(2021)]%
        {agesic_estrategia_2021}
\bibfield{author}{\bibinfo{person}{AGESIC}.} \bibinfo{year}{2021}\natexlab{}.
\newblock \bibinfo{title}{Estrategia de {Inteligencia} {Artificial}}.
\newblock
\newblock
\urldef\tempurl%
\url{https://www.gub.uy/agencia-gobierno-electronico-sociedad-informacion-conocimiento/comunicacion/publicaciones/estrategia-inteligencia-artificial}
\showURL{%
\tempurl}


\bibitem[AGESIC(2023)]%
        {agesic_proceso_2023}
\bibfield{author}{\bibinfo{person}{AGESIC}.} \bibinfo{year}{2023}\natexlab{}.
\newblock \bibinfo{title}{Proceso de revisión {Estrategia} de {Inteligencia} {Artificial} y creación de la {Estrategia} {Nacional} de {Datos} - {Plataforma} de {Participación} {Ciudadana} {Digital}}.
\newblock
\newblock
\urldef\tempurl%
\url{https://plataformaparticipacionciudadana.gub.uy/processes/estrategia-ia-datos}
\showURL{%
\tempurl}


\bibitem[Aguerre and Bonina(2024)]%
        {aguerre_open_2024}
\bibfield{author}{\bibinfo{person}{Carolina Aguerre} {and} \bibinfo{person}{Carla Bonina}.} \bibinfo{year}{2024}\natexlab{}.
\newblock \showarticletitle{Open government, civic tech and digital platforms in {Latin} {America}: {A} governance study of {Montevideo}'s urban app ‘{Por} {Mi} {Barrio}’}.
\newblock \bibinfo{journal}{\emph{Information Systems Journal}} \bibinfo{volume}{34}, \bibinfo{number}{4} (\bibinfo{year}{2024}), \bibinfo{pages}{1037--1067}.
\newblock
\showISSN{1365-2575}
\urldef\tempurl%
\url{https://doi.org/10.1111/isj.12468}
\showDOI{\tempurl}
\newblock
\shownote{\_eprint: https://onlinelibrary.wiley.com/doi/pdf/10.1111/isj.12468}.


\bibitem[Aitamurto and Chen(2017)]%
        {aitamurto_value_2017}
\bibfield{author}{\bibinfo{person}{Tanja Aitamurto} {and} \bibinfo{person}{Kaiping Chen}.} \bibinfo{year}{2017}\natexlab{}.
\newblock \showarticletitle{The value of crowdsourcing in public policymaking: epistemic, democratic and economic value}.
\newblock \bibinfo{journal}{\emph{The Theory and Practice of Legislation}} \bibinfo{volume}{5}, \bibinfo{number}{1} (\bibinfo{date}{Jan.} \bibinfo{year}{2017}), \bibinfo{pages}{55--72}.
\newblock
\showISSN{2050-8840}
\urldef\tempurl%
\url{https://doi.org/10.1080/20508840.2017.1282665}
\showDOI{\tempurl}


\bibitem[Aitamurto and Landemore(2016)]%
        {aitamurto_crowdsourced_2016}
\bibfield{author}{\bibinfo{person}{Tanja Aitamurto} {and} \bibinfo{person}{Hélène Landemore}.} \bibinfo{year}{2016}\natexlab{}.
\newblock \showarticletitle{Crowdsourced {Deliberation}: {The} {Case} of the {Law} on {Off}-{Road} {Traffic} in {Finland}}.
\newblock \bibinfo{journal}{\emph{Policy \& Internet}} \bibinfo{volume}{8}, \bibinfo{number}{2} (\bibinfo{year}{2016}), \bibinfo{pages}{174--196}.
\newblock
\showISSN{1944-2866}
\urldef\tempurl%
\url{https://doi.org/10.1002/poi3.115}
\showDOI{\tempurl}


\bibitem[Aitamurto et~al\mbox{.}(2017)]%
        {aitamurto_unmasking_2017}
\bibfield{author}{\bibinfo{person}{Tanja Aitamurto}, \bibinfo{person}{Hélène Landemore}, {and} \bibinfo{person}{Jorge Saldivar~Galli}.} \bibinfo{year}{2017}\natexlab{}.
\newblock \showarticletitle{Unmasking the crowd: participants’ motivation factors, expectations, and profile in a crowdsourced law reform}.
\newblock \bibinfo{journal}{\emph{Information, Communication \& Society}} \bibinfo{volume}{20}, \bibinfo{number}{8} (\bibinfo{date}{Aug.} \bibinfo{year}{2017}), \bibinfo{pages}{1239--1260}.
\newblock
\showISSN{1369-118X}
\urldef\tempurl%
\url{https://doi.org/10.1080/1369118X.2016.1228993}
\showDOI{\tempurl}
\newblock
\shownote{Publisher: Routledge \_eprint: https://doi.org/10.1080/1369118X.2016.1228993}.


\bibitem[Alvarado~Garcia et~al\mbox{.}(2020)]%
        {alvarado_garcia_fostering_2020}
\bibfield{author}{\bibinfo{person}{Adriana Alvarado~Garcia}, \bibinfo{person}{Karla Badillo-Urquiola}, \bibinfo{person}{Mayra~D. Barrera~Machuca}, \bibinfo{person}{Franceli~L. Cibrian}, \bibinfo{person}{Marianela Ciolfi~Felice}, \bibinfo{person}{Laura~S. Gaytán-Lugo}, \bibinfo{person}{Diego Gómez-Zará}, \bibinfo{person}{Carla~F. Griggio}, \bibinfo{person}{Monica Perusquia-Hernandez}, \bibinfo{person}{Soraia Silva-Prietch}, \bibinfo{person}{Carlos~E. Tejada}, {and} \bibinfo{person}{Marisol Wong-Villacres}.} \bibinfo{year}{2020}\natexlab{}.
\newblock \showarticletitle{Fostering {HCI} {Research} in, by, and for {Latin} {America}}. In \bibinfo{booktitle}{\emph{Extended {Abstracts} of the 2020 {CHI} {Conference} on {Human} {Factors} in {Computing} {Systems}}} \emph{(\bibinfo{series}{{CHI} {EA} '20})}. \bibinfo{publisher}{Association for Computing Machinery}, \bibinfo{address}{New York, NY, USA}, \bibinfo{pages}{1--4}.
\newblock
\showISBNx{978-1-4503-6819-3}
\urldef\tempurl%
\url{https://doi.org/10.1145/3334480.3381055}
\showDOI{\tempurl}


\bibitem[Aragon et~al\mbox{.}(2020)]%
        {aragon_civic_2020}
\bibfield{author}{\bibinfo{person}{Pablo Aragon}, \bibinfo{person}{Adriana Alvarado~Garcia}, \bibinfo{person}{Christopher~A. Le~Dantec}, \bibinfo{person}{Claudia Flores-Saviaga}, {and} \bibinfo{person}{Jorge Saldivar}.} \bibinfo{year}{2020}\natexlab{}.
\newblock \showarticletitle{Civic {Technologies}: {Research}, {Practice} and {Open} {Challenges}}. In \bibinfo{booktitle}{\emph{Conference {Companion} {Publication} of the 2020 on {Computer} {Supported} {Cooperative} {Work} and {Social} {Computing}}} \emph{(\bibinfo{series}{{CSCW} '20 {Companion}})}. \bibinfo{publisher}{Association for Computing Machinery}, \bibinfo{address}{New York, NY, USA}, \bibinfo{pages}{537--545}.
\newblock
\showISBNx{978-1-4503-8059-1}
\urldef\tempurl%
\url{https://doi.org/10.1145/3406865.3430888}
\showDOI{\tempurl}


\bibitem[Arana-Catania et~al\mbox{.}(2021)]%
        {arana-catania_citizen_2021}
\bibfield{author}{\bibinfo{person}{Miguel Arana-Catania}, \bibinfo{person}{Felix-Anselm~Van Lier}, \bibinfo{person}{Rob Procter}, \bibinfo{person}{Nataliya Tkachenko}, \bibinfo{person}{Yulan He}, \bibinfo{person}{Arkaitz Zubiaga}, {and} \bibinfo{person}{Maria Liakata}.} \bibinfo{year}{2021}\natexlab{}.
\newblock \showarticletitle{Citizen {Participation} and {Machine} {Learning} for a {Better} {Democracy}}.
\newblock \bibinfo{journal}{\emph{Digital Government: Research and Practice}} \bibinfo{volume}{2}, \bibinfo{number}{3} (\bibinfo{date}{July} \bibinfo{year}{2021}), \bibinfo{pages}{1--22}.
\newblock
\showISSN{2691-199X, 2639-0175}
\urldef\tempurl%
\url{https://doi.org/10.1145/3452118}
\showDOI{\tempurl}


\bibitem[Arnstein(1969)]%
        {arnstein_ladder_1969}
\bibfield{author}{\bibinfo{person}{Sherry~R. Arnstein}.} \bibinfo{year}{1969}\natexlab{}.
\newblock \showarticletitle{A {Ladder} {Of} {Citizen} {Participation}}.
\newblock \bibinfo{journal}{\emph{Journal of the American Institute of Planners}} \bibinfo{volume}{35}, \bibinfo{number}{4} (\bibinfo{date}{July} \bibinfo{year}{1969}), \bibinfo{pages}{216--224}.
\newblock
\showISSN{0002-8991}
\urldef\tempurl%
\url{https://doi.org/10.1080/01944366908977225}
\showDOI{\tempurl}
\newblock
\shownote{Publisher: Routledge \_eprint: https://doi.org/10.1080/01944366908977225}.


\bibitem[Arya et~al\mbox{.}(2019)]%
        {arya_one_2019}
\bibfield{author}{\bibinfo{person}{Vijay Arya}, \bibinfo{person}{Rachel K.~E. Bellamy}, \bibinfo{person}{Pin-Yu Chen}, \bibinfo{person}{Amit Dhurandhar}, \bibinfo{person}{Michael Hind}, \bibinfo{person}{Samuel~C. Hoffman}, \bibinfo{person}{Stephanie Houde}, \bibinfo{person}{Q.~Vera Liao}, \bibinfo{person}{Ronny Luss}, \bibinfo{person}{Aleksandra Mojsilović}, \bibinfo{person}{Sami Mourad}, \bibinfo{person}{Pablo Pedemonte}, \bibinfo{person}{Ramya Raghavendra}, \bibinfo{person}{John Richards}, \bibinfo{person}{Prasanna Sattigeri}, \bibinfo{person}{Karthikeyan Shanmugam}, \bibinfo{person}{Moninder Singh}, \bibinfo{person}{Kush~R. Varshney}, \bibinfo{person}{Dennis Wei}, {and} \bibinfo{person}{Yunfeng Zhang}.} \bibinfo{year}{2019}\natexlab{}.
\newblock \bibinfo{title}{One {Explanation} {Does} {Not} {Fit} {All}: {A} {Toolkit} and {Taxonomy} of {AI} {Explainability} {Techniques}}.
\newblock
\newblock
\urldef\tempurl%
\url{https://doi.org/10.48550/arXiv.1909.03012}
\showDOI{\tempurl}
\newblock
\shownote{arXiv:1909.03012 [cs, stat]}.


\bibitem[Ash et~al\mbox{.}(2020)]%
        {ash_deep_2020}
\bibfield{author}{\bibinfo{person}{Jordan~T. Ash}, \bibinfo{person}{Chicheng Zhang}, \bibinfo{person}{Akshay Krishnamurthy}, \bibinfo{person}{John Langford}, {and} \bibinfo{person}{Alekh Agarwal}.} \bibinfo{year}{2020}\natexlab{}.
\newblock \bibinfo{title}{Deep {Batch} {Active} {Learning} by {Diverse}, {Uncertain} {Gradient} {Lower} {Bounds}}.
\newblock
\newblock
\urldef\tempurl%
\url{https://doi.org/10.48550/arXiv.1906.03671}
\showDOI{\tempurl}
\newblock
\shownote{arXiv:1906.03671 [cs, stat]}.


\bibitem[Balayn et~al\mbox{.}(2023)]%
        {balayn_fairness_2023}
\bibfield{author}{\bibinfo{person}{Agathe Balayn}, \bibinfo{person}{Mireia Yurrita}, \bibinfo{person}{Jie Yang}, {and} \bibinfo{person}{Ujwal Gadiraju}.} \bibinfo{year}{2023}\natexlab{}.
\newblock \showarticletitle{“{Fairness} {Toolkits}, {A} {Checkbox} {Culture}?” {On} the {Factors} that {Fragment} {Developer} {Practices} in {Handling} {Algorithmic} {Harms}}. In \bibinfo{booktitle}{\emph{Proceedings of the 2023 {AAAI}/{ACM} {Conference} on {AI}, {Ethics}, and {Society}}} \emph{(\bibinfo{series}{{AIES} '23})}. \bibinfo{publisher}{Association for Computing Machinery}, \bibinfo{address}{New York, NY, USA}, \bibinfo{pages}{482--495}.
\newblock
\showISBNx{9798400702310}
\urldef\tempurl%
\url{https://doi.org/10.1145/3600211.3604674}
\showDOI{\tempurl}


\bibitem[Bank(2023)]%
        {world_bank_world_2023}
\bibfield{author}{\bibinfo{person}{World Bank}.} \bibinfo{year}{2023}\natexlab{}.
\newblock \bibinfo{title}{World {Bank} {Open} {Data}}.
\newblock
\newblock
\urldef\tempurl%
\url{https://data.worldbank.org}
\showURL{%
\tempurl}


\bibitem[Baum(2015)]%
        {baum_citizen_2015}
\bibfield{author}{\bibinfo{person}{Howell~S. Baum}.} \bibinfo{year}{2015}\natexlab{}.
\newblock \showarticletitle{Citizen {Participation}}.
\newblock In \bibinfo{booktitle}{\emph{International {Encyclopedia} of the {Social} \& {Behavioral} {Sciences} ({Second} {Edition})}}, \bibfield{editor}{\bibinfo{person}{James~D. Wright}} (Ed.). \bibinfo{publisher}{Elsevier}, \bibinfo{address}{Oxford}, \bibinfo{pages}{625--630}.
\newblock
\showISBNx{978-0-08-097087-5}
\urldef\tempurl%
\url{https://doi.org/10.1016/B978-0-08-097086-8.74005-0}
\showDOI{\tempurl}


\bibitem[Behrendt et~al\mbox{.}(2024)]%
        {behrendt_aqua_2024}
\bibfield{author}{\bibinfo{person}{Maike Behrendt}, \bibinfo{person}{Stefan~Sylvius Wagner}, \bibinfo{person}{Marc Ziegele}, \bibinfo{person}{Lena Wilms}, \bibinfo{person}{Anke Stoll}, \bibinfo{person}{Dominique Heinbach}, {and} \bibinfo{person}{Stefan Harmeling}.} \bibinfo{year}{2024}\natexlab{}.
\newblock \bibinfo{title}{{AQuA} -- {Combining} {Experts}' and {Non}-{Experts}' {Views} {To} {Assess} {Deliberation} {Quality} in {Online} {Discussions} {Using} {LLMs}}.
\newblock
\newblock
\urldef\tempurl%
\url{https://doi.org/10.48550/arXiv.2404.02761}
\showDOI{\tempurl}
\newblock
\shownote{arXiv:2404.02761 [cs]}.


\bibitem[Birhane(2021)]%
        {birhane_algorithmic_2021}
\bibfield{author}{\bibinfo{person}{Abeba Birhane}.} \bibinfo{year}{2021}\natexlab{}.
\newblock \showarticletitle{Algorithmic injustice: a relational ethics approach}.
\newblock \bibinfo{journal}{\emph{Patterns}} \bibinfo{volume}{2}, \bibinfo{number}{2} (\bibinfo{date}{Feb.} \bibinfo{year}{2021}), \bibinfo{pages}{100205}.
\newblock
\showISSN{2666-3899}
\urldef\tempurl%
\url{https://doi.org/10.1016/j.patter.2021.100205}
\showDOI{\tempurl}


\bibitem[Birhane et~al\mbox{.}(2022)]%
        {birhane_power_2022}
\bibfield{author}{\bibinfo{person}{Abeba Birhane}, \bibinfo{person}{William Isaac}, \bibinfo{person}{Vinodkumar Prabhakaran}, \bibinfo{person}{Mark Diaz}, \bibinfo{person}{Madeleine~Clare Elish}, \bibinfo{person}{Iason Gabriel}, {and} \bibinfo{person}{Shakir Mohamed}.} \bibinfo{year}{2022}\natexlab{}.
\newblock \showarticletitle{Power to the {People}? {Opportunities} and {Challenges} for {Participatory} {AI}}. In \bibinfo{booktitle}{\emph{Equity and {Access} in {Algorithms}, {Mechanisms}, and {Optimization}}} \emph{(\bibinfo{series}{{EAAMO} '22})}. \bibinfo{publisher}{Association for Computing Machinery}, \bibinfo{address}{New York, NY, USA}, \bibinfo{pages}{1--8}.
\newblock
\showISBNx{978-1-4503-9477-2}
\urldef\tempurl%
\url{https://doi.org/10.1145/3551624.3555290}
\showDOI{\tempurl}


\bibitem[Bonina and Cordella(2009)]%
        {bonina_public_2009}
\bibfield{author}{\bibinfo{person}{Carla Bonina} {and} \bibinfo{person}{Antonio Cordella}.} \bibinfo{year}{2009}\natexlab{}.
\newblock \showarticletitle{Public {Sector} {Reforms} and the {Notion} of '{Public} {Value}': {Implications} for {eGovernment} {Deployment}}.
\newblock \bibinfo{journal}{\emph{AMCIS 2009 Proceedings}} (\bibinfo{date}{Jan.} \bibinfo{year}{2009}).
\newblock
\urldef\tempurl%
\url{https://aisel.aisnet.org/amcis2009/15}
\showURL{%
\tempurl}


\bibitem[Boyne(2002)]%
        {boyne_public_2002}
\bibfield{author}{\bibinfo{person}{George~A. Boyne}.} \bibinfo{year}{2002}\natexlab{}.
\newblock \showarticletitle{Public and {Private} {Management}: {What}’s the {Difference}?}
\newblock \bibinfo{journal}{\emph{Journal of Management Studies}} \bibinfo{volume}{39}, \bibinfo{number}{1} (\bibinfo{year}{2002}), \bibinfo{pages}{97--122}.
\newblock
\showISSN{1467-6486}
\urldef\tempurl%
\url{https://doi.org/10.1111/1467-6486.00284}
\showDOI{\tempurl}
\newblock
\shownote{\_eprint: https://onlinelibrary.wiley.com/doi/pdf/10.1111/1467-6486.00284}.


\bibitem[Bretschneider(1990)]%
        {bretschneider_management_1990}
\bibfield{author}{\bibinfo{person}{Stuart Bretschneider}.} \bibinfo{year}{1990}\natexlab{}.
\newblock \showarticletitle{Management {Information} {Systems} in {Public} and {Private} {Organizations}: {An} {Empirical} {Test}}.
\newblock \bibinfo{journal}{\emph{Public Administration Review}} \bibinfo{volume}{50}, \bibinfo{number}{5} (\bibinfo{year}{1990}), \bibinfo{pages}{536--545}.
\newblock
\showISSN{0033-3352}
\urldef\tempurl%
\url{https://doi.org/10.2307/976784}
\showDOI{\tempurl}
\newblock
\shownote{Publisher: [American Society for Public Administration, Wiley]}.


\bibitem[Cai et~al\mbox{.}(2018)]%
        {cai_interactive_2018}
\bibfield{author}{\bibinfo{person}{Guoray Cai}, \bibinfo{person}{Feng Sun}, {and} \bibinfo{person}{Yongzhong Sha}.} \bibinfo{year}{2018}\natexlab{}.
\newblock \showarticletitle{Interactive {Visualization} for {Topic} {Model} {Curation}}. In \bibinfo{booktitle}{\emph{{ESIDA} '18}}.
\newblock


\bibitem[Chen and Aitamurto(2019)]%
        {chen_barriers_2019}
\bibfield{author}{\bibinfo{person}{Kaiping Chen} {and} \bibinfo{person}{Tanja Aitamurto}.} \bibinfo{year}{2019}\natexlab{}.
\newblock \showarticletitle{Barriers for {Crowd}’s {Impact} in {Crowdsourced} {Policymaking}: {Civic} {Data} {Overload} and {Filter} {Hierarchy}}.
\newblock \bibinfo{journal}{\emph{International Public Management Journal}} \bibinfo{volume}{22}, \bibinfo{number}{1} (\bibinfo{date}{Jan.} \bibinfo{year}{2019}), \bibinfo{pages}{99--126}.
\newblock
\showISSN{1096-7494}
\urldef\tempurl%
\url{https://doi.org/10.1080/10967494.2018.1488780}
\showDOI{\tempurl}


\bibitem[Chun et~al\mbox{.}(2010)]%
        {chun_government_2010}
\bibfield{author}{\bibinfo{person}{Soon Chun}, \bibinfo{person}{Stuart Shulman}, \bibinfo{person}{Rodrigo Sandoval~Almazan}, {and} \bibinfo{person}{Eduard Hovy}.} \bibinfo{year}{2010}\natexlab{}.
\newblock \showarticletitle{Government 2.0: {Making} {Connections} {Between} {Citizens}, {Data} and {Government}}.
\newblock \bibinfo{journal}{\emph{Information Polity}}  \bibinfo{volume}{15} (\bibinfo{date}{April} \bibinfo{year}{2010}), \bibinfo{pages}{1--9}.
\newblock
\urldef\tempurl%
\url{https://doi.org/10.3233/IP-2010-0205}
\showDOI{\tempurl}


\bibitem[Corbin and Strauss(2014)]%
        {corbin_basics_2014}
\bibfield{author}{\bibinfo{person}{Juliet~M. Corbin} {and} \bibinfo{person}{Anselm~L. Strauss}.} \bibinfo{year}{2014}\natexlab{}.
\newblock \bibinfo{booktitle}{\emph{Basics of qualitative research: techniques and procedures for developing grounded theory} (\bibinfo{edition}{fourth edition} ed.)}.
\newblock \bibinfo{publisher}{SAGE}, \bibinfo{address}{Los Angeles}.
\newblock
\showISBNx{978-1-4129-9746-1}


\bibitem[Cruz et~al\mbox{.}(2023)]%
        {cruz_measuring_2023}
\bibfield{author}{\bibinfo{person}{Andrés Cruz}, \bibinfo{person}{Zachary Elkins}, \bibinfo{person}{Roy Gardner}, \bibinfo{person}{Matthew Martin}, {and} \bibinfo{person}{Ashley Moran}.} \bibinfo{year}{2023}\natexlab{}.
\newblock \showarticletitle{Measuring constitutional preferences: {A} new method for analyzing public consultation data}.
\newblock \bibinfo{journal}{\emph{PLOS ONE}} \bibinfo{volume}{18}, \bibinfo{number}{12} (\bibinfo{year}{2023}), \bibinfo{pages}{e0295396}.
\newblock
\showISSN{1932-6203}
\urldef\tempurl%
\url{https://doi.org/10.1371/journal.pone.0295396}
\showDOI{\tempurl}
\newblock
\shownote{Publisher: Public Library of Science}.


\bibitem[de~Inteligencia~Artificial(2023)]%
        {centro_nacional_de_inteligencia_artificial_indice_2023}
\bibfield{author}{\bibinfo{person}{Centro~Nacional de Inteligencia~Artificial}.} \bibinfo{year}{2023}\natexlab{}.
\newblock \bibinfo{booktitle}{\emph{Índice latinoamericano de inteligencia artificial}}.
\newblock \bibinfo{type}{{T}echnical {R}eport}. \bibinfo{institution}{Centro Nacional de Inteligencia Artificial}, \bibinfo{address}{Santiago, Chile}.
\newblock


\bibitem[Dean(2023)]%
        {dean_deliberating_2023}
\bibfield{author}{\bibinfo{person}{Rikki Dean}.} \bibinfo{year}{2023}\natexlab{}.
\newblock \showarticletitle{Deliberating {Like} a {State}: {Locating} {Public} {Administration} {Within} the {Deliberative} {System}}.
\newblock \bibinfo{journal}{\emph{Political Studies}} (\bibinfo{date}{April} \bibinfo{year}{2023}), \bibinfo{pages}{00323217231166285}.
\newblock
\showISSN{0032-3217}
\urldef\tempurl%
\url{https://doi.org/10.1177/00323217231166285}
\showDOI{\tempurl}
\newblock
\shownote{Publisher: SAGE Publications Ltd}.


\bibitem[Delgado et~al\mbox{.}(2023)]%
        {delgado_participatory_2023}
\bibfield{author}{\bibinfo{person}{Fernando Delgado}, \bibinfo{person}{Stephen Yang}, \bibinfo{person}{Michael Madaio}, {and} \bibinfo{person}{Qian Yang}.} \bibinfo{year}{2023}\natexlab{}.
\newblock \showarticletitle{The {Participatory} {Turn} in {AI} {Design}: {Theoretical} {Foundations} and the {Current} {State} of {Practice}}. In \bibinfo{booktitle}{\emph{Proceedings of the 3rd {ACM} {Conference} on {Equity} and {Access} in {Algorithms}, {Mechanisms}, and {Optimization}}} \emph{(\bibinfo{series}{{EAAMO} '23})}. \bibinfo{publisher}{Association for Computing Machinery}, \bibinfo{address}{New York, NY, USA}, \bibinfo{pages}{1--23}.
\newblock
\showISBNx{9798400703812}
\urldef\tempurl%
\url{https://doi.org/10.1145/3617694.3623261}
\showDOI{\tempurl}


\bibitem[De’ and Sarkar(2010)]%
        {de_rituals_2010}
\bibfield{author}{\bibinfo{person}{Rahul De’} {and} \bibinfo{person}{Sandeep Sarkar}.} \bibinfo{year}{2010}\natexlab{}.
\newblock \showarticletitle{Rituals in {E}-{Government} {Implementation}: {An} {Analysis} of {Failure}}. In \bibinfo{booktitle}{\emph{Electronic {Government}}} \emph{(\bibinfo{series}{Lecture {Notes} in {Computer} {Science}})}, \bibfield{editor}{\bibinfo{person}{Maria~A. Wimmer}, \bibinfo{person}{Jean-Loup Chappelet}, \bibinfo{person}{Marijn Janssen}, {and} \bibinfo{person}{Hans~J. Scholl}} (Eds.). \bibinfo{publisher}{Springer}, \bibinfo{address}{Berlin, Heidelberg}, \bibinfo{pages}{226--237}.
\newblock
\showISBNx{978-3-642-14799-9}
\urldef\tempurl%
\url{https://doi.org/10.1007/978-3-642-14799-9_20}
\showDOI{\tempurl}


\bibitem[Dor and Coglianese(2021)]%
        {dor_procurement_2021}
\bibfield{author}{\bibinfo{person}{Lavi M.~Ben Dor} {and} \bibinfo{person}{Cary Coglianese}.} \bibinfo{year}{2021}\natexlab{}.
\newblock \showarticletitle{Procurement as {AI} {Governance}}.
\newblock \bibinfo{journal}{\emph{IEEE Transactions on Technology and Society}} \bibinfo{volume}{2}, \bibinfo{number}{4} (\bibinfo{date}{Dec.} \bibinfo{year}{2021}), \bibinfo{pages}{192--199}.
\newblock
\showISSN{2637-6415}
\urldef\tempurl%
\url{https://doi.org/10.1109/TTS.2021.3111764}
\showDOI{\tempurl}
\newblock
\shownote{Conference Name: IEEE Transactions on Technology and Society}.


\bibitem[Edwards and Veale(2017)]%
        {edwards_slave_2017}
\bibfield{author}{\bibinfo{person}{Lilian Edwards} {and} \bibinfo{person}{Michael Veale}.} \bibinfo{year}{2017}\natexlab{}.
\newblock \bibinfo{booktitle}{\emph{Slave to the {Algorithm}? {Why} a 'right to an explanation' is probably not the remedy you are looking for}}.
\newblock \bibinfo{type}{preprint}. \bibinfo{institution}{LawArXiv}.
\newblock
\urldef\tempurl%
\url{https://doi.org/10.31228/osf.io/97upg}
\showDOI{\tempurl}


\bibitem[Ehsan et~al\mbox{.}(2021a)]%
        {ehsan_expanding_2021}
\bibfield{author}{\bibinfo{person}{Upol Ehsan}, \bibinfo{person}{Q.~Vera Liao}, \bibinfo{person}{Michael Muller}, \bibinfo{person}{Mark~O. Riedl}, {and} \bibinfo{person}{Justin~D. Weisz}.} \bibinfo{year}{2021}\natexlab{a}.
\newblock \showarticletitle{Expanding {Explainability}: {Towards} {Social} {Transparency} in {AI} systems}. In \bibinfo{booktitle}{\emph{Proceedings of the 2021 {CHI} {Conference} on {Human} {Factors} in {Computing} {Systems}}} \emph{(\bibinfo{series}{{CHI} '21})}. \bibinfo{publisher}{Association for Computing Machinery}, \bibinfo{address}{New York, NY, USA}, \bibinfo{pages}{1--19}.
\newblock
\showISBNx{978-1-4503-8096-6}
\urldef\tempurl%
\url{https://doi.org/10.1145/3411764.3445188}
\showDOI{\tempurl}


\bibitem[Ehsan et~al\mbox{.}(2021b)]%
        {ehsan_who_2021}
\bibfield{author}{\bibinfo{person}{Upol Ehsan}, \bibinfo{person}{Samir Passi}, \bibinfo{person}{Q.~Vera Liao}, \bibinfo{person}{Larry Chan}, \bibinfo{person}{I.-Hsiang Lee}, \bibinfo{person}{Michael Muller}, {and} \bibinfo{person}{Mark~O. Riedl}.} \bibinfo{year}{2021}\natexlab{b}.
\newblock \bibinfo{title}{The {Who} in {Explainable} {AI}: {How} {AI} {Background} {Shapes} {Perceptions} of {AI} {Explanations}}.
\newblock
\newblock
\urldef\tempurl%
\url{https://doi.org/10.48550/arXiv.2107.13509}
\showDOI{\tempurl}
\newblock
\shownote{arXiv:2107.13509 [cs]}.


\bibitem[Esaiasson(2010)]%
        {esaiasson_will_2010}
\bibfield{author}{\bibinfo{person}{Peter Esaiasson}.} \bibinfo{year}{2010}\natexlab{}.
\newblock \showarticletitle{Will citizens take no for an answer? {What} government officials can do to enhance decision acceptance}.
\newblock \bibinfo{journal}{\emph{European Political Science Review}} \bibinfo{volume}{2}, \bibinfo{number}{3} (\bibinfo{date}{Nov.} \bibinfo{year}{2010}), \bibinfo{pages}{351--371}.
\newblock
\showISSN{1755-7747, 1755-7739}
\urldef\tempurl%
\url{https://doi.org/10.1017/S1755773910000238}
\showDOI{\tempurl}
\newblock
\shownote{Publisher: Cambridge University Press}.


\bibitem[Fares(2023)]%
        {fares_role_2023}
\bibfield{author}{\bibinfo{person}{Dina Fares}.} \bibinfo{year}{2023}\natexlab{}.
\newblock \showarticletitle{The {Role} of {Large} {Language} {Models} ({LLMs}) {Driven} {Chatbots} in {Shaping} the {Future} of {Government} {Services} and {Communication} with {Citizens} in {UAE}}.
\newblock \bibinfo{journal}{\emph{Theses}} (\bibinfo{date}{Dec.} \bibinfo{year}{2023}).
\newblock
\urldef\tempurl%
\url{https://repository.rit.edu/theses/11694}
\showURL{%
\tempurl}


\bibitem[Fierro et~al\mbox{.}(2017)]%
        {fierro_200k_2017}
\bibfield{author}{\bibinfo{person}{Constanza Fierro}, \bibinfo{person}{Claudio Fuentes}, \bibinfo{person}{Jorge Pérez}, {and} \bibinfo{person}{Mauricio Quezada}.} \bibinfo{year}{2017}\natexlab{}.
\newblock \showarticletitle{{200K}+ {Crowdsourced} {Political} {Arguments} for a {New} {Chilean} {Constitution}}. In \bibinfo{booktitle}{\emph{Proceedings of the 4th {Workshop} on {Argument} {Mining}}}. \bibinfo{publisher}{Association for Computational Linguistics}, \bibinfo{address}{Copenhagen, Denmark}, \bibinfo{pages}{1--10}.
\newblock
\urldef\tempurl%
\url{https://doi.org/10.18653/v1/W17-5101}
\showDOI{\tempurl}


\bibitem[Flak and Nordheim(2006)]%
        {flak_stakeholders_2006}
\bibfield{author}{\bibinfo{person}{L.S. Flak} {and} \bibinfo{person}{S. Nordheim}.} \bibinfo{year}{2006}\natexlab{}.
\newblock \showarticletitle{Stakeholders, {Contradictions} and {Salience}: {An} {Empirical} {Study} of a {Norwegian} {G2G} {Effort}}. In \bibinfo{booktitle}{\emph{Proceedings of the 39th {Annual} {Hawaii} {International} {Conference} on {System} {Sciences} ({HICSS}'06)}}, Vol.~\bibinfo{volume}{4}. \bibinfo{pages}{75a--75a}.
\newblock
\urldef\tempurl%
\url{https://doi.org/10.1109/HICSS.2006.436}
\showDOI{\tempurl}
\newblock
\shownote{ISSN: 1530-1605}.


\bibitem[Font et~al\mbox{.}(2016)]%
        {font_tracing_2016}
\bibfield{author}{\bibinfo{person}{Joan Font}, \bibinfo{person}{Sara Pasadas~del Amo}, {and} \bibinfo{person}{Graham Smith}.} \bibinfo{year}{2016}\natexlab{}.
\newblock \showarticletitle{Tracing the {Impact} of {Proposals} from {Participatory} {Processes}: {Methodological} {Challenges} and {Substantive} {Lessons}}.
\newblock \bibinfo{journal}{\emph{Journal of Deliberative Democracy}} \bibinfo{volume}{12}, \bibinfo{number}{1} (\bibinfo{date}{June} \bibinfo{year}{2016}).
\newblock
\showISSN{2634-0488}
\urldef\tempurl%
\url{https://doi.org/10.16997/jdd.243}
\showDOI{\tempurl}
\newblock
\shownote{Number: 1 Publisher: University of Westminster Press}.


\bibitem[Giest(2017)]%
        {giest_big_2017}
\bibfield{author}{\bibinfo{person}{Sarah Giest}.} \bibinfo{year}{2017}\natexlab{}.
\newblock \showarticletitle{Big data for policymaking: fad or fasttrack?}
\newblock \bibinfo{journal}{\emph{Policy Sciences}} \bibinfo{volume}{50}, \bibinfo{number}{3} (\bibinfo{date}{Sept.} \bibinfo{year}{2017}), \bibinfo{pages}{367--382}.
\newblock
\showISSN{1573-0891}
\urldef\tempurl%
\url{https://doi.org/10.1007/s11077-017-9293-1}
\showDOI{\tempurl}


\bibitem[Goñi et~al\mbox{.}(2023)]%
        {goni_experiential_2023}
\bibfield{author}{\bibinfo{person}{Julian~“Iñaki” Goñi}, \bibinfo{person}{Claudio Fuentes}, {and} \bibinfo{person}{Maria~Paz Raveau}.} \bibinfo{year}{2023}\natexlab{}.
\newblock \showarticletitle{An experiential account of a large-scale interdisciplinary data analysis of public engagement}.
\newblock \bibinfo{journal}{\emph{AI \& SOCIETY}} \bibinfo{volume}{38}, \bibinfo{number}{2} (\bibinfo{date}{April} \bibinfo{year}{2023}), \bibinfo{pages}{581--593}.
\newblock
\showISSN{1435-5655}
\urldef\tempurl%
\url{https://doi.org/10.1007/s00146-022-01457-4}
\showDOI{\tempurl}


\bibitem[Goñi et~al\mbox{.}(2024)]%
        {goni_analytical_2024}
\bibfield{author}{\bibinfo{person}{Julian~“Iñaki” Goñi}, \bibinfo{person}{Maria~Paz Raveau}, {and} \bibinfo{person}{Claudio Fuentes~Bravo}.} \bibinfo{year}{2024}\natexlab{}.
\newblock \showarticletitle{Analytical categories to describe imaginations about the collective futures: {From} theory to linguistics to computational analysis}.
\newblock \bibinfo{journal}{\emph{Futures}}  \bibinfo{volume}{156} (\bibinfo{date}{Feb.} \bibinfo{year}{2024}), \bibinfo{pages}{103324}.
\newblock
\showISSN{0016-3287}
\urldef\tempurl%
\url{https://doi.org/10.1016/j.futures.2024.103324}
\showDOI{\tempurl}


\bibitem[Gray and Chivukula(2019)]%
        {gray_ethical_2019}
\bibfield{author}{\bibinfo{person}{Colin~M. Gray} {and} \bibinfo{person}{Shruthi~Sai Chivukula}.} \bibinfo{year}{2019}\natexlab{}.
\newblock \showarticletitle{Ethical {Mediation} in {UX} {Practice}}. In \bibinfo{booktitle}{\emph{Proceedings of the 2019 {CHI} {Conference} on {Human} {Factors} in {Computing} {Systems}}}. \bibinfo{publisher}{ACM}, \bibinfo{address}{Glasgow Scotland Uk}, \bibinfo{pages}{1--11}.
\newblock
\showISBNx{978-1-4503-5970-2}
\urldef\tempurl%
\url{https://doi.org/10.1145/3290605.3300408}
\showDOI{\tempurl}


\bibitem[Grudin and Poltrock(2012)]%
        {kozlowski_taxonomy_2012}
\bibfield{author}{\bibinfo{person}{Jonathan Grudin} {and} \bibinfo{person}{Steven Poltrock}.} \bibinfo{year}{2012}\natexlab{}.
\newblock \showarticletitle{Taxonomy and {Theory} in {Computer} {Supported} {Cooperative} {Work}}.
\newblock In \bibinfo{booktitle}{\emph{The {Oxford} {Handbook} of {Organizational} {Psychology}, {Volume} 2} (\bibinfo{edition}{1} ed.)}, \bibfield{editor}{\bibinfo{person}{Steve W.~J. Kozlowski}} (Ed.). \bibinfo{publisher}{Oxford University Press}, \bibinfo{pages}{1323--1348}.
\newblock
\showISBNx{978-0-19-992828-6 978-0-19-996883-1}
\urldef\tempurl%
\url{https://doi.org/10.1093/oxfordhb/9780199928286.013.0040}
\showDOI{\tempurl}


\bibitem[Guridi et~al\mbox{.}(2024a)]%
        {guridi_image_2024}
\bibfield{author}{\bibinfo{person}{Jose~A. Guridi}, \bibinfo{person}{Cristobal Cheyre}, \bibinfo{person}{Maria Goula}, \bibinfo{person}{Duarte Santo}, \bibinfo{person}{Lee Humphreys}, \bibinfo{person}{Aishwarya Shankar}, {and} \bibinfo{person}{Achilleas Souras}.} \bibinfo{year}{2024}\natexlab{a}.
\newblock \showarticletitle{Image {Generative} {AI} to {Design} {Public} {Spaces}: a {Reflection} of how {AI} {Could} {Improve} {Co}-{Design} of {Public} {Parks}}.
\newblock \bibinfo{journal}{\emph{Digital Government: Research and Practice}} (\bibinfo{year}{2024}).
\newblock
\urldef\tempurl%
\url{https://doi.org/10.1145/3656588}
\showDOI{\tempurl}
\newblock
\shownote{Just Accepted}.


\bibitem[Guridi et~al\mbox{.}(2024b)]%
        {guridi_supporting_2024}
\bibfield{author}{\bibinfo{person}{Jose~A. Guridi}, \bibinfo{person}{Julio~A. Pertuze}, {and} \bibinfo{person}{Catalina Zamora}.} \bibinfo{year}{2024}\natexlab{b}.
\newblock \showarticletitle{Supporting {Participation} {Processes} {Using} {NLP} in {Constrained} {Resources} {Settings}}. In \bibinfo{booktitle}{\emph{Proceedings of the {Ongoing} {Research}, {Practitioners}, {Posters}, {Workshops}, and {Projects} of the {International} {Conference} {EGOV}-{CeDEM}-{ePart} 2024}}, Vol.~\bibinfo{volume}{3737}. \bibinfo{address}{Ghent University and KU Leuven, Belgium}.
\newblock
\urldef\tempurl%
\url{https://ceur-ws.org/Vol-3737/paper30.pdf}
\showURL{%
\tempurl}


\bibitem[Hagen(2018)]%
        {hagen_content_2018}
\bibfield{author}{\bibinfo{person}{Loni Hagen}.} \bibinfo{year}{2018}\natexlab{}.
\newblock \showarticletitle{Content analysis of e-petitions with topic modeling: {How} to train and evaluate {LDA} models?}
\newblock \bibinfo{journal}{\emph{Information Processing \& Management}} \bibinfo{volume}{54}, \bibinfo{number}{6} (\bibinfo{date}{Nov.} \bibinfo{year}{2018}), \bibinfo{pages}{1292--1307}.
\newblock
\showISSN{0306-4573}
\urldef\tempurl%
\url{https://doi.org/10.1016/j.ipm.2018.05.006}
\showDOI{\tempurl}


\bibitem[Irvin and Stansbury(2004)]%
        {irvin_citizen_2004}
\bibfield{author}{\bibinfo{person}{Renée~A. Irvin} {and} \bibinfo{person}{John Stansbury}.} \bibinfo{year}{2004}\natexlab{}.
\newblock \showarticletitle{Citizen {Participation} in {Decision} {Making}: {Is} {It} {Worth} the {Effort}?}
\newblock \bibinfo{journal}{\emph{Public Administration Review}} \bibinfo{volume}{64}, \bibinfo{number}{1} (\bibinfo{year}{2004}), \bibinfo{pages}{55--65}.
\newblock
\showISSN{1540-6210}
\urldef\tempurl%
\url{https://doi.org/10.1111/j.1540-6210.2004.00346.x}
\showDOI{\tempurl}
\newblock
\shownote{\_eprint: https://onlinelibrary.wiley.com/doi/pdf/10.1111/j.1540-6210.2004.00346.x}.


\bibitem[Jackson et~al\mbox{.}(2014)]%
        {jackson_policy_2014}
\bibfield{author}{\bibinfo{person}{Steven~J. Jackson}, \bibinfo{person}{Tarleton Gillespie}, {and} \bibinfo{person}{Sandy Payette}.} \bibinfo{year}{2014}\natexlab{}.
\newblock \showarticletitle{The policy knot: re-integrating policy, practice and design in cscw studies of social computing}. In \bibinfo{booktitle}{\emph{Proceedings of the 17th {ACM} conference on {Computer} supported cooperative work \& social computing}} \emph{(\bibinfo{series}{{CSCW} '14})}. \bibinfo{publisher}{Association for Computing Machinery}, \bibinfo{address}{New York, NY, USA}, \bibinfo{pages}{588--602}.
\newblock
\showISBNx{978-1-4503-2540-0}
\urldef\tempurl%
\url{https://doi.org/10.1145/2531602.2531674}
\showDOI{\tempurl}


\bibitem[Jacobs et~al\mbox{.}(2020)]%
        {jacobs_meaning_2020}
\bibfield{author}{\bibinfo{person}{Abigail~Z. Jacobs}, \bibinfo{person}{Su~Lin Blodgett}, \bibinfo{person}{Solon Barocas}, \bibinfo{person}{Hal Daumé}, {and} \bibinfo{person}{Hanna Wallach}.} \bibinfo{year}{2020}\natexlab{}.
\newblock \showarticletitle{The meaning and measurement of bias: lessons from natural language processing}. In \bibinfo{booktitle}{\emph{Proceedings of the 2020 {Conference} on {Fairness}, {Accountability}, and {Transparency}}} \emph{(\bibinfo{series}{{FAT}* '20})}. \bibinfo{publisher}{Association for Computing Machinery}, \bibinfo{address}{New York, NY, USA}, \bibinfo{pages}{706}.
\newblock
\showISBNx{978-1-4503-6936-7}
\urldef\tempurl%
\url{https://doi.org/10.1145/3351095.3375671}
\showDOI{\tempurl}


\bibitem[Janssen et~al\mbox{.}(2012)]%
        {janssen_benefits_2012}
\bibfield{author}{\bibinfo{person}{Marijn Janssen}, \bibinfo{person}{Yannis Charalabidis}, {and} \bibinfo{person}{Anneke Zuiderwijk}.} \bibinfo{year}{2012}\natexlab{}.
\newblock \showarticletitle{Benefits, {Adoption} {Barriers} and {Myths} of {Open} {Data} and {Open} {Government}}.
\newblock \bibinfo{journal}{\emph{Information Systems Management}} \bibinfo{volume}{29}, \bibinfo{number}{4} (\bibinfo{date}{Sept.} \bibinfo{year}{2012}), \bibinfo{pages}{258--268}.
\newblock
\showISSN{1058-0530}
\urldef\tempurl%
\url{https://doi.org/10.1080/10580530.2012.716740}
\showDOI{\tempurl}


\bibitem[Jasim et~al\mbox{.}(2021)]%
        {jasim_communitypulse_2021}
\bibfield{author}{\bibinfo{person}{Mahmood Jasim}, \bibinfo{person}{Enamul Hoque}, \bibinfo{person}{Ali Sarvghad}, {and} \bibinfo{person}{Narges Mahyar}.} \bibinfo{year}{2021}\natexlab{}.
\newblock \showarticletitle{{CommunityPulse}: {Facilitating} {Community} {Input} {Analysis} by {Surfacing} {Hidden} {Insights}, {Reflections}, and {Priorities}}. In \bibinfo{booktitle}{\emph{Proceedings of the 2021 {ACM} {Designing} {Interactive} {Systems} {Conference}}} \emph{(\bibinfo{series}{{DIS} '21})}. \bibinfo{publisher}{Association for Computing Machinery}, \bibinfo{address}{New York, NY, USA}, \bibinfo{pages}{846--863}.
\newblock
\showISBNx{978-1-4503-8476-6}
\urldef\tempurl%
\url{https://doi.org/10.1145/3461778.3462132}
\showDOI{\tempurl}


\bibitem[Kaur et~al\mbox{.}(2022)]%
        {kaur_sensible_2022}
\bibfield{author}{\bibinfo{person}{Harmanpreet Kaur}, \bibinfo{person}{Eytan Adar}, \bibinfo{person}{Eric Gilbert}, {and} \bibinfo{person}{Cliff Lampe}.} \bibinfo{year}{2022}\natexlab{}.
\newblock \showarticletitle{Sensible {AI}: {Re}-imagining {Interpretability} and {Explainability} using {Sensemaking} {Theory}}. In \bibinfo{booktitle}{\emph{2022 {ACM} {Conference} on {Fairness}, {Accountability}, and {Transparency}}} \emph{(\bibinfo{series}{{FAccT} '22})}. \bibinfo{publisher}{Association for Computing Machinery}, \bibinfo{address}{New York, NY, USA}, \bibinfo{pages}{702--714}.
\newblock
\showISBNx{978-1-4503-9352-2}
\urldef\tempurl%
\url{https://doi.org/10.1145/3531146.3533135}
\showDOI{\tempurl}


\bibitem[Kawakami et~al\mbox{.}(2024a)]%
        {kawakami_studying_2024}
\bibfield{author}{\bibinfo{person}{Anna Kawakami}, \bibinfo{person}{Amanda Coston}, \bibinfo{person}{Hoda Heidari}, \bibinfo{person}{Kenneth Holstein}, {and} \bibinfo{person}{Haiyi Zhu}.} \bibinfo{year}{2024}\natexlab{a}.
\newblock \bibinfo{title}{Studying {Up} {Public} {Sector} {AI}: {How} {Networks} of {Power} {Relations} {Shape} {Agency} {Decisions} {Around} {AI} {Design} and {Use}}.
\newblock
\newblock
\urldef\tempurl%
\url{http://arxiv.org/abs/2405.12458}
\showURL{%
\tempurl}
\newblock
\shownote{arXiv:2405.12458 [cs]}.


\bibitem[Kawakami et~al\mbox{.}(2024b)]%
        {kawakami_situate_2024}
\bibfield{author}{\bibinfo{person}{Anna Kawakami}, \bibinfo{person}{Amanda Coston}, \bibinfo{person}{Haiyi Zhu}, \bibinfo{person}{Hoda Heidari}, {and} \bibinfo{person}{Kenneth Holstein}.} \bibinfo{year}{2024}\natexlab{b}.
\newblock \bibinfo{title}{The {Situate} {AI} {Guidebook}: {Co}-{Designing} a {Toolkit} to {Support} {Multi}-{Stakeholder} {Early}-stage {Deliberations} {Around} {Public} {Sector} {AI} {Proposals}}.
\newblock
\newblock
\urldef\tempurl%
\url{https://doi.org/10.1145/3613904.3642849}
\showDOI{\tempurl}
\newblock
\shownote{arXiv:2402.18774 [cs]}.


\bibitem[Kawakami et~al\mbox{.}(2022)]%
        {kawakami_improving_2022}
\bibfield{author}{\bibinfo{person}{Anna Kawakami}, \bibinfo{person}{Venkatesh Sivaraman}, \bibinfo{person}{Hao-Fei Cheng}, \bibinfo{person}{Logan Stapleton}, \bibinfo{person}{Yanghuidi Cheng}, \bibinfo{person}{Diana Qing}, \bibinfo{person}{Adam Perer}, \bibinfo{person}{Zhiwei~Steven Wu}, \bibinfo{person}{Haiyi Zhu}, {and} \bibinfo{person}{Kenneth Holstein}.} \bibinfo{year}{2022}\natexlab{}.
\newblock \showarticletitle{Improving {Human}-{AI} {Partnerships} in {Child} {Welfare}: {Understanding} {Worker} {Practices}, {Challenges}, and {Desires} for {Algorithmic} {Decision} {Support}}. In \bibinfo{booktitle}{\emph{Proceedings of the 2022 {CHI} {Conference} on {Human} {Factors} in {Computing} {Systems}}} \emph{(\bibinfo{series}{{CHI} '22})}. \bibinfo{publisher}{Association for Computing Machinery}, \bibinfo{address}{New York, NY, USA}, \bibinfo{pages}{1--18}.
\newblock
\showISBNx{978-1-4503-9157-3}
\urldef\tempurl%
\url{https://doi.org/10.1145/3491102.3517439}
\showDOI{\tempurl}


\bibitem[Kim(2024)]%
        {kim_2024_2024}
\bibfield{author}{\bibinfo{person}{Anthony Kim}.} \bibinfo{year}{2024}\natexlab{}.
\newblock \bibinfo{booktitle}{\emph{2024 {Index} of {Economic} {Freedom}}}.
\newblock \bibinfo{type}{{T}echnical {R}eport}. \bibinfo{institution}{The Heritage Foundation}, \bibinfo{address}{Washington DC, USA}.
\newblock


\bibitem[Kim et~al\mbox{.}(2021b)]%
        {kim_value_2021}
\bibfield{author}{\bibinfo{person}{Byungjun Kim}, \bibinfo{person}{Minjoo Yoo}, \bibinfo{person}{Keon~Chul Park}, \bibinfo{person}{Kyeo~Re Lee}, {and} \bibinfo{person}{Jang~Hyun Kim}.} \bibinfo{year}{2021}\natexlab{b}.
\newblock \showarticletitle{A value of civic voices for smart city: {A} big data analysis of civic queries posed by {Seoul} citizens}.
\newblock \bibinfo{journal}{\emph{Cities}}  \bibinfo{volume}{108} (\bibinfo{date}{Jan.} \bibinfo{year}{2021}), \bibinfo{pages}{102941}.
\newblock
\showISSN{0264-2751}
\urldef\tempurl%
\url{https://doi.org/10.1016/j.cities.2020.102941}
\showDOI{\tempurl}


\bibitem[Kim et~al\mbox{.}(2021a)]%
        {kim_improving_2021}
\bibfield{author}{\bibinfo{person}{Taewook Kim}, \bibinfo{person}{Hyunwoo Kim}, \bibinfo{person}{Juho Kim}, {and} \bibinfo{person}{Xiaojuan Ma}.} \bibinfo{year}{2021}\natexlab{a}.
\newblock \showarticletitle{Improving {Readers}’ {Awareness} of {Divergent} {Viewpoints} by {Displaying} {Agendas} of {Comments} in {Online} {News} {Discussions}}. In \bibinfo{booktitle}{\emph{Companion {Publication} of the 2021 {Conference} on {Computer} {Supported} {Cooperative} {Work} and {Social} {Computing}}} \emph{(\bibinfo{series}{{CSCW} '21})}. \bibinfo{publisher}{Association for Computing Machinery}, \bibinfo{address}{New York, NY, USA}, \bibinfo{pages}{99--103}.
\newblock
\showISBNx{978-1-4503-8479-7}
\urldef\tempurl%
\url{https://doi.org/10.1145/3462204.3481761}
\showDOI{\tempurl}


\bibitem[Koc-Michalska and Lilleker(2017)]%
        {koc-michalska_digital_2017}
\bibfield{author}{\bibinfo{person}{Karolina Koc-Michalska} {and} \bibinfo{person}{Darren Lilleker}.} \bibinfo{year}{2017}\natexlab{}.
\newblock \showarticletitle{Digital {Politics}: {Mobilization}, {Engagement}, and {Participation}}.
\newblock \bibinfo{journal}{\emph{Political Communication}} \bibinfo{volume}{34}, \bibinfo{number}{1} (\bibinfo{date}{Jan.} \bibinfo{year}{2017}), \bibinfo{pages}{1--5}.
\newblock
\showISSN{1058-4609}
\urldef\tempurl%
\url{https://doi.org/10.1080/10584609.2016.1243178}
\showDOI{\tempurl}


\bibitem[Lam et~al\mbox{.}(2023)]%
        {lam_sociotechnical_2023}
\bibfield{author}{\bibinfo{person}{Michelle~S. Lam}, \bibinfo{person}{Ayush Pandit}, \bibinfo{person}{Colin~H. Kalicki}, \bibinfo{person}{Rachit Gupta}, \bibinfo{person}{Poonam Sahoo}, {and} \bibinfo{person}{Danaë Metaxa}.} \bibinfo{year}{2023}\natexlab{}.
\newblock \showarticletitle{Sociotechnical {Audits}: {Broadening} the {Algorithm} {Auditing} {Lens} to {Investigate} {Targeted} {Advertising}}.
\newblock \bibinfo{journal}{\emph{Proceedings of the ACM on Human-Computer Interaction}} \bibinfo{volume}{7}, \bibinfo{number}{CSCW2} (\bibinfo{date}{Oct.} \bibinfo{year}{2023}), \bibinfo{pages}{360:1--360:37}.
\newblock
\urldef\tempurl%
\url{https://doi.org/10.1145/3610209}
\showDOI{\tempurl}


\bibitem[Lam et~al\mbox{.}(2024)]%
        {lam_concept_2024}
\bibfield{author}{\bibinfo{person}{Michelle~S. Lam}, \bibinfo{person}{Janice Teoh}, \bibinfo{person}{James~A. Landay}, \bibinfo{person}{Jeffrey Heer}, {and} \bibinfo{person}{Michael~S. Bernstein}.} \bibinfo{year}{2024}\natexlab{}.
\newblock \showarticletitle{Concept {Induction}: {Analyzing} {Unstructured} {Text} with {High}-{Level} {Concepts} {Using} {LLooM}}. In \bibinfo{booktitle}{\emph{Proceedings of the {CHI} {Conference} on {Human} {Factors} in {Computing} {Systems}}} \emph{(\bibinfo{series}{{CHI} '24})}. \bibinfo{publisher}{Association for Computing Machinery}, \bibinfo{address}{New York, NY, USA}, \bibinfo{pages}{1--28}.
\newblock
\showISBNx{9798400703300}
\urldef\tempurl%
\url{https://doi.org/10.1145/3613904.3642830}
\showDOI{\tempurl}


\bibitem[Levy et~al\mbox{.}(2021)]%
        {levy_algorithms_2021}
\bibfield{author}{\bibinfo{person}{Karen Levy}, \bibinfo{person}{Kyla~E. Chasalow}, {and} \bibinfo{person}{Sarah Riley}.} \bibinfo{year}{2021}\natexlab{}.
\newblock \showarticletitle{Algorithms and {Decision}-{Making} in the {Public} {Sector}}.
\newblock \bibinfo{journal}{\emph{Annual Review of Law and Social Science}} \bibinfo{volume}{17}, \bibinfo{number}{1} (\bibinfo{year}{2021}), \bibinfo{pages}{309--334}.
\newblock
\urldef\tempurl%
\url{https://doi.org/10.1146/annurev-lawsocsci-041221-023808}
\showDOI{\tempurl}
\newblock
\shownote{\_eprint: https://doi.org/10.1146/annurev-lawsocsci-041221-023808}.


\bibitem[Liao and Sundar(2022)]%
        {liao_designing_2022}
\bibfield{author}{\bibinfo{person}{Q.Vera Liao} {and} \bibinfo{person}{S.~Shyam Sundar}.} \bibinfo{year}{2022}\natexlab{}.
\newblock \showarticletitle{Designing for {Responsible} {Trust} in {AI} {Systems}: {A} {Communication} {Perspective}}. In \bibinfo{booktitle}{\emph{Proceedings of the 2022 {ACM} {Conference} on {Fairness}, {Accountability}, and {Transparency}}} \emph{(\bibinfo{series}{{FAccT} '22})}. \bibinfo{publisher}{Association for Computing Machinery}, \bibinfo{address}{New York, NY, USA}, \bibinfo{pages}{1257--1268}.
\newblock
\showISBNx{978-1-4503-9352-2}
\urldef\tempurl%
\url{https://doi.org/10.1145/3531146.3533182}
\showDOI{\tempurl}


\bibitem[Liu(2017)]%
        {liu_crowdsourcing_2017}
\bibfield{author}{\bibinfo{person}{Helen~K. Liu}.} \bibinfo{year}{2017}\natexlab{}.
\newblock \showarticletitle{Crowdsourcing {Government}: {Lessons} from {Multiple} {Disciplines}}.
\newblock \bibinfo{journal}{\emph{Public Administration Review}} \bibinfo{volume}{77}, \bibinfo{number}{5} (\bibinfo{year}{2017}), \bibinfo{pages}{656--667}.
\newblock
\showISSN{1540-6210}
\urldef\tempurl%
\url{https://doi.org/10.1111/puar.12808}
\showDOI{\tempurl}
\newblock
\shownote{\_eprint: https://onlinelibrary.wiley.com/doi/pdf/10.1111/puar.12808}.


\bibitem[Livermore et~al\mbox{.}(2017)]%
        {livermore_computationally_2017}
\bibfield{author}{\bibinfo{person}{Michael~A. Livermore}, \bibinfo{person}{Vladimir Eidelman}, {and} \bibinfo{person}{Brian Grom}.} \bibinfo{year}{2017}\natexlab{}.
\newblock \showarticletitle{Computationally {Assisted} {Regulatory} {Participation}}.
\newblock \bibinfo{journal}{\emph{Notre Dame Law Review}} \bibinfo{volume}{93}, \bibinfo{number}{3} (\bibinfo{year}{2017}), \bibinfo{pages}{977--1034}.
\newblock
\urldef\tempurl%
\url{https://heinonline.org/HOL/P?h=hein.journals/tndl93&i=1011}
\showURL{%
\tempurl}


\bibitem[Ma et~al\mbox{.}(2016)]%
        {ma_semantic_2016}
\bibfield{author}{\bibinfo{person}{Baojun Ma}, \bibinfo{person}{Nan Zhang}, \bibinfo{person}{Guannan Liu}, \bibinfo{person}{Liangqiang Li}, {and} \bibinfo{person}{Hua Yuan}.} \bibinfo{year}{2016}\natexlab{}.
\newblock \showarticletitle{Semantic search for public opinions on urban affairs: {A} probabilistic topic modeling-based approach}.
\newblock \bibinfo{journal}{\emph{Information Processing \& Management}} \bibinfo{volume}{52}, \bibinfo{number}{3} (\bibinfo{date}{May} \bibinfo{year}{2016}), \bibinfo{pages}{430--445}.
\newblock
\showISSN{0306-4573}
\urldef\tempurl%
\url{https://doi.org/10.1016/j.ipm.2015.10.004}
\showDOI{\tempurl}


\bibitem[Madaio et~al\mbox{.}(2020)]%
        {madaio_co-designing_2020}
\bibfield{author}{\bibinfo{person}{Michael~A. Madaio}, \bibinfo{person}{Luke Stark}, \bibinfo{person}{Jennifer Wortman~Vaughan}, {and} \bibinfo{person}{Hanna Wallach}.} \bibinfo{year}{2020}\natexlab{}.
\newblock \showarticletitle{Co-{Designing} {Checklists} to {Understand} {Organizational} {Challenges} and {Opportunities} around {Fairness} in {AI}}. In \bibinfo{booktitle}{\emph{Proceedings of the 2020 {CHI} {Conference} on {Human} {Factors} in {Computing} {Systems}}}. \bibinfo{publisher}{ACM}, \bibinfo{address}{Honolulu HI USA}, \bibinfo{pages}{1--14}.
\newblock
\showISBNx{978-1-4503-6708-0}
\urldef\tempurl%
\url{https://doi.org/10.1145/3313831.3376445}
\showDOI{\tempurl}


\bibitem[Madan and Ashok(2023)]%
        {madan_ai_2023}
\bibfield{author}{\bibinfo{person}{Rohit Madan} {and} \bibinfo{person}{Mona Ashok}.} \bibinfo{year}{2023}\natexlab{}.
\newblock \showarticletitle{{AI} adoption and diffusion in public administration: {A} systematic literature review and future research agenda}.
\newblock \bibinfo{journal}{\emph{Government Information Quarterly}} \bibinfo{volume}{40}, \bibinfo{number}{1} (\bibinfo{date}{Jan.} \bibinfo{year}{2023}), \bibinfo{pages}{101774}.
\newblock
\showISSN{0740-624X}
\urldef\tempurl%
\url{https://doi.org/10.1016/j.giq.2022.101774}
\showDOI{\tempurl}


\bibitem[Mahyar et~al\mbox{.}(2019)]%
        {mahyar_civic_2019}
\bibfield{author}{\bibinfo{person}{Narges Mahyar}, \bibinfo{person}{Diana~V. Nguyen}, \bibinfo{person}{Maggie Chan}, \bibinfo{person}{Jiayi Zheng}, {and} \bibinfo{person}{Steven~P. Dow}.} \bibinfo{year}{2019}\natexlab{}.
\newblock \showarticletitle{The {Civic} {Data} {Deluge}: {Understanding} the {Challenges} of {Analyzing} {Large}-{Scale} {Community} {Input}}. In \bibinfo{booktitle}{\emph{Proceedings of the 2019 on {Designing} {Interactive} {Systems} {Conference}}} \emph{(\bibinfo{series}{{DIS} '19})}. \bibinfo{publisher}{Association for Computing Machinery}, \bibinfo{address}{New York, NY, USA}, \bibinfo{pages}{1171--1181}.
\newblock
\showISBNx{978-1-4503-5850-7}
\urldef\tempurl%
\url{https://doi.org/10.1145/3322276.3322354}
\showDOI{\tempurl}


\bibitem[Maslej et~al\mbox{.}(2023)]%
        {maslej_ai_2023}
\bibfield{author}{\bibinfo{person}{Nestor Maslej}, \bibinfo{person}{Loredana Fattorini}, \bibinfo{person}{Erik Brynjolfsson}, \bibinfo{person}{John Etchemendy}, \bibinfo{person}{Katrina Ligett}, \bibinfo{person}{Terah Lyons}, \bibinfo{person}{James Manyika}, \bibinfo{person}{Helen Ngo}, \bibinfo{person}{Juan~Carlos Niebles}, \bibinfo{person}{Vanessa Parli}, \bibinfo{person}{Yoav Shoham}, \bibinfo{person}{Russell Wald}, \bibinfo{person}{Jack Clark}, {and} \bibinfo{person}{Raymond Perrault}.} \bibinfo{year}{2023}\natexlab{}.
\newblock \bibinfo{booktitle}{\emph{The {AI} {Index} 2023 {Annual} {Report}}}.
\newblock \bibinfo{type}{{T}echnical {R}eport}. \bibinfo{institution}{Stanford University}.
\newblock
\urldef\tempurl%
\url{https://aiindex.stanford.edu/wp-content/uploads/2023/04/HAI_AI-Index-Report_2023.pdf}
\showURL{%
\tempurl}


\bibitem[Meisner et~al\mbox{.}(2022)]%
        {meisner_labor_2022}
\bibfield{author}{\bibinfo{person}{Colten Meisner}, \bibinfo{person}{Brooke~Erin Duffy}, {and} \bibinfo{person}{Malte Ziewitz}.} \bibinfo{year}{2022}\natexlab{}.
\newblock \showarticletitle{The labor of search engine evaluation: {Making} algorithms more human or humans more algorithmic?}
\newblock \bibinfo{journal}{\emph{New Media \& Society}} (\bibinfo{date}{Jan.} \bibinfo{year}{2022}), \bibinfo{pages}{14614448211063860}.
\newblock
\showISSN{1461-4448}
\urldef\tempurl%
\url{https://doi.org/10.1177/14614448211063860}
\showDOI{\tempurl}
\newblock
\shownote{Publisher: SAGE Publications}.


\bibitem[Mendaro(2020)]%
        {mendaro_uruguayan_2020}
\bibfield{author}{\bibinfo{person}{Maria Laura~Rodríguez Mendaro}.} \bibinfo{year}{2020}\natexlab{}.
\newblock \showarticletitle{The {Uruguayan} {Digital} {Data} {Journey}}.
\newblock \bibinfo{journal}{\emph{Patterns}} \bibinfo{volume}{1}, \bibinfo{number}{3} (\bibinfo{date}{June} \bibinfo{year}{2020}).
\newblock
\showISSN{2666-3899}
\urldef\tempurl%
\url{https://doi.org/10.1016/j.patter.2020.100047}
\showDOI{\tempurl}
\newblock
\shownote{Publisher: Elsevier}.


\bibitem[Mendelson(2012)]%
        {mendelson_should_2012}
\bibfield{author}{\bibinfo{person}{Nina~A. Mendelson}.} \bibinfo{year}{2012}\natexlab{}.
\newblock \showarticletitle{Should {Mass} {Comments} {Count}? {Response} {Essay}}.
\newblock \bibinfo{journal}{\emph{Michigan Journal of Environmental \& Administrative Law}} \bibinfo{volume}{2}, \bibinfo{number}{1} (\bibinfo{year}{2012}), \bibinfo{pages}{173--184}.
\newblock
\urldef\tempurl%
\url{https://heinonline.org/HOL/P?h=hein.journals/michjo2&i=173}
\showURL{%
\tempurl}


\bibitem[{MinCTCI}(2021)]%
        {minctci_politica_2021}
\bibfield{author}{\bibinfo{person}{{MinCTCI}}.} \bibinfo{year}{2021}\natexlab{}.
\newblock \bibinfo{booktitle}{\emph{Politica {Nacional} de {Inteligencia} {Artificial}}}.
\newblock \bibinfo{type}{{T}echnical {R}eport}. \bibinfo{institution}{Ministerio de Ciencia Tecnologia Conocimiento e Innovacion}, \bibinfo{address}{Chile}.
\newblock
\urldef\tempurl%
\url{https://minciencia.gob.cl/uploads/filer_public/bc/38/bc389daf-4514-4306-867c-760ae7686e2c/documento_politica_ia_digital_.pdf}
\showURL{%
\tempurl}


\bibitem[MinCTCI(2024)]%
        {minctci_politica_2024}
\bibfield{author}{\bibinfo{person}{MinCTCI}.} \bibinfo{year}{2024}\natexlab{}.
\newblock \bibinfo{booktitle}{\emph{Política {Nacional} de {Inteligencia} {Artificial} - {Actualización} 2024}}.
\newblock \bibinfo{type}{{T}echnical {R}eport}. \bibinfo{institution}{Ministerio de Ciencia Tecnologia Conocimiento e Innovacion}.
\newblock
\urldef\tempurl%
\url{https://drive.google.com/file/d/11OLxLp8NyKgpeRFLe45X0zStY7SFEJIC/view?usp=sharing&usp=embed_facebook}
\showURL{%
\tempurl}


\bibitem[Méndez et~al\mbox{.}(2022a)]%
        {mendez_exploring_2022}
\bibfield{author}{\bibinfo{person}{Gonzalo~Gabriel Méndez}, \bibinfo{person}{Katherine Chiluiza}, \bibinfo{person}{Javier Tibau}, \bibinfo{person}{Vanessa~Ines Cedeno-Mieles}, \bibinfo{person}{Oscar Moreno}, \bibinfo{person}{Miguel Murillo}, {and} \bibinfo{person}{Marisol Wong-Villacres}.} \bibinfo{year}{2022}\natexlab{a}.
\newblock \showarticletitle{Exploring {Open} {Parliament} {Initiatives} in {Ecuador} {Through} {Technology}}. In \bibinfo{booktitle}{\emph{Extended {Abstracts} of the 2022 {CHI} {Conference} on {Human} {Factors} in {Computing} {Systems}}} \emph{(\bibinfo{series}{{CHI} {EA} '22})}. \bibinfo{publisher}{Association for Computing Machinery}, \bibinfo{address}{New York, NY, USA}, \bibinfo{pages}{1--8}.
\newblock
\showISBNx{978-1-4503-9156-6}
\urldef\tempurl%
\url{https://doi.org/10.1145/3491101.3519763}
\showDOI{\tempurl}


\bibitem[Méndez et~al\mbox{.}(2022b)]%
        {mendez_legislatio_2022}
\bibfield{author}{\bibinfo{person}{Gonzalo~Gabriel Méndez}, \bibinfo{person}{Oscar Moreno}, {and} \bibinfo{person}{Patricio Mendoza}.} \bibinfo{year}{2022}\natexlab{b}.
\newblock \showarticletitle{{LegisLatio}: {A} visualization {Tool} for {Legislative} {Roll}-call {Vote} {Data}}. In \bibinfo{booktitle}{\emph{Proceedings of the 15th {International} {Symposium} on {Visual} {Information} {Communication} and {Interaction}}} \emph{(\bibinfo{series}{{VINCI} '22})}. \bibinfo{publisher}{Association for Computing Machinery}, \bibinfo{address}{New York, NY, USA}, \bibinfo{pages}{1--8}.
\newblock
\showISBNx{978-1-4503-9806-0}
\urldef\tempurl%
\url{https://doi.org/10.1145/3554944.3554957}
\showDOI{\tempurl}


\bibitem[Nagitta et~al\mbox{.}(2022)]%
        {nagitta_human-centered_2022}
\bibfield{author}{\bibinfo{person}{Pross~Oluka Nagitta}, \bibinfo{person}{Godfrey Mugurusi}, \bibinfo{person}{Peter~Adoko Obicci}, {and} \bibinfo{person}{Emmanuel Awuor}.} \bibinfo{year}{2022}\natexlab{}.
\newblock \showarticletitle{Human-centered artificial intelligence for the public sector: {The} gate keeping role of the public procurement professional}.
\newblock \bibinfo{journal}{\emph{Procedia Computer Science}}  \bibinfo{volume}{200} (\bibinfo{date}{Jan.} \bibinfo{year}{2022}), \bibinfo{pages}{1084--1092}.
\newblock
\showISSN{1877-0509}
\urldef\tempurl%
\url{https://doi.org/10.1016/j.procs.2022.01.308}
\showDOI{\tempurl}


\bibitem[Orange et~al\mbox{.}(2007)]%
        {orange_care_2007}
\bibfield{author}{\bibinfo{person}{Graham Orange}, \bibinfo{person}{Alan Burke}, \bibinfo{person}{Tony Elliman}, {and} \bibinfo{person}{Ah~Lian Kor}.} \bibinfo{year}{2007}\natexlab{}.
\newblock \showarticletitle{{CARE}: {An} {Integrated} {Framework} to {Support} {Continuous}, {Adaptable}, {Reflective} {Evaluation} of {E}-{Government} {Systems}}.
\newblock \bibinfo{journal}{\emph{International Journal of Cases on Electronic Commerce (IJCEC)}} \bibinfo{volume}{3}, \bibinfo{number}{3} (\bibinfo{date}{July} \bibinfo{year}{2007}), \bibinfo{pages}{18--32}.
\newblock
\showISSN{1548-0623}
\urldef\tempurl%
\url{https://doi.org/10.4018/jcec.2007070102}
\showDOI{\tempurl}
\newblock
\shownote{Publisher: IGI Global}.


\bibitem[Panopoulou et~al\mbox{.}(2010)]%
        {panopoulou_eparticipation_2010}
\bibfield{author}{\bibinfo{person}{Eleni Panopoulou}, \bibinfo{person}{Efthimios Tambouris}, {and} \bibinfo{person}{Konstantinos Tarabanis}.} \bibinfo{year}{2010}\natexlab{}.
\newblock \showarticletitle{{eParticipation} {Initiatives} in {Europe}: {Learning} from {Practitioners}}. In \bibinfo{booktitle}{\emph{Electronic {Participation}}} \emph{(\bibinfo{series}{Lecture {Notes} in {Computer} {Science}})}, \bibfield{editor}{\bibinfo{person}{Efthimios Tambouris}, \bibinfo{person}{Ann Macintosh}, {and} \bibinfo{person}{Olivier Glassey}} (Eds.). \bibinfo{publisher}{Springer}, \bibinfo{address}{Berlin, Heidelberg}, \bibinfo{pages}{54--65}.
\newblock
\showISBNx{978-3-642-15158-3}
\urldef\tempurl%
\url{https://doi.org/10.1007/978-3-642-15158-3_5}
\showDOI{\tempurl}


\bibitem[Passi and Jackson(2018)]%
        {passi_trust_2018}
\bibfield{author}{\bibinfo{person}{Samir Passi} {and} \bibinfo{person}{Steven~J. Jackson}.} \bibinfo{year}{2018}\natexlab{}.
\newblock \showarticletitle{Trust in {Data} {Science}: {Collaboration}, {Translation}, and {Accountability} in {Corporate} {Data} {Science} {Projects}}.
\newblock \bibinfo{journal}{\emph{Proceedings of the ACM on Human-Computer Interaction}} \bibinfo{volume}{2}, \bibinfo{number}{CSCW} (\bibinfo{date}{Nov.} \bibinfo{year}{2018}), \bibinfo{pages}{136:1--136:28}.
\newblock
\urldef\tempurl%
\url{https://doi.org/10.1145/3274405}
\showDOI{\tempurl}


\bibitem[Perez(2008)]%
        {perez_complexity_2008}
\bibfield{author}{\bibinfo{person}{Oren Perez}.} \bibinfo{year}{2008}\natexlab{}.
\newblock \showarticletitle{Complexity, {Information} {Overload}, and {Online} {Deliberation} {Online} {Consultation} and {Democratic} {Communication}}.
\newblock \bibinfo{journal}{\emph{I/S: A Journal of Law and Policy for the Information Society}} \bibinfo{volume}{5}, \bibinfo{number}{1} (\bibinfo{year}{2008}), \bibinfo{pages}{43--86}.
\newblock
\urldef\tempurl%
\url{https://heinonline.org/HOL/P?h=hein.journals/isjlpsoc5&i=53}
\showURL{%
\tempurl}


\bibitem[Persson and Goldkuhl(2010)]%
        {persson_government_2010}
\bibfield{author}{\bibinfo{person}{Anders Persson} {and} \bibinfo{person}{Göran Goldkuhl}.} \bibinfo{year}{2010}\natexlab{}.
\newblock \showarticletitle{Government {Value} {Paradigms}—{Bureaucracy}, {New} {Public} {Management}, and {E}-{Government}}.
\newblock \bibinfo{journal}{\emph{Communications of the Association for Information Systems}} \bibinfo{volume}{27}, \bibinfo{number}{1} (\bibinfo{date}{July} \bibinfo{year}{2010}).
\newblock
\showISSN{1529-3181}
\urldef\tempurl%
\url{https://doi.org/10.17705/1CAIS.02704}
\showDOI{\tempurl}


\bibitem[Pečarič(2017)]%
        {pecaric_can_2017}
\bibfield{author}{\bibinfo{person}{Mirko Pečarič}.} \bibinfo{year}{2017}\natexlab{}.
\newblock \showarticletitle{Can a group of people be smarter than experts?}
\newblock \bibinfo{journal}{\emph{The Theory and Practice of Legislation}} \bibinfo{volume}{5}, \bibinfo{number}{1} (\bibinfo{date}{Jan.} \bibinfo{year}{2017}), \bibinfo{pages}{5--29}.
\newblock
\showISSN{2050-8840}
\urldef\tempurl%
\url{https://doi.org/10.1080/20508840.2016.1259823}
\showDOI{\tempurl}


\bibitem[Poel et~al\mbox{.}(2018)]%
        {poel_big_2018}
\bibfield{author}{\bibinfo{person}{Martijn Poel}, \bibinfo{person}{Eric~T. Meyer}, {and} \bibinfo{person}{Ralph Schroeder}.} \bibinfo{year}{2018}\natexlab{}.
\newblock \showarticletitle{Big {Data} for {Policymaking}: {Great} {Expectations}, but with {Limited} {Progress}?}
\newblock \bibinfo{journal}{\emph{Policy \& Internet}} \bibinfo{volume}{10}, \bibinfo{number}{3} (\bibinfo{year}{2018}), \bibinfo{pages}{347--367}.
\newblock
\showISSN{1944-2866}
\urldef\tempurl%
\url{https://doi.org/10.1002/poi3.176}
\showDOI{\tempurl}
\newblock
\shownote{\_eprint: https://onlinelibrary.wiley.com/doi/pdf/10.1002/poi3.176}.


\bibitem[Pogrebinschi(2021)]%
        {pogrebinschi_thirty_2021}
\bibfield{author}{\bibinfo{person}{Thamy Pogrebinschi}.} \bibinfo{year}{2021}\natexlab{}.
\newblock \showarticletitle{Thirty {Years} of {Democratic} {Innovations} in {Latin} {America}}.
\newblock  (\bibinfo{year}{2021}).
\newblock
\urldef\tempurl%
\url{https://www.econstor.eu/handle/10419/235143}
\showURL{%
\tempurl}
\newblock
\shownote{Publisher: Berlin: WZB Berlin Social Science Center}.


\bibitem[Porwol et~al\mbox{.}(2022)]%
        {porwol_transforming_2022}
\bibfield{author}{\bibinfo{person}{Lukasz Porwol}, \bibinfo{person}{Agustin Garcia~Pereira}, {and} \bibinfo{person}{Catherine Dumas}.} \bibinfo{year}{2022}\natexlab{}.
\newblock \showarticletitle{Transforming e-participation: {VR}-dialogue – building and evaluating an {AI}-supported framework for next-gen {VR}-enabled e-participation research}.
\newblock \bibinfo{journal}{\emph{Transforming Government: People, Process and Policy}} \bibinfo{volume}{17}, \bibinfo{number}{2} (\bibinfo{date}{Jan.} \bibinfo{year}{2022}), \bibinfo{pages}{233--250}.
\newblock
\showISSN{1750-6166}
\urldef\tempurl%
\url{https://doi.org/10.1108/TG-12-2021-0205}
\showDOI{\tempurl}
\newblock
\shownote{Publisher: Emerald Publishing Limited}.


\bibitem[Quinones(2014)]%
        {quinones_cultivating_2014}
\bibfield{author}{\bibinfo{person}{Pablo-Alejandro Quinones}.} \bibinfo{year}{2014}\natexlab{}.
\newblock \showarticletitle{Cultivating practice \& shepherding technology use: supporting appropriation among unanticipated users}. In \bibinfo{booktitle}{\emph{Proceedings of the 17th {ACM} conference on {Computer} supported cooperative work \& social computing}} \emph{(\bibinfo{series}{{CSCW} '14})}. \bibinfo{publisher}{Association for Computing Machinery}, \bibinfo{address}{New York, NY, USA}, \bibinfo{pages}{305--318}.
\newblock
\showISBNx{978-1-4503-2540-0}
\urldef\tempurl%
\url{https://doi.org/10.1145/2531602.2531698}
\showDOI{\tempurl}


\bibitem[Rappaport(2020)]%
        {Rappaport2020}
\bibfield{author}{\bibinfo{person}{J Rappaport}.} \bibinfo{year}{2020}\natexlab{}.
\newblock \bibinfo{booktitle}{\emph{{Cowards Don't Make History: Orlando Fals Borda and the Origins of Participatory Action Research}}}.
\newblock \bibinfo{publisher}{Duke University Press}, \bibinfo{address}{Durnham}.
\newblock
\showISBNx{9781478012542}
\urldef\tempurl%
\url{https://books.google.cl/books?id=p2n9DwAAQBAJ}
\showURL{%
\tempurl}


\bibitem[Raveau et~al\mbox{.}(2022)]%
        {raveau_citizens_2022}
\bibfield{author}{\bibinfo{person}{María~Paz Raveau}, \bibinfo{person}{Juan~Pablo Couyoumdjian}, \bibinfo{person}{Claudio Fuentes-Bravo}, \bibinfo{person}{Carlos Rodriguez-Sickert}, {and} \bibinfo{person}{Cristian Candia}.} \bibinfo{year}{2022}\natexlab{}.
\newblock \showarticletitle{Citizens at the forefront of the constitutional debate: {Voluntary} citizen participation determinants and emergent content in {Chile}}.
\newblock \bibinfo{journal}{\emph{PLOS ONE}} \bibinfo{volume}{17}, \bibinfo{number}{6} (\bibinfo{date}{June} \bibinfo{year}{2022}), \bibinfo{pages}{e0267443}.
\newblock
\showISSN{1932-6203}
\urldef\tempurl%
\url{https://doi.org/10.1371/journal.pone.0267443}
\showDOI{\tempurl}
\newblock
\shownote{Publisher: Public Library of Science}.


\bibitem[Rechkemmer and Yin(2022)]%
        {rechkemmer_when_2022}
\bibfield{author}{\bibinfo{person}{Amy Rechkemmer} {and} \bibinfo{person}{Ming Yin}.} \bibinfo{year}{2022}\natexlab{}.
\newblock \showarticletitle{When {Confidence} {Meets} {Accuracy}: {Exploring} the {Effects} of {Multiple} {Performance} {Indicators} on {Trust} in {Machine} {Learning} {Models}}. In \bibinfo{booktitle}{\emph{Proceedings of the 2022 {CHI} {Conference} on {Human} {Factors} in {Computing} {Systems}}} \emph{(\bibinfo{series}{{CHI} '22})}. \bibinfo{publisher}{Association for Computing Machinery}, \bibinfo{address}{New York, NY, USA}, \bibinfo{pages}{1--14}.
\newblock
\showISBNx{978-1-4503-9157-3}
\urldef\tempurl%
\url{https://doi.org/10.1145/3491102.3501967}
\showDOI{\tempurl}


\bibitem[Reynante et~al\mbox{.}(2021)]%
        {reynante_framework_2021}
\bibfield{author}{\bibinfo{person}{Brandon Reynante}, \bibinfo{person}{Steven~P. Dow}, {and} \bibinfo{person}{Narges Mahyar}.} \bibinfo{year}{2021}\natexlab{}.
\newblock \showarticletitle{A {Framework} for {Open} {Civic} {Design}: {Integrating} {Public} {Participation}, {Crowdsourcing}, and {Design} {Thinking}}.
\newblock \bibinfo{journal}{\emph{Digital Government: Research and Practice}} \bibinfo{volume}{2}, \bibinfo{number}{4} (\bibinfo{date}{Oct.} \bibinfo{year}{2021}), \bibinfo{pages}{1--22}.
\newblock
\showISSN{2691-199X, 2639-0175}
\urldef\tempurl%
\url{https://doi.org/10.1145/3487607}
\showDOI{\tempurl}


\bibitem[Reynolds-Cu{\'{e}}llar et~al\mbox{.}(2022)]%
        {Reynolds-Cuellar2022}
\bibfield{author}{\bibinfo{person}{Pedro Reynolds-Cu{\'{e}}llar}, \bibinfo{person}{Claudia Grisales}, \bibinfo{person}{Marisol Wong-Villacr{\'{e}}s}, \bibinfo{person}{Bibiana Serpa}, \bibinfo{person}{Julian~I{\~{n}}aki Go{\~{n}}i}, {and} \bibinfo{person}{Oscar~A. Lemus}.} \bibinfo{year}{2022}\natexlab{}.
\newblock \showarticletitle{{Reviews Gone South: A Subversive Experiment on Participatory Design Canons}}. In \bibinfo{booktitle}{\emph{Participatory Design Conference 2022: Volume 1}}. \bibinfo{publisher}{ACM}, \bibinfo{address}{New York, NY, USA}, \bibinfo{pages}{206--217}.
\newblock
\showISBNx{9781450393881}
\urldef\tempurl%
\url{https://doi.org/10.1145/3536169.3537794}
\showDOI{\tempurl}


\bibitem[Reynolds-Cuéllar et~al\mbox{.}(2023)]%
        {reynolds-cuellar_para_2023}
\bibfield{author}{\bibinfo{person}{Pedro Reynolds-Cuéllar}, \bibinfo{person}{Marisol Wong-Villacres}, \bibinfo{person}{Karla Badillo-Urquiola}, \bibinfo{person}{Mayra~Donaji Barrera~Machuca}, \bibinfo{person}{Franceli~L. Cibrian}, \bibinfo{person}{Marianela Ciolfi~Felice}, \bibinfo{person}{Carolina Fuentes}, \bibinfo{person}{Laura~Sanely Gaytán-Lugo}, \bibinfo{person}{Vivian Genaro~Motti}, \bibinfo{person}{Monica Perusquia-Hernandez}, {and} \bibinfo{person}{Oscar~A Lemus}.} \bibinfo{year}{2023}\natexlab{}.
\newblock \showarticletitle{Para {Cima} y {Pa}’ {Abajo}: {Building} {Bridges} {Between} {HCI} {Research} in {Latin} {America} and in the {Global} {North}}. In \bibinfo{booktitle}{\emph{Proceedings of the 2023 {CHI} {Conference} on {Human} {Factors} in {Computing} {Systems}}}. \bibinfo{publisher}{ACM}, \bibinfo{address}{Hamburg Germany}, \bibinfo{pages}{1--19}.
\newblock
\showISBNx{978-1-4503-9421-5}
\urldef\tempurl%
\url{https://doi.org/10.1145/3544548.3581138}
\showDOI{\tempurl}


\bibitem[Ricaurte et~al\mbox{.}(2024)]%
        {ricaurte_algorithmic_2024}
\bibfield{author}{\bibinfo{person}{Paola Ricaurte}, \bibinfo{person}{Edgar Gómez-Cruz}, {and} \bibinfo{person}{Ignacio Siles}.} \bibinfo{year}{2024}\natexlab{}.
\newblock \showarticletitle{Algorithmic governmentality in {Latin} {America}: {Sociotechnical} imaginaries, neocolonial soft power, and authoritarianism}.
\newblock \bibinfo{journal}{\emph{Big Data \& Society}} \bibinfo{volume}{11}, \bibinfo{number}{1} (\bibinfo{date}{March} \bibinfo{year}{2024}), \bibinfo{pages}{20539517241229697}.
\newblock
\showISSN{2053-9517}
\urldef\tempurl%
\url{https://doi.org/10.1177/20539517241229697}
\showDOI{\tempurl}
\newblock
\shownote{Publisher: SAGE Publications Ltd}.


\bibitem[Rivoir and Landinelli(2017)]%
        {rivoir_ict-mediated_2017}
\bibfield{author}{\bibinfo{person}{Ana Rivoir} {and} \bibinfo{person}{Javier Landinelli}.} \bibinfo{year}{2017}\natexlab{}.
\newblock \showarticletitle{{ICT}-mediated {Citizen} {Engagement} - {Case} {Study}: {Open} {Government} {National} {Action} {Plan} in {Uruguay}}. In \bibinfo{booktitle}{\emph{Proceedings of the 10th {International} {Conference} on {Theory} and {Practice} of {Electronic} {Governance}}} \emph{(\bibinfo{series}{{ICEGOV} '17})}. \bibinfo{publisher}{Association for Computing Machinery}, \bibinfo{address}{New York, NY, USA}, \bibinfo{pages}{214--217}.
\newblock
\showISBNx{978-1-4503-4825-6}
\urldef\tempurl%
\url{https://doi.org/10.1145/3047273.3047359}
\showDOI{\tempurl}


\bibitem[Roetzel(2019)]%
        {roetzel_information_2019}
\bibfield{author}{\bibinfo{person}{Peter~Gordon Roetzel}.} \bibinfo{year}{2019}\natexlab{}.
\newblock \showarticletitle{Information overload in the information age: a review of the literature from business administration, business psychology, and related disciplines with a bibliometric approach and framework development}.
\newblock \bibinfo{journal}{\emph{Business Research}} \bibinfo{volume}{12}, \bibinfo{number}{2} (\bibinfo{date}{Dec.} \bibinfo{year}{2019}), \bibinfo{pages}{479--522}.
\newblock
\showISSN{2198-2627}
\urldef\tempurl%
\url{https://doi.org/10.1007/s40685-018-0069-z}
\showDOI{\tempurl}


\bibitem[Rogerson et~al\mbox{.}(2022)]%
        {rogerson_government_2022}
\bibfield{author}{\bibinfo{person}{Annys Rogerson}, \bibinfo{person}{Emma Hankins}, \bibinfo{person}{Pablo Fuentes}, {and} \bibinfo{person}{Sulamaan Rahim}.} \bibinfo{year}{2022}\natexlab{}.
\newblock \bibinfo{booktitle}{\emph{Government {AI} {Readiness} {Index} 2022}}.
\newblock \bibinfo{type}{{T}echnical {R}eport}. \bibinfo{institution}{Oxford Insights}.
\newblock


\bibitem[Romberg(2022)]%
        {romberg_is_2022}
\bibfield{author}{\bibinfo{person}{Julia Romberg}.} \bibinfo{year}{2022}\natexlab{}.
\newblock \showarticletitle{Is {Your} {Perspective} {Also} {My} {Perspective}? {Enriching} {Prediction} with {Subjectivity}}. In \bibinfo{booktitle}{\emph{Proceedings of the 9th {Workshop} on {Argument} {Mining}}}. \bibinfo{publisher}{International Conference on Computational Linguistics}, \bibinfo{address}{Online and in Gyeongju, Republic of Korea}, \bibinfo{pages}{115--125}.
\newblock
\urldef\tempurl%
\url{https://aclanthology.org/2022.argmining-1.11}
\showURL{%
\tempurl}


\bibitem[Romberg and Escher(2022)]%
        {romberg_automated_2022}
\bibfield{author}{\bibinfo{person}{Julia Romberg} {and} \bibinfo{person}{Tobias Escher}.} \bibinfo{year}{2022}\natexlab{}.
\newblock \showarticletitle{Automated {Topic} {Categorisation} of {Citizens}’ {Contributions}: {Reducing} {Manual} {Labelling} {Efforts} {Through} {Active} {Learning}}. In \bibinfo{booktitle}{\emph{Electronic {Government}}} \emph{(\bibinfo{series}{Lecture {Notes} in {Computer} {Science}})}, \bibfield{editor}{\bibinfo{person}{Marijn Janssen}, \bibinfo{person}{Csaba Csáki}, \bibinfo{person}{Ida Lindgren}, \bibinfo{person}{Euripidis Loukis}, \bibinfo{person}{Ulf Melin}, \bibinfo{person}{Gabriela Viale~Pereira}, \bibinfo{person}{Manuel~Pedro Rodríguez~Bolívar}, {and} \bibinfo{person}{Efthimios Tambouris}} (Eds.). \bibinfo{publisher}{Springer International Publishing}, \bibinfo{address}{Cham}, \bibinfo{pages}{369--385}.
\newblock
\showISBNx{978-3-031-15086-9}
\urldef\tempurl%
\url{https://doi.org/10.1007/978-3-031-15086-9_24}
\showDOI{\tempurl}


\bibitem[Romberg and Escher(2023)]%
        {romberg_making_2023}
\bibfield{author}{\bibinfo{person}{Julia Romberg} {and} \bibinfo{person}{Tobias Escher}.} \bibinfo{year}{2023}\natexlab{}.
\newblock \showarticletitle{Making {Sense} of {Citizens}’ {Input} through {Artificial} {Intelligence}: {A} {Review} of {Methods} for {Computational} {Text} {Analysis} to {Support} the {Evaluation} of {Contributions} in {Public} {Participation}}.
\newblock \bibinfo{journal}{\emph{Digital Government: Research and Practice}} (\bibinfo{date}{June} \bibinfo{year}{2023}).
\newblock
\showISSN{2691-199X}
\urldef\tempurl%
\url{https://doi.org/10.1145/3603254}
\showDOI{\tempurl}
\newblock
\shownote{Just Accepted}.


\bibitem[Rose and Tenenberg(2016)]%
        {rose_arguing_2016}
\bibfield{author}{\bibinfo{person}{Emma Rose} {and} \bibinfo{person}{Josh Tenenberg}.} \bibinfo{year}{2016}\natexlab{}.
\newblock \showarticletitle{Arguing about design: {A} taxonomy of rhetorical strategies deployed by user experience practitioners}. In \bibinfo{booktitle}{\emph{Proceedings of the 34th {ACM} {International} {Conference} on the {Design} of {Communication}}} \emph{(\bibinfo{series}{{SIGDOC} '16})}. \bibinfo{publisher}{Association for Computing Machinery}, \bibinfo{address}{New York, NY, USA}, \bibinfo{pages}{1--10}.
\newblock
\showISBNx{978-1-4503-4495-1}
\urldef\tempurl%
\url{https://doi.org/10.1145/2987592.2987608}
\showDOI{\tempurl}


\bibitem[Rose et~al\mbox{.}(2018)]%
        {rose_stakeholder_2018}
\bibfield{author}{\bibinfo{person}{Jeremy Rose}, \bibinfo{person}{Leif~Skiftenes Flak}, {and} \bibinfo{person}{Øystein Sæbø}.} \bibinfo{year}{2018}\natexlab{}.
\newblock \showarticletitle{Stakeholder theory for the {E}-government context: {Framing} a value-oriented normative core}.
\newblock \bibinfo{journal}{\emph{Government Information Quarterly}} \bibinfo{volume}{35}, \bibinfo{number}{3} (\bibinfo{date}{Sept.} \bibinfo{year}{2018}), \bibinfo{pages}{362--374}.
\newblock
\showISSN{0740-624X}
\urldef\tempurl%
\url{https://doi.org/10.1016/j.giq.2018.06.005}
\showDOI{\tempurl}


\bibitem[Rose et~al\mbox{.}(2015)]%
        {rose_managing_2015}
\bibfield{author}{\bibinfo{person}{Jeremy Rose}, \bibinfo{person}{John~Stouby Persson}, \bibinfo{person}{Lise~Tordrup Heeager}, {and} \bibinfo{person}{Zahir Irani}.} \bibinfo{year}{2015}\natexlab{}.
\newblock \showarticletitle{Managing e-{Government}: value positions and relationships}.
\newblock \bibinfo{journal}{\emph{Information Systems Journal}} \bibinfo{volume}{25}, \bibinfo{number}{5} (\bibinfo{year}{2015}), \bibinfo{pages}{531--571}.
\newblock
\showISSN{1365-2575}
\urldef\tempurl%
\url{https://doi.org/10.1111/isj.12052}
\showDOI{\tempurl}


\bibitem[Rose and Sæbø(2010)]%
        {rose_designing_2010}
\bibfield{author}{\bibinfo{person}{Jeremy Rose} {and} \bibinfo{person}{Øystein Sæbø}.} \bibinfo{year}{2010}\natexlab{}.
\newblock \showarticletitle{Designing {Deliberation} {Systems}}.
\newblock \bibinfo{journal}{\emph{The Information Society}} \bibinfo{volume}{26}, \bibinfo{number}{3} (\bibinfo{date}{April} \bibinfo{year}{2010}), \bibinfo{pages}{228--240}.
\newblock
\showISSN{0197-2243}
\urldef\tempurl%
\url{https://doi.org/10.1080/01972241003712298}
\showDOI{\tempurl}
\newblock
\shownote{Publisher: Routledge \_eprint: https://doi.org/10.1080/01972241003712298}.


\bibitem[Rowley(2011)]%
        {rowley_e-government_2011}
\bibfield{author}{\bibinfo{person}{Jennifer Rowley}.} \bibinfo{year}{2011}\natexlab{}.
\newblock \showarticletitle{e-{Government} stakeholders—{Who} are they and what do they want?}
\newblock \bibinfo{journal}{\emph{International Journal of Information Management}} \bibinfo{volume}{31}, \bibinfo{number}{1} (\bibinfo{date}{Feb.} \bibinfo{year}{2011}), \bibinfo{pages}{53--62}.
\newblock
\showISSN{0268-4012}
\urldef\tempurl%
\url{https://doi.org/10.1016/j.ijinfomgt.2010.05.005}
\showDOI{\tempurl}


\bibitem[Saxena et~al\mbox{.}(2021)]%
        {saxena_framework_2021}
\bibfield{author}{\bibinfo{person}{Devansh Saxena}, \bibinfo{person}{Karla Badillo-Urquiola}, \bibinfo{person}{Pamela~J. Wisniewski}, {and} \bibinfo{person}{Shion Guha}.} \bibinfo{year}{2021}\natexlab{}.
\newblock \showarticletitle{A {Framework} of {High}-{Stakes} {Algorithmic} {Decision}-{Making} for the {Public} {Sector} {Developed} through a {Case} {Study} of {Child}-{Welfare}}.
\newblock \bibinfo{journal}{\emph{Proc. ACM Hum.-Comput. Interact.}} \bibinfo{volume}{5}, \bibinfo{number}{CSCW2} (\bibinfo{date}{Oct.} \bibinfo{year}{2021}), \bibinfo{pages}{348:1--348:41}.
\newblock
\urldef\tempurl%
\url{https://doi.org/10.1145/3476089}
\showDOI{\tempurl}


\bibitem[Schmidt(2013)]%
        {schmidt_democracy_2013}
\bibfield{author}{\bibinfo{person}{Vivien~A. Schmidt}.} \bibinfo{year}{2013}\natexlab{}.
\newblock \showarticletitle{Democracy and {Legitimacy} in the {European} {Union} {Revisited}: {Input}, {Output} and ‘{Throughput}’}.
\newblock \bibinfo{journal}{\emph{Political Studies}} \bibinfo{volume}{61}, \bibinfo{number}{1} (\bibinfo{date}{March} \bibinfo{year}{2013}), \bibinfo{pages}{2--22}.
\newblock
\showISSN{0032-3217}
\urldef\tempurl%
\url{https://doi.org/10.1111/j.1467-9248.2012.00962.x}
\showDOI{\tempurl}
\newblock
\shownote{Publisher: SAGE Publications Ltd}.


\bibitem[Scholl(2004)]%
        {scholl_involving_2004}
\bibfield{author}{\bibinfo{person}{Hans~J. Scholl}.} \bibinfo{year}{2004}\natexlab{}.
\newblock \showarticletitle{Involving {Salient} {Stakeholders}: {Beyond} the {Technocratic} {View} on {Change}}.
\newblock \bibinfo{journal}{\emph{Action Research}} \bibinfo{volume}{2}, \bibinfo{number}{3} (\bibinfo{date}{Sept.} \bibinfo{year}{2004}), \bibinfo{pages}{277--304}.
\newblock
\showISSN{1476-7503}
\urldef\tempurl%
\url{https://doi.org/10.1177/1476750304045940}
\showDOI{\tempurl}
\newblock
\shownote{Publisher: SAGE Publications}.


\bibitem[Shen et~al\mbox{.}(2022)]%
        {shen_model_2022}
\bibfield{author}{\bibinfo{person}{Hong Shen}, \bibinfo{person}{Leijie Wang}, \bibinfo{person}{Wesley~H. Deng}, \bibinfo{person}{Ciell Brusse}, \bibinfo{person}{Ronald Velgersdijk}, {and} \bibinfo{person}{Haiyi Zhu}.} \bibinfo{year}{2022}\natexlab{}.
\newblock \showarticletitle{The {Model} {Card} {Authoring} {Toolkit}: {Toward} {Community}-centered, {Deliberation}-driven {AI} {Design}}. In \bibinfo{booktitle}{\emph{2022 {ACM} {Conference} on {Fairness}, {Accountability}, and {Transparency}}} \emph{(\bibinfo{series}{{FAccT} '22})}. \bibinfo{publisher}{Association for Computing Machinery}, \bibinfo{address}{New York, NY, USA}, \bibinfo{pages}{440--451}.
\newblock
\showISBNx{978-1-4503-9352-2}
\urldef\tempurl%
\url{https://doi.org/10.1145/3531146.3533110}
\showDOI{\tempurl}


\bibitem[Simonofski et~al\mbox{.}(2021)]%
        {simonofski_supporting_2021}
\bibfield{author}{\bibinfo{person}{Anthony Simonofski}, \bibinfo{person}{Jerôme Fink}, {and} \bibinfo{person}{Corentin Burnay}.} \bibinfo{year}{2021}\natexlab{}.
\newblock \showarticletitle{Supporting policy-making with social media and e-participation platforms data: {A} policy analytics framework}.
\newblock \bibinfo{journal}{\emph{Government Information Quarterly}} \bibinfo{volume}{38}, \bibinfo{number}{3} (\bibinfo{date}{July} \bibinfo{year}{2021}), \bibinfo{pages}{101590}.
\newblock
\showISSN{0740-624X}
\urldef\tempurl%
\url{https://doi.org/10.1016/j.giq.2021.101590}
\showDOI{\tempurl}


\bibitem[Sloane et~al\mbox{.}(2021)]%
        {sloane_ai_2021}
\bibfield{author}{\bibinfo{person}{Mona Sloane}, \bibinfo{person}{Rumman Chowdhury}, \bibinfo{person}{John~C. Havens}, \bibinfo{person}{Tomo Lazovich}, {and} \bibinfo{person}{Luis Rincon~Alba}.} \bibinfo{year}{2021}\natexlab{}.
\newblock \bibinfo{booktitle}{\emph{{AI} and {Procurement} - {A} {Primer}}}.
\newblock \bibinfo{type}{Working {Paper}}.
\newblock
\urldef\tempurl%
\url{https://doi.org/10.17609/bxzf-df18}
\showDOI{\tempurl}
\newblock
\shownote{Accepted: 2021-06-25T16:16:18Z}.


\bibitem[Sloane et~al\mbox{.}(2020)]%
        {sloane_participation_2020}
\bibfield{author}{\bibinfo{person}{Mona Sloane}, \bibinfo{person}{Emanuel Moss}, \bibinfo{person}{Olaitan Awomolo}, {and} \bibinfo{person}{Laura Forlano}.} \bibinfo{year}{2020}\natexlab{}.
\newblock \bibinfo{title}{Participation is not a {Design} {Fix} for {Machine} {Learning}}.
\newblock
\newblock
\urldef\tempurl%
\url{https://doi.org/10.48550/arXiv.2007.02423}
\showDOI{\tempurl}
\newblock
\shownote{arXiv:2007.02423 [cs]}.


\bibitem[Sloane et~al\mbox{.}(2022)]%
        {sloane_participation_2022}
\bibfield{author}{\bibinfo{person}{Mona Sloane}, \bibinfo{person}{Emanuel Moss}, \bibinfo{person}{Olaitan Awomolo}, {and} \bibinfo{person}{Laura Forlano}.} \bibinfo{year}{2022}\natexlab{}.
\newblock \showarticletitle{Participation {Is} not a {Design} {Fix} for {Machine} {Learning}}. In \bibinfo{booktitle}{\emph{Proceedings of the 2nd {ACM} {Conference} on {Equity} and {Access} in {Algorithms}, {Mechanisms}, and {Optimization}}} \emph{(\bibinfo{series}{{EAAMO} '22})}. \bibinfo{publisher}{Association for Computing Machinery}, \bibinfo{address}{New York, NY, USA}, \bibinfo{pages}{1--6}.
\newblock
\showISBNx{978-1-4503-9477-2}
\urldef\tempurl%
\url{https://doi.org/10.1145/3551624.3555285}
\showDOI{\tempurl}


\bibitem[Small et~al\mbox{.}(2023)]%
        {small_opportunities_2023}
\bibfield{author}{\bibinfo{person}{Christopher~T. Small}, \bibinfo{person}{Ivan Vendrov}, \bibinfo{person}{Esin Durmus}, \bibinfo{person}{Hadjar Homaei}, \bibinfo{person}{Elizabeth Barry}, \bibinfo{person}{Julien Cornebise}, \bibinfo{person}{Ted Suzman}, \bibinfo{person}{Deep Ganguli}, {and} \bibinfo{person}{Colin Megill}.} \bibinfo{year}{2023}\natexlab{}.
\newblock \bibinfo{title}{Opportunities and {Risks} of {LLMs} for {Scalable} {Deliberation} with {Polis}}.
\newblock
\newblock
\urldef\tempurl%
\url{https://doi.org/10.48550/arXiv.2306.11932}
\showDOI{\tempurl}
\newblock
\shownote{arXiv:2306.11932 [cs]}.


\bibitem[Stapleton et~al\mbox{.}(2022)]%
        {stapleton_who_2022}
\bibfield{author}{\bibinfo{person}{Logan Stapleton}, \bibinfo{person}{Devansh Saxena}, \bibinfo{person}{Anna Kawakami}, \bibinfo{person}{Tonya Nguyen}, \bibinfo{person}{Asbjørn Ammitzbøll~Flügge}, \bibinfo{person}{Motahhare Eslami}, \bibinfo{person}{Naja Holten~Møller}, \bibinfo{person}{Min~Kyung Lee}, \bibinfo{person}{Shion Guha}, \bibinfo{person}{Kenneth Holstein}, {and} \bibinfo{person}{Haiyi Zhu}.} \bibinfo{year}{2022}\natexlab{}.
\newblock \showarticletitle{Who {Has} an {Interest} in “{Public} {Interest} {Technology}”?: {Critical} {Questions} for {Working} with {Local} {Governments} \& {Impacted} {Communities}}. In \bibinfo{booktitle}{\emph{Companion {Publication} of the 2022 {Conference} on {Computer} {Supported} {Cooperative} {Work} and {Social} {Computing}}} \emph{(\bibinfo{series}{{CSCW}'22 {Companion}})}. \bibinfo{publisher}{Association for Computing Machinery}, \bibinfo{address}{New York, NY, USA}, \bibinfo{pages}{282--286}.
\newblock
\showISBNx{978-1-4503-9190-0}
\urldef\tempurl%
\url{https://doi.org/10.1145/3500868.3560484}
\showDOI{\tempurl}


\bibitem[Strebel et~al\mbox{.}(2019)]%
        {strebel_importance_2019}
\bibfield{author}{\bibinfo{person}{Michael~Andrea Strebel}, \bibinfo{person}{Daniel Kübler}, {and} \bibinfo{person}{Frank Marcinkowski}.} \bibinfo{year}{2019}\natexlab{}.
\newblock \showarticletitle{The importance of input and output legitimacy in democratic governance: {Evidence} from a population-based survey experiment in four {West} {European} countries}.
\newblock \bibinfo{journal}{\emph{European Journal of Political Research}} \bibinfo{volume}{58}, \bibinfo{number}{2} (\bibinfo{year}{2019}), \bibinfo{pages}{488--513}.
\newblock
\showISSN{1475-6765}
\urldef\tempurl%
\url{https://doi.org/10.1111/1475-6765.12293}
\showDOI{\tempurl}
\newblock
\shownote{\_eprint: https://onlinelibrary.wiley.com/doi/pdf/10.1111/1475-6765.12293}.


\bibitem[Susser(2019)]%
        {susser_invisible_2019}
\bibfield{author}{\bibinfo{person}{Daniel Susser}.} \bibinfo{year}{2019}\natexlab{}.
\newblock \showarticletitle{Invisible {Influence}: {Artificial} {Intelligence} and the {Ethics} of {Adaptive} {Choice} {Architectures}}. In \bibinfo{booktitle}{\emph{Proceedings of the 2019 {AAAI}/{ACM} {Conference} on {AI}, {Ethics}, and {Society}}} \emph{(\bibinfo{series}{{AIES} '19})}. \bibinfo{publisher}{Association for Computing Machinery}, \bibinfo{address}{New York, NY, USA}, \bibinfo{pages}{403--408}.
\newblock
\showISBNx{978-1-4503-6324-2}
\urldef\tempurl%
\url{https://doi.org/10.1145/3306618.3314286}
\showDOI{\tempurl}


\bibitem[Sæbø et~al\mbox{.}(2011)]%
        {saebo_understanding_2011}
\bibfield{author}{\bibinfo{person}{Øystein Sæbø}, \bibinfo{person}{Leif~Skiftenes Flak}, {and} \bibinfo{person}{Maung~K. Sein}.} \bibinfo{year}{2011}\natexlab{}.
\newblock \showarticletitle{Understanding the dynamics in e-{Participation} initiatives: {Looking} through the genre and stakeholder lenses}.
\newblock \bibinfo{journal}{\emph{Government Information Quarterly}} \bibinfo{volume}{28}, \bibinfo{number}{3} (\bibinfo{date}{July} \bibinfo{year}{2011}), \bibinfo{pages}{416--425}.
\newblock
\showISSN{0740-624X}
\urldef\tempurl%
\url{https://doi.org/10.1016/j.giq.2010.10.005}
\showDOI{\tempurl}


\bibitem[Talhouk et~al\mbox{.}(2023)]%
        {talhouk_re-articulating_2023}
\bibfield{author}{\bibinfo{person}{Reem Talhouk}, \bibinfo{person}{Ebtisam Alabdulqader}, \bibinfo{person}{Cat Kutay}, \bibinfo{person}{Kagonya Awori}, \bibinfo{person}{Marisol Wong-Villacres}, \bibinfo{person}{Neha Kumar}, \bibinfo{person}{Tariq Zaman}, \bibinfo{person}{Volker Wulf}, \bibinfo{person}{Zainab Almeraj}, {and} \bibinfo{person}{Shaimaa Lazem}.} \bibinfo{year}{2023}\natexlab{}.
\newblock \showarticletitle{Re-articulating {North}-{South} {Collaborations} in {HCI}}. In \bibinfo{booktitle}{\emph{Extended {Abstracts} of the 2023 {CHI} {Conference} on {Human} {Factors} in {Computing} {Systems}}} \emph{(\bibinfo{series}{{CHI} {EA} '23})}. \bibinfo{publisher}{Association for Computing Machinery}, \bibinfo{address}{New York, NY, USA}, \bibinfo{pages}{1--4}.
\newblock
\showISBNx{978-1-4503-9422-2}
\urldef\tempurl%
\url{https://doi.org/10.1145/3544549.3583752}
\showDOI{\tempurl}


\bibitem[Tangi et~al\mbox{.}(2023)]%
        {tangi_challenges_2023}
\bibfield{author}{\bibinfo{person}{Luca Tangi}, \bibinfo{person}{Colin van Noordt}, {and} \bibinfo{person}{A.~Paula Rodriguez~Müller}.} \bibinfo{year}{2023}\natexlab{}.
\newblock \showarticletitle{The challenges of {AI} implementation in the public sector. {An} in-depth case studies analysis}. In \bibinfo{booktitle}{\emph{Proceedings of the 24th {Annual} {International} {Conference} on {Digital} {Government} {Research}}} \emph{(\bibinfo{series}{{DGO} '23})}. \bibinfo{publisher}{Association for Computing Machinery}, \bibinfo{address}{New York, NY, USA}, \bibinfo{pages}{414--422}.
\newblock
\showISBNx{9798400708374}
\urldef\tempurl%
\url{https://doi.org/10.1145/3598469.3598516}
\showDOI{\tempurl}


\bibitem[Tolmie et~al\mbox{.}(2002)]%
        {tolmie_unremarkable_2002}
\bibfield{author}{\bibinfo{person}{Peter Tolmie}, \bibinfo{person}{James Pycock}, \bibinfo{person}{Tim Diggins}, \bibinfo{person}{Allan MacLean}, {and} \bibinfo{person}{Alain Karsenty}.} \bibinfo{year}{2002}\natexlab{}.
\newblock \showarticletitle{Unremarkable computing}. In \bibinfo{booktitle}{\emph{Proceedings of the {SIGCHI} {Conference} on {Human} {Factors} in {Computing} {Systems}}} \emph{(\bibinfo{series}{{CHI} '02})}. \bibinfo{publisher}{Association for Computing Machinery}, \bibinfo{address}{New York, NY, USA}, \bibinfo{pages}{399--406}.
\newblock
\showISBNx{978-1-58113-453-7}
\urldef\tempurl%
\url{https://doi.org/10.1145/503376.503448}
\showDOI{\tempurl}


\bibitem[UNESCO(2023)]%
        {unesco_chile_2023}
\bibfield{author}{\bibinfo{person}{UNESCO}.} \bibinfo{year}{2023}\natexlab{}.
\newblock \bibinfo{booktitle}{\emph{Chile: artificial intelligence readiness assessment report - {UNESCO} {Biblioteca} {Digital}}}.
\newblock \bibinfo{type}{{T}echnical {R}eport}. \bibinfo{institution}{UNESCO}, \bibinfo{address}{Paris}.
\newblock
\urldef\tempurl%
\url{https://unesdoc.unesco.org/ark:/48223/pf0000387216}
\showURL{%
\tempurl}


\bibitem[Unit(2024)]%
        {economist_intelligence_unit_democracy_2024}
\bibfield{author}{\bibinfo{person}{Economist~Intelligence Unit}.} \bibinfo{year}{2024}\natexlab{}.
\newblock \bibinfo{booktitle}{\emph{Democracy {Index} 2023 - {Age} of {Conflict}}}.
\newblock \bibinfo{type}{{T}echnical {R}eport}. \bibinfo{institution}{Economist Intelligence Unit}.
\newblock


\bibitem[Vecchione et~al\mbox{.}(2021)]%
        {vecchione_algorithmic_2021}
\bibfield{author}{\bibinfo{person}{Briana Vecchione}, \bibinfo{person}{Karen Levy}, {and} \bibinfo{person}{Solon Barocas}.} \bibinfo{year}{2021}\natexlab{}.
\newblock \showarticletitle{Algorithmic {Auditing} and {Social} {Justice}: {Lessons} from the {History} of {Audit} {Studies}}. In \bibinfo{booktitle}{\emph{Proceedings of the 1st {ACM} {Conference} on {Equity} and {Access} in {Algorithms}, {Mechanisms}, and {Optimization}}} \emph{(\bibinfo{series}{{EAAMO} '21})}. \bibinfo{publisher}{Association for Computing Machinery}, \bibinfo{address}{New York, NY, USA}, \bibinfo{pages}{1--9}.
\newblock
\showISBNx{978-1-4503-8553-4}
\urldef\tempurl%
\url{https://doi.org/10.1145/3465416.3483294}
\showDOI{\tempurl}


\bibitem[von Brackel-Schmidt et~al\mbox{.}(2024)]%
        {von_brackel-schmidt_equipping_2024}
\bibfield{author}{\bibinfo{person}{Constantin von Brackel-Schmidt}, \bibinfo{person}{Emir Kučević}, \bibinfo{person}{Stephan Leible}, \bibinfo{person}{Dejan Simic}, \bibinfo{person}{Gian-Luca Gücük}, {and} \bibinfo{person}{Felix~N. Schmidt}.} \bibinfo{year}{2024}\natexlab{}.
\newblock \showarticletitle{Equipping {Participation} {Formats} with {Generative} {AI}: {A} {Case} {Study} {Predicting} the {Future} of a {Metropolitan} {City} in the {Year} 2040}. In \bibinfo{booktitle}{\emph{{HCI} in {Business}, {Government} and {Organizations}}}, \bibfield{editor}{\bibinfo{person}{Fiona Fui-Hoon Nah} {and} \bibinfo{person}{Keng~Leng Siau}} (Eds.). \bibinfo{publisher}{Springer Nature Switzerland}, \bibinfo{address}{Cham}, \bibinfo{pages}{270--285}.
\newblock
\showISBNx{978-3-031-61315-9}
\urldef\tempurl%
\url{https://doi.org/10.1007/978-3-031-61315-9_19}
\showDOI{\tempurl}


\bibitem[Wang et~al\mbox{.}(2019)]%
        {wang_designing_2019}
\bibfield{author}{\bibinfo{person}{Danding Wang}, \bibinfo{person}{Qian Yang}, \bibinfo{person}{Ashraf Abdul}, {and} \bibinfo{person}{Brian~Y. Lim}.} \bibinfo{year}{2019}\natexlab{}.
\newblock \showarticletitle{Designing {Theory}-{Driven} {User}-{Centric} {Explainable} {AI}}. In \bibinfo{booktitle}{\emph{Proceedings of the 2019 {CHI} {Conference} on {Human} {Factors} in {Computing} {Systems}}} \emph{(\bibinfo{series}{{CHI} '19})}. \bibinfo{publisher}{Association for Computing Machinery}, \bibinfo{address}{New York, NY, USA}, \bibinfo{pages}{1--15}.
\newblock
\showISBNx{978-1-4503-5970-2}
\urldef\tempurl%
\url{https://doi.org/10.1145/3290605.3300831}
\showDOI{\tempurl}


\bibitem[Weidinger et~al\mbox{.}(2022)]%
        {weidinger_taxonomy_2022}
\bibfield{author}{\bibinfo{person}{Laura Weidinger}, \bibinfo{person}{Jonathan Uesato}, \bibinfo{person}{Maribeth Rauh}, \bibinfo{person}{Conor Griffin}, \bibinfo{person}{Po-Sen Huang}, \bibinfo{person}{John Mellor}, \bibinfo{person}{Amelia Glaese}, \bibinfo{person}{Myra Cheng}, \bibinfo{person}{Borja Balle}, \bibinfo{person}{Atoosa Kasirzadeh}, \bibinfo{person}{Courtney Biles}, \bibinfo{person}{Sasha Brown}, \bibinfo{person}{Zac Kenton}, \bibinfo{person}{Will Hawkins}, \bibinfo{person}{Tom Stepleton}, \bibinfo{person}{Abeba Birhane}, \bibinfo{person}{Lisa~Anne Hendricks}, \bibinfo{person}{Laura Rimell}, \bibinfo{person}{William Isaac}, \bibinfo{person}{Julia Haas}, \bibinfo{person}{Sean Legassick}, \bibinfo{person}{Geoffrey Irving}, {and} \bibinfo{person}{Iason Gabriel}.} \bibinfo{year}{2022}\natexlab{}.
\newblock \showarticletitle{Taxonomy of {Risks} posed by {Language} {Models}}. In \bibinfo{booktitle}{\emph{Proceedings of the 2022 {ACM} {Conference} on {Fairness}, {Accountability}, and {Transparency}}} \emph{(\bibinfo{series}{{FAccT} '22})}. \bibinfo{publisher}{Association for Computing Machinery}, \bibinfo{address}{New York, NY, USA}, \bibinfo{pages}{214--229}.
\newblock
\showISBNx{978-1-4503-9352-2}
\urldef\tempurl%
\url{https://doi.org/10.1145/3531146.3533088}
\showDOI{\tempurl}


\bibitem[Weng et~al\mbox{.}(2021)]%
        {weng_ai_2021}
\bibfield{author}{\bibinfo{person}{Min-Hsien Weng}, \bibinfo{person}{Shaoqun Wu}, {and} \bibinfo{person}{Mark Dyer}.} \bibinfo{year}{2021}\natexlab{}.
\newblock \showarticletitle{{AI} {Augmented} {Approach} to {Identify} {Shared} {Ideas} from {Large} {Format} {Public} {Consultation}}.
\newblock \bibinfo{journal}{\emph{Sustainability}} \bibinfo{volume}{13}, \bibinfo{number}{16} (\bibinfo{date}{Jan.} \bibinfo{year}{2021}), \bibinfo{pages}{9310}.
\newblock
\showISSN{2071-1050}
\urldef\tempurl%
\url{https://doi.org/10.3390/su13169310}
\showDOI{\tempurl}
\newblock
\shownote{Number: 16 Publisher: Multidisciplinary Digital Publishing Institute}.


\bibitem[Wirtz et~al\mbox{.}(2019)]%
        {wirtz_artificial_2019}
\bibfield{author}{\bibinfo{person}{Bernd~W. Wirtz}, \bibinfo{person}{Jan~C. Weyerer}, {and} \bibinfo{person}{Carolin Geyer}.} \bibinfo{year}{2019}\natexlab{}.
\newblock \showarticletitle{Artificial {Intelligence} and the {Public} {Sector}—{Applications} and {Challenges}}.
\newblock \bibinfo{journal}{\emph{International Journal of Public Administration}} \bibinfo{volume}{42}, \bibinfo{number}{7} (\bibinfo{date}{May} \bibinfo{year}{2019}), \bibinfo{pages}{596--615}.
\newblock
\showISSN{0190-0692}
\urldef\tempurl%
\url{https://doi.org/10.1080/01900692.2018.1498103}
\showDOI{\tempurl}
\newblock
\shownote{Publisher: Routledge \_eprint: https://doi.org/10.1080/01900692.2018.1498103}.


\bibitem[Wong(2021)]%
        {wong_tactics_2021}
\bibfield{author}{\bibinfo{person}{Richmond~Y. Wong}.} \bibinfo{year}{2021}\natexlab{}.
\newblock \showarticletitle{Tactics of {Soft} {Resistance} in {User} {Experience} {Professionals}’ {Values} {Work}}.
\newblock \bibinfo{journal}{\emph{Proceedings of the ACM Human-Computer Interaction}} \bibinfo{volume}{5}, \bibinfo{number}{CSCW2} (\bibinfo{date}{Oct.} \bibinfo{year}{2021}).
\newblock
\urldef\tempurl%
\url{https://doi.org/10.1145/3479499}
\showDOI{\tempurl}


\bibitem[Wong et~al\mbox{.}(2023)]%
        {wong_seeing_2023}
\bibfield{author}{\bibinfo{person}{Richmond~Y. Wong}, \bibinfo{person}{Michael~A. Madaio}, {and} \bibinfo{person}{Nick Merrill}.} \bibinfo{year}{2023}\natexlab{}.
\newblock \showarticletitle{Seeing {Like} a {Toolkit}: {How} {Toolkits} {Envision} the {Work} of {AI} {Ethics}}.
\newblock \bibinfo{journal}{\emph{Proceedings of the ACM on Human-Computer Interaction}} \bibinfo{volume}{7}, \bibinfo{number}{CSCW1} (\bibinfo{year}{2023}), \bibinfo{pages}{145:1--145:27}.
\newblock
\urldef\tempurl%
\url{https://doi.org/10.1145/3579621}
\showDOI{\tempurl}


\bibitem[Wong-Villacres et~al\mbox{.}(2021)]%
        {wong-villacres_lessons_2021}
\bibfield{author}{\bibinfo{person}{Marisol Wong-Villacres}, \bibinfo{person}{Adriana~Alvarado Garcia}, \bibinfo{person}{Karla Badillo-Urquiola}, \bibinfo{person}{Mayra Donaji~Barrera Machuca}, \bibinfo{person}{Marianela~Ciolfi Felice}, \bibinfo{person}{Laura~S. Gaytán-Lugo}, \bibinfo{person}{Oscar~A. Lemus}, \bibinfo{person}{Pedro Reynolds-Cuéllar}, {and} \bibinfo{person}{Monica Perusquía-Hernández}.} \bibinfo{year}{2021}\natexlab{}.
\newblock \showarticletitle{Lessons from {Latin} {America}: embracing horizontality to reconstruct {HCI} as a pluriverse}.
\newblock \bibinfo{journal}{\emph{Interactions}} \bibinfo{volume}{28}, \bibinfo{number}{2} (\bibinfo{date}{March} \bibinfo{year}{2021}), \bibinfo{pages}{56--63}.
\newblock
\showISSN{1072-5520}
\urldef\tempurl%
\url{https://doi.org/10.1145/3447794}
\showDOI{\tempurl}


\bibitem[Yang et~al\mbox{.}(2019a)]%
        {yang_sketching_2019}
\bibfield{author}{\bibinfo{person}{Qian Yang}, \bibinfo{person}{Justin Cranshaw}, \bibinfo{person}{Saleema Amershi}, \bibinfo{person}{Shamsi~T. Iqbal}, {and} \bibinfo{person}{Jaime Teevan}.} \bibinfo{year}{2019}\natexlab{a}.
\newblock \showarticletitle{Sketching {NLP}: {A} {Case} {Study} of {Exploring} the {Right} {Things} {To} {Design} with {Language} {Intelligence}}. In \bibinfo{booktitle}{\emph{Proceedings of the 2019 {CHI} {Conference} on {Human} {Factors} in {Computing} {Systems}}} \emph{(\bibinfo{series}{{CHI} '19})}. \bibinfo{publisher}{Association for Computing Machinery}, \bibinfo{address}{New York, NY, USA}, \bibinfo{pages}{1--12}.
\newblock
\showISBNx{978-1-4503-5970-2}
\urldef\tempurl%
\url{https://doi.org/10.1145/3290605.3300415}
\showDOI{\tempurl}


\bibitem[Yang et~al\mbox{.}(2020)]%
        {yang_re-examining_2020}
\bibfield{author}{\bibinfo{person}{Qian Yang}, \bibinfo{person}{Aaron Steinfeld}, \bibinfo{person}{Carolyn Rosé}, {and} \bibinfo{person}{John Zimmerman}.} \bibinfo{year}{2020}\natexlab{}.
\newblock \showarticletitle{Re-examining {Whether}, {Why}, and {How} {Human}-{AI} {Interaction} {Is} {Uniquely} {Difficult} to {Design}}. In \bibinfo{booktitle}{\emph{Proceedings of the 2020 {CHI} {Conference} on {Human} {Factors} in {Computing} {Systems}}} \emph{(\bibinfo{series}{{CHI} '20})}. \bibinfo{publisher}{Association for Computing Machinery}, \bibinfo{address}{New York, NY, USA}, \bibinfo{pages}{1--13}.
\newblock
\showISBNx{978-1-4503-6708-0}
\urldef\tempurl%
\url{https://doi.org/10.1145/3313831.3376301}
\showDOI{\tempurl}


\bibitem[Yang et~al\mbox{.}(2019b)]%
        {yang_unremarkable_2019}
\bibfield{author}{\bibinfo{person}{Qian Yang}, \bibinfo{person}{Aaron Steinfeld}, {and} \bibinfo{person}{John Zimmerman}.} \bibinfo{year}{2019}\natexlab{b}.
\newblock \showarticletitle{Unremarkable {AI}: {Fitting} {Intelligent} {Decision} {Support} into {Critical}, {Clinical} {Decision}-{Making} {Processes}}. In \bibinfo{booktitle}{\emph{Proceedings of the 2019 {CHI} {Conference} on {Human} {Factors} in {Computing} {Systems}}} \emph{(\bibinfo{series}{{CHI} '19})}. \bibinfo{publisher}{Association for Computing Machinery}, \bibinfo{address}{New York, NY, USA}, \bibinfo{pages}{1--11}.
\newblock
\showISBNx{978-1-4503-5970-2}
\urldef\tempurl%
\url{https://doi.org/10.1145/3290605.3300468}
\showDOI{\tempurl}


\bibitem[Yang et~al\mbox{.}(2023)]%
        {yang_designing_2023}
\bibfield{author}{\bibinfo{person}{Qian Yang}, \bibinfo{person}{Richmond Wong}, \bibinfo{person}{Thomas Gilbert}, \bibinfo{person}{Margaret Hagan}, \bibinfo{person}{Steven Jackson}, \bibinfo{person}{Sabine Junginger}, {and} \bibinfo{person}{John Zimmerman}.} \bibinfo{year}{2023}\natexlab{}.
\newblock \showarticletitle{Designing {Technology} and {Policy} {Simultaneously}: {Towards} {A} {Research} {Agenda} and {New} {Practice}}.
\newblock
\urldef\tempurl%
\url{https://doi.org/10.1145/3544549.3573827}
\showDOI{\tempurl}


\bibitem[Yang et~al\mbox{.}(2024)]%
        {Yang_HCIPolicy_CHI24}
\bibfield{author}{\bibinfo{person}{Qian Yang}, \bibinfo{person}{Richmond~Y. Wong}, \bibinfo{person}{Steven Jackson}, \bibinfo{person}{Sabine Junginger}, \bibinfo{person}{Margaret~D. Hagan}, \bibinfo{person}{Thomas Gilbert}, {and} \bibinfo{person}{John Zimmerman}.} \bibinfo{year}{2024}\natexlab{}.
\newblock \showarticletitle{The Future of HCI-Policy Collaboration}. In \bibinfo{booktitle}{\emph{Proceedings of the 2024 CHI Conference on Human Factors in Computing Systems}} (Honolulu, HI, USA) \emph{(\bibinfo{series}{CHI '24})}. \bibinfo{publisher}{Association for Computing Machinery}, \bibinfo{address}{New York, NY, USA}, Article \bibinfo{articleno}{820}, \bibinfo{numpages}{15}~pages.
\newblock
\showISBNx{9798400703300}
\urldef\tempurl%
\url{https://doi.org/10.1145/3613904.3642771}
\showDOI{\tempurl}


\bibitem[Yildirim et~al\mbox{.}(2022)]%
        {yildirim_how_2022}
\bibfield{author}{\bibinfo{person}{Nur Yildirim}, \bibinfo{person}{Alex Kass}, \bibinfo{person}{Teresa Tung}, \bibinfo{person}{Connor Upton}, \bibinfo{person}{Donnacha Costello}, \bibinfo{person}{Robert Giusti}, \bibinfo{person}{Sinem Lacin}, \bibinfo{person}{Sara Lovic}, \bibinfo{person}{James~M O'Neill}, \bibinfo{person}{Rudi~O'Reilly Meehan}, \bibinfo{person}{Eoin Ó~Loideáin}, \bibinfo{person}{Azzurra Pini}, \bibinfo{person}{Medb Corcoran}, \bibinfo{person}{Jeremiah Hayes}, \bibinfo{person}{Diarmuid~J Cahalane}, \bibinfo{person}{Gaurav Shivhare}, \bibinfo{person}{Luigi Castoro}, \bibinfo{person}{Giovanni Caruso}, \bibinfo{person}{Changhoon Oh}, \bibinfo{person}{James McCann}, \bibinfo{person}{Jodi Forlizzi}, {and} \bibinfo{person}{John Zimmerman}.} \bibinfo{year}{2022}\natexlab{}.
\newblock \showarticletitle{How {Experienced} {Designers} of {Enterprise} {Applications} {Engage} {AI} as a {Design} {Material}}. In \bibinfo{booktitle}{\emph{{CHI} {Conference} on {Human} {Factors} in {Computing} {Systems}}}. \bibinfo{publisher}{ACM}, \bibinfo{address}{New Orleans LA USA}, \bibinfo{pages}{1--13}.
\newblock
\showISBNx{978-1-4503-9157-3}
\urldef\tempurl%
\url{https://doi.org/10.1145/3491102.3517491}
\showDOI{\tempurl}


\bibitem[Yovanovic et~al\mbox{.}(2021)]%
        {yovanovic_remote_2021}
\bibfield{author}{\bibinfo{person}{Ivania Yovanovic}, \bibinfo{person}{Iñaki Goñi}, {and} \bibinfo{person}{Constanza Miranda}.} \bibinfo{year}{2021}\natexlab{}.
\newblock \showarticletitle{Remote {Usability} {Assessment} of {Topic} {Visualization} {Interfaces} with {Public} {Participation} {Data}: {A} {Case} {Study}}.
\newblock \bibinfo{journal}{\emph{JeDEM - eJournal of eDemocracy and Open Government}} \bibinfo{volume}{13}, \bibinfo{number}{1} (\bibinfo{date}{Aug.} \bibinfo{year}{2021}), \bibinfo{pages}{101--126}.
\newblock
\showISSN{2075-9517}
\urldef\tempurl%
\url{https://doi.org/10.29379/jedem.v13i1.640}
\showDOI{\tempurl}
\newblock
\shownote{Number: 1}.


\end{thebibliography}

%\appendix

\end{document}